\makeatletter
\def\input@path{{latex/}}
\makeatother
\documentclass[cernpreprint,atlasdraft=false,texlive=2016,USenglish,texmf,texmf]{atlasdoc}
 
\usepackage[subcaption]{atlaspackage}
\usepackage[backref=false]{atlasbiblatex}
 
\usepackage{atlasphysics}
 
\graphicspath{{logos/}{figures/}}
 
\usepackage{ANA-EXOT-2018-06-PAPER-defs}

\usepackage{hhline}
\usepackage{multirow}
\usepackage{tabularx}

\newcolumntype{L}{>{\raggedright\arraybackslash}X}
\newcolumntype{C}{>{\centering\arraybackslash}X}
 
\usepackage{lscape}

\AtlasTitle{Search for new phenomena in events with an energetic jet and missing transverse momentum in $pp$ collisions at $\sqrt{s} = 13$~TeV with the ATLAS detector}
\AtlasAbstract{
Results of a search for new physics in final states with an
energetic jet and large missing transverse momentum are reported.
The search uses proton--proton collision data corresponding to an
integrated luminosity of 139~fb$^{-1}$ at
a center-of-mass energy of 13~TeV collected
in the period 2015--2018 with the ATLAS detector at the Large Hadron Collider.
Compared to previous publications, in addition to an increase of almost a factor of four in the data size, the analysis implements a number of improvements in the signal selection and the background determination leading to enhanced sensitivity.
Events are required to have at least one jet with transverse
momentum above 150~GeV and no reconstructed leptons ($e$, $\mu$ or $\tau$) or photons.  Several signal regions are
considered with increasing requirements on the missing transverse momentum
starting at 200~GeV.
Overall agreement is observed between the number of events in data and the
Standard Model predictions. Model-independent $95\%$  confidence-level limits on
visible cross sections for new processes are obtained in the range between 736~fb and 0.3~fb. 
Results are also translated into improved exclusion limits in models with
pair-produced weakly interacting dark-matter candidates, large extra
spatial dimensions, supersymmetric particles in several
compressed scenarios, axion-like particles, and new scalar particles in dark-energy-inspired models.
In addition, the data are translated into bounds on the invisible branching ratio of the Higgs boson.
}
 
\author{The ATLAS Collaboration}
 
\AtlasJournalRef{Phys. Rev. D 103 (2021) 112006}
\PreprintIdNumber{CERN-EP-2020-238}
\AtlasDOI{10.1103/PhysRevD.103.112006}

\AtlasJournal{Physical Review D}
\hypersetup{pdftitle={ATLAS document},pdfauthor={The ATLAS Collaboration}}
 
\begin{document}
 
\maketitle

\section{Introduction}
\label{sec:introduction}
 
\begin{figure}[!ht]
\begin{center}
 
\begin{subfigure}{0.25\textwidth}
\includegraphics[width=\textwidth]{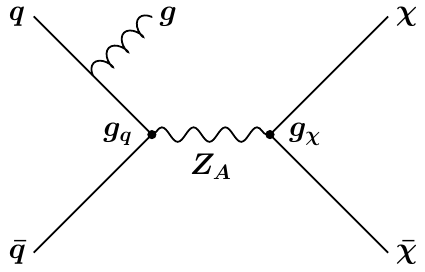}
\caption{}
\label{fig1a}
\end{subfigure}
\begin{subfigure}{0.25\textwidth}
\includegraphics[width=\textwidth]{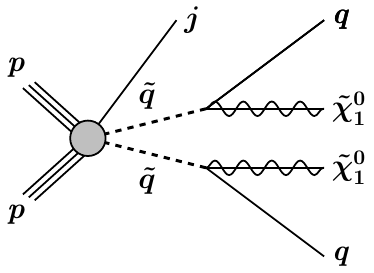}
\caption{}
\label{fig1b}
\end{subfigure}
\begin{subfigure}{0.25\textwidth}
\includegraphics[width=\textwidth]{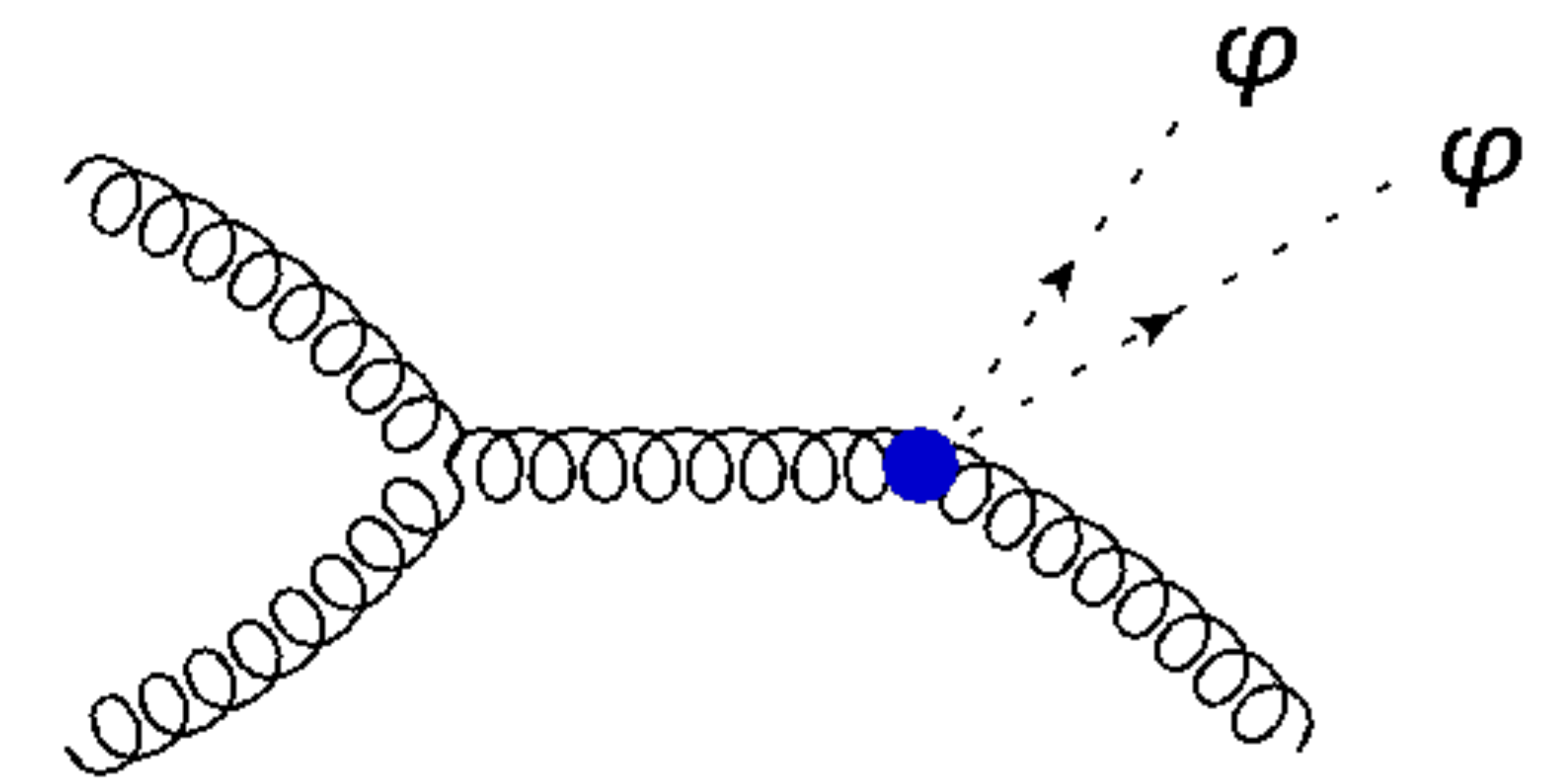}
\caption{}
\label{fig1c}
\end{subfigure}
\end{center}
\caption{
\protect\subref{fig1a} Diagram for the pair production of weakly interacting massive particles $\chi$,
with a mediator $Z_A$ with axial-vector couplings exchanged
in the $s$-channel.
\protect\subref{fig1b}
A generic diagram for the pair production of squarks
with the decay mode $\tilde{q} \to q  + \ninoone$.
The presence of a jet from initial-state radiation 
is indicated for illustration purposes.
\protect\subref{fig1c} Diagram for the pair production of dark-energy scalar fields $\varphi$ in association with an energetic jet in the final state.
}
\label{fig:feynman}
\end{figure}

This paper
presents the results of a new search for new phenomena in events
containing an energetic jet and large missing transverse momentum \ptmiss (with magnitude $\met$) in proton--proton collisions at a centre-of-mass energy $\sqrt{s}=\SI{13}{TeV}$ recorded by the ATLAS detector at the Large Hadron Collider (LHC).  The final-state monojet signature
of at least one energetic jet, 
large \met and no leptons
constitutes a distinctive signature for new physics beyond the Standard Model
(SM) at colliders.
This signature has been extensively studied at the LHC in the context of searches for large extra spatial dimensions (LED), supersymmetry (SUSY),  weakly interacting massive particles (WIMPs) as candidates for dark matter (DM)~\cite{Aaboud:2016tnv,Sirunyan:2017hci,Aaboud:2017buf,Aaboud:2017phn,Khachatryan:2014rra}, and signals from models inspired by dark energy (DE) with new scalar particles in the final state~\cite{Aaboud:2019yqu}. In addition,  experimental results have been reinterpreted in terms of new theoretical scenarios with axion-like particles~\cite{Mimasu:2014nea}. Finally, the monojet final-state results have been used to constrain the invisible branching fraction of the Higgs boson~\cite{EXOT-2013-13,Khachatryan:2016whc}. In the following, the different models are discussed briefly. Figure~\ref{fig:feynman} shows diagrams for some of the models.

The existence of a non-baryonic form of
matter is well established from a number of astronomical
observations~\cite{Trimble:1987ee,Bertone:2004pz,Feng:2010gw}.  The existence of a new,
weakly interacting massive particle is often hypothesized~\cite{Steigman:1984ac},
as it can lead to the correct relic density for nonrelativistic matter in the early
universe~\cite{Kolb:1990vq} as measured
from data from the
Planck~\cite{Planck2018} and
WMAP~\cite{2013ApJS..208...19H} Collaborations.
For WIMP masses below \SI{1}{TeV},
WIMPs may be pair-produced at the LHC.  Traditionally, a monojet final state has been considered a golden channel
for the discovery of WIMPs at colliders. In this case,  the WIMP pair is produced in association with a jet of particles
from initial-state radiation, leading to the  signature of a jet and missing transverse momentum (see Figure~\ref{fig:feynman}(a)).
Results are presented for simplified  DM models~\cite{Abdallah:2015ter,Abercrombie:2015wmb,Buchmueller:2014yoa} where Dirac fermion WIMPs ($\chi$) are pair-produced
from quarks via $s$-channel exchange of a spin-1 mediator particle ($Z_A$) or a spin-0 mediator particle  ($Z_P$) with
axial-vector or pseudoscalar couplings, respectively.
In the case of the axial-vector mediator model with couplings of the mediator to WIMPs and SM quarks set to $g_\chi=1$ and $g_q=1/4$, respectively,
mediator masses below $1.55~\TeV$ have been already excluded at 95$\%$ confidence level (CL) for very light WIMPs in previous analyses~\cite{Aaboud:2017phn}.

Supersymmetry is a theory of  physics beyond the SM which can
solve the hierarchy problem in a natural way and can provide candidates for dark
matter~\cite{Golfand:1971iw,Volkov:1973ix,Wess:1974tw,Wess:1974jb,Ferrara:1974pu,Salam:1974ig}. SUSY introduces a new supersymmetric partner (sparticle)
for each particle in the SM. Specifically, a new scalar field is
associated with each quark chirality state.  Two squark
mass eigenstates $\tilde{q}_1$ and $\tilde{q}_2$ result from the
mixing of the scalar fields for a particular flavor.
In supersymmetric extensions of the SM that assume R-parity
conservation~\cite{Farrar:1978xj,Goldberg:1983nd,Ellis:1983ew},
sparticles are produced in pairs and the lightest supersymmetric particle
(LSP) is stable.
The LSP is assumed to be the lightest
neutralino $\ninoone$.
The results are interpreted in terms of
searches for squark production using simplified models
in scenarios for which the mass difference $\Delta m \equiv m_{\tilde{q}} - m_{\ninoone}$
is small (compressed-mass scenario).
In this case, the $\pt$ of the resulting quark jets and the $\met$ in the final state are both small, making it difficult to reconstruct the SUSY signal.  The monojet signature provides  unique access to this parameter space, for which the presence of jets from initial-state radiation is used to identify signal events, leading to larger $\met$ (see Figure~\ref{fig:feynman}(b)).  In the case of bottom-squark (sbottom) and top-squark (stop) pair production in a compressed-mass supersymmetric scenario, squark masses
below about $430~\GeV$ have been already excluded at 95$\%$~CL~\cite{Aaboud:2017phn}.

The origin of the accelerating expansion of the universe~\cite{Riess:1998cb,Perlmutter:1998np} is,  together with the nature of the dark matter,  a major open question in cosmology.
The theoretical understanding of the accelerating expansion of the universe in terms of fundamental physics,  beyond the ad hoc adoption of a cosmological constant in general relativity,  often involves the introduction of additional scalars interacting with both the gravity and matter fields~\cite{Brax:2016did}. Here an effective field theory implementation of  the Horndeski theories~\cite{Horndeski1974} is considered~\cite{Georgi1993}, introducing a new dark-energy scalar field $\varphi$, governed by an effective mass $M_2$ and a coupling $g_{*}$ to matter, which is considered universal.
For the model relevant for this case, the new scalar particle is stable and is produced in pairs, leaving the experiment undetected.  When they are produced in association with an energetic gluon,  it leads to a monojet final-state topology (see Figure~\ref{fig:feynman}(c)).
Previous results~\cite{Aaboud:2019yqu} indicate no sensitivity for $g_{*} \leq 1.8$, and values of $M_2$ below 1.2~\TeV\
have been excluded at $95\%$ CL for $g_{*} \geq 3.5$.

Large extra spatial dimensions have been postulated to explain the large difference between the
electroweak unification scale at $O(10^2)~\GeV$ and the Planck scale
$M_{\mathrm{Pl}}$ at $O(10^{19})~\GeV$.  In the Arkani-Hamed, Dimopoulos, and
Dvali (ADD) model of LED~\cite{ArkaniHamed:1998rs}, the presence of $n$
extra spatial dimensions of size $R$ leads to a fundamental
Planck scale in $4+n$ dimensions given by ${M_{\mathrm{Pl}}}^2 \sim {M_D}^{2+n}R^n$,
where $M_D$ is the fundamental scale of the $4+n$-dimensional theory.
The extra spatial dimensions are compactified, resulting in a
Kaluza--Klein tower of massive graviton modes (KK graviton).  If produced in high-energy
proton--proton collisions in association with a jet of hadrons, a KK graviton escaping into the extra dimensions can
be inferred from $\met$, and can lead to a monojet event signature.
Values of $M_D$ below $7.7~\TeV$ at $n=2$ and below $4.8~\TeV$ at $n=6$ have been already excluded at
95$\%$~CL~\cite{Aaboud:2017phn}.

New pseudoscalar bosons, referred to as axion-like particles (ALPs), are introduced in different models involving the breaking of  additional U(1) symmetries. The existence of axions was postulated~\cite{Peccei:1977hh} in order to address the strong CP problem for which an anomalous global U(1) symmetry is spontaneously broken. Axions are candidates for explaining the dark-matter content of the universe. The strength of the coupling between the axions and the ordinary matter is governed by the electroweak symmetry breaking scale,  leading to rather strong constraints.  Other models with ALPs, going beyond the minimal QCD axion realization, consider much weaker interactions suppressed by a much higher scale~\cite{Brivio:2017ije,Mimasu:2014nea}. In this paper, an effective implementation of an ALP model (with an effective scale $f_a$) is considered,  in which ALPs are   produced in association with a gluon in a final state governed by an ALP--gluon coupling $\CG$. By construction,  ALP decays are suppressed and the ALP leaves the detector undetected, leading to a monojet final-state topology.

A variety of models of WIMP dark matter at the LHC involve the Higgs boson acting as a portal between the dark sector and the SM sector,
either via direct Yukawa couplings to fermionic dark-matter candidates or via other mechanisms.  The decay of the Higgs boson into  dark-matter particles translates into a signature of  $\met$ in the final state.  Searches for invisible Higgs boson decays have been carried out at ATLAS and CMS, considering different SM Higgs production processes and different center-of-mass energies,  leading to a 95$\%$~CL upper limit on the invisible Higgs boson branching ratio of 0.26~\cite{HIGG-2018-54} and 0.19~\cite{CMS-HIG-17-023}, respectively.

In this publication, a
data sample corresponding to a total integrated luminosity of $139\,\ifb$ is used, and the analysis strategy closely follows that of the previous publication based on  $36.1\,\ifb$~\cite{Aaboud:2017phn}.  In addition, a number of improvements are implemented leading to  enhanced sensitivity to new phenomena.  The $\pt$ requirements for identifying electrons and muons in the final state are lowered,  translating into tighter lepton vetoes and a larger background reduction, which is also complemented with the inclusion of  $\tau$-lepton and photon  vetoes.  The kinematic range covered by the analysis is extended towards lower values of $\met$ and leading-jet $\pt$, and new control regions are defined for a better determination of backgrounds related to top-quark and $Z$-boson production processes. Finally, the analysis profits from improved theoretical predictions for $W$+jets and $Z$+jets production, including higher-order corrections at next-to-next-to-leading order in QCD and next-to-leading order in electroweak couplings supplemented by Sudakov logarithms at two loops.
 
The paper
is organized as follows. The ATLAS detector is described in the
next Section.  Section~\ref{sec:simulation} provides details of the Monte Carlo simulations
used in the analysis for background and signal processes.
Section~\ref{sec:reconstruction} discusses the reconstruction
and identification
of jets, leptons,
and missing transverse momentum, while Section~\ref{sec:selection} describes the
event selection.  The estimation of background contributions and the
study of systematic uncertainties are discussed in Sections~\ref{sec:background}
and~\ref{sec:systematics}.  The results are presented in Section~\ref{sec:results}
and are interpreted in terms of limits in models of WIMP-pair production,
ADD, SUSY in compressed scenarios, axion-like particles, new bosons in DE-inspired models, and limits on the Higgs boson invisible branching fraction.
Finally, Section~\ref{sec:conclusions} is devoted to the conclusions.


\section{ATLAS detector}
 
 
\newcommand{\AtlasCoordFootnote}{
ATLAS uses a right-handed coordinate system with its origin at the nominal interaction point (IP)
in the center of the detector and the \(z\)-axis along the beam pipe.
The \(x\)-axis points from the IP to the center of the LHC ring,
and the \(y\)-axis points upwards.
Cylindrical coordinates \((r,\phi)\) are used in the transverse plane,
\(\phi\) being the azimuthal angle around the \(z\)-axis.
The pseudorapidity is defined in terms of the polar angle \(\theta\) as \(\eta = -\ln \tan(\theta/2)\).
Angular distance is measured in units of \(\Delta R \equiv \sqrt{(\Delta\eta)^{2} + (\Delta\phi)^{2}}\).}
 
The ATLAS detector~\cite{PERF-2007-01} at the LHC covers nearly the entire solid angle around the collision point.\footnote{\AtlasCoordFootnote}
It consists of an inner tracking detector surrounded by a thin superconducting solenoid, electromagnetic and hadronic calorimeters,
and a muon spectrometer incorporating three large superconducting toroidal magnets.
 
The inner-detector system 
is immersed in a \SI{2}{\tesla} axial magnetic field
and provides charged-particle tracking in the range \(|\eta| < 2.5\).
The high-granularity silicon pixel detector covers the vertex region and typically provides four measurements per track,
the first hit normally being in the insertable B-layer 
installed before Run~2~\cite{ATLAS-TDR-2010-19,PIX-2018-001}.
It is followed by the silicon microstrip tracker, 
which usually provides eight measurements per track.
These silicon detectors are complemented by the transition radiation tracker (TRT),
which enables radially extended track reconstruction up to \(|\eta| = 2.0\).
The TRT also provides electron identification information
based on the fraction of hits (typically 30 in total) above a higher energy-deposit threshold corresponding to transition radiation.

The calorimeter system covers the pseudorapidity range \(|\eta| < 4.9\).
Within the region \(|\eta|< 3.2\), electromagnetic calorimetry is provided by barrel and
endcap high-granularity lead/liquid-argon (LAr) calorimeters,
with an additional thin LAr presampler covering \(|\eta| < 1.8\)
to correct for energy loss in material upstream of the calorimeters.
Hadronic calorimetry is provided by the steel/scintillator-tile calorimeter,
segmented into three barrel structures within \(|\eta| < 1.7\), and two copper/LAr hadronic endcap calorimeters.
The solid angle coverage is completed with forward copper/LAr and tungsten/LAr calorimeter modules
optimized for electromagnetic and hadronic measurements respectively.
 
The muon spectrometer 
comprises separate trigger and
high-precision tracking chambers measuring the deflection of muons in a magnetic field generated by the superconducting air-core toroids.
The field integral of the toroids ranges between \num{2.0} and \SI{6.0}{\tesla\metre}
across most of the detector.
A set of precision chambers covers the region \(|\eta| < 2.7\) with three layers of monitored drift tubes,
complemented by cathode-strip chambers in the forward region, where the background is highest.
The muon trigger system covers the range \(|\eta| < 2.4\) with resistive-plate chambers in the barrel, and thin-gap chambers in the endcap regions.
 
Events of interest are selected to be recorded by the first-level trigger system implemented in custom hardware,
followed by selections made by algorithms implemented in software in the high-level trigger~\cite{TRIG-2016-01}.
The first-level trigger accepts events from the \SI{40}{\MHz} bunch crossings at a rate below \SI{100}{\kHz},
which the high-level trigger reduces in order to record events to disk at about \SI{1}{\kHz}.
 
\section{Monte Carlo simulation}
\label{sec:simulation}
 
 
Monte Carlo (MC) simulated event samples are used to compute detector
acceptance and reconstruction efficiencies, determine signal and
background contributions, and estimate systematic uncertainties in the
final results.  The SM background samples were processed with the
full ATLAS detector
simulation~\cite{Aad:2010ah} based on $\GEANT4$~\cite{Agostinelli:2002hh}.
Signal simulated samples, with the exception of those for Higgs production for which full simulation is used,
were processed with
a fast simulation using a parameterization of the calorimeter
response and $\GEANT4$ for the other parts of the detector.
Simulated events are then reconstructed and analyzed with the same
analysis chain as for the data, using the same trigger and event
selection criteria.  The effects of multiple proton--proton
interactions in the same or neighboring bunch-crossings (pileup) were taken into
account by overlaying the hard-scattering process with
simulated minimum-bias events, distributed according to the frequency in data and generated by
{\PYTHIA}~8.186~\cite{pythia} with the A3 set of tuned parameters (tune)~\cite{ATL-PHYS-PUB-2016-017} and the NNPDF2.3LO parton distribution function (PDF) set~\cite{Ball:2014uwa}.  Correction factors are applied to the Monte Carlo simulation to account for differences between simulation and the data in pileup, the energy and momentum scales, and reconstruction and identification efficiencies of physics objects.
 
\subsection{Signal simulation}

WIMP $s$-channel signal samples were simulated in
\POWHEGBOXV{v2}~\cite{Alioli:2010xd,Frixione:2007vw,Nason:2004rx} (revision 3049)
using two implementations of simplified models, introduced in
Ref.~\cite{Haisch:2013ata}.  The DMV model of WIMP-pair production was
used for $s$-channel spin-1 axial-vector mediator exchange at NLO in the strong coupling constant, and the {DMS\_tloop} model was used for
WIMP-pair production with $s$-channel spin-0 pseudoscalar mediator
exchange with the full quark-loop calculation at leading order (LO)~\cite{Haisch:2015ioa}.
Renormalization and
factorization scales were set to $H_\mathrm{T}/2$ on an event-by-event
basis, where $H_\mathrm{T}=\sqrt{m_{\chi\chi}^2+p_{\mathrm{T}, j1}^2}+p_{\mathrm{T}, j1}$
is defined by the invariant mass of the WIMP pair ($m_{\chi\chi}$) and the transverse momentum of the
highest-$p_{\mathrm{T}}$ parton-level jet
($p_{\mathrm{T}, j1}$).  The mediator propagator is described by a
Breit--Wigner distribution.
Events were generated using the NNPDF30~\cite{Ball:2014uwa} PDFs and interfaced to {\PYTHIA}~8.205 with
the A14 tune~\cite{ATL-PHYS-PUB-2014-021} for parton showering,
hadronization and the underlying event.  Couplings of the
mediator to WIMP particles and those of the SM quarks were set to $g_\chi=1$ and
$g_q=1/4$, respectively, for the axial-vector  mediator model whereas both couplings were set to one in the
case of the pseudoscalar mediator model, following the conventions of the LHC DM Working Group~\cite{Abdallah:2015ter,Abercrombie:2015wmb}.
Each model was simulated for a range of possible WIMP and mediator masses, with  WIMP masses ranging from $1~\GeV$ to $1~\TeV$ and mediator masses between
$10~\GeV$ and $10~\TeV$.

SUSY signals for squark-pair production
were generated with \MGMCatNLO~v2.2.3~\cite{Alwall:2014hca}
and interfaced to {\PYTHIA}~8.186 with the A14 tune for modeling of
the squark decay, parton showering, hadronization, and the underlying
event. The PDF set used for the generation was NNPDF23LO, and the
renormalization and factorization scales were set to
$\mu = \sum_i \sqrt{m_i^2 + p_{\mathrm{T},i}^2}$, where the sum runs
over all final-state particles from the hard-scatter process.
The matrix-element calculation was performed at tree level, and
includes the emission of up to two
additional partons. Matching
to parton-shower calculations was accomplished by using the
CKKW-L prescription~\cite{ckkw-l}, with a matching scale
set to one quarter of the pair-produced superpartner mass.
All signal cross sections were calculated to approximate
next-to-next-to-leading order (NNLO) in the strong coupling constant, adding the resummation of soft
gluon emission at next-to-next-to-leading-logarithm accuracy (approximate NNLO+NNLL)~\cite{Beenakker:1997ut,Beenakker:2010nq,Beenakker:2016lwe,Beenakker:2016gmf}.
The
nominal cross section and its uncertainty were taken from an envelope
of cross-section predictions using different PDF sets and
factorization and renormalization scales, as discussed in
Ref.~\cite{Borschensky:2014cia}.
Simulated samples were produced with squark masses in the range between
$250~\GeV$ and $1.2~\TeV$, and squark--neutralino mass differences $\Delta m$ varying between $5~\GeV$ and
$50~\GeV$.

Simulated events for the dark-energy model were generated using an effective field theory implementation~\cite{Brax:2016did} in \MGMCatNLO~v2.6.1 at
LO accuracy in the strong coupling constant.
Following the work in Ref.~\cite{Aaboud:2019yqu}, only terms corresponding to the $\mathcal{L}_2$ operator, relevant for the monojet final-state topology,  were considered, with the Wilson coefficient $c_2 =1$ and  the
rest of the Wilson coefficients ($c_i$) set to zero. Electroweak terms were vetoed and only one insertion of a $\mathcal{L}_2$ operator in each diagram was allowed.  The generated events were
interfaced to {\PYTHIA}~8.240 with the A14 tune for modeling of parton showering, hadronization, and the underlying
event. The PDF set used for the generation was NNPDF23LO, and the
renormalization and factorization scales were set to
$0.5 \times H_\mathrm{T}=\sqrt{m_{\varphi\varphi}^2+p_{\mathrm{T}, j1}^2}+p_{\mathrm{T}, j1}$, where $m_{\varphi\varphi}$ is the invariant mass of the two scalar particles in the final state.
The dark-energy field mass and the coupling to gluons were set to $m_\varphi = 100$~\MeV\ and $g_{*}=4\pi$, respectively.
Effective scales $M_2$ up to 3~\TeV\ are explored.

Simulated samples for the ADD LED model with different numbers of extra
dimensions in the range $n = 2$--$6$ and a fundamental scale $M_D$ in the range $3$--$12~\TeV$
were generated using {\PYTHIA}~8.205 with the A14 tune and NNPDF2.3LO
PDFs. The cross section is computed at next-to-leading order (NLO) accuracy in the strong coupling constant.
The renormalization scale was set to the geometric mean of the squared
transverse masses of the two produced particles, $\sqrt{(p_{\mathrm{T},G}^2+ m_G^2)(p_{\mathrm{T},p}^2 + m_p^2)}$, where  $p_{\mathrm{T},G}$ and $m_G$
($p_{\mathrm{T},p}$ and $m_p$) denote, respectively, the
transverse momentum and the mass of the KK graviton (parton) in the final state. The
factorization scale was set to the smaller of the transverse masses, $\sqrt{\pt^2 + m^2}$, of the KK graviton and the parton.

Samples of simulated events for ALP production in association with a jet~\cite{Brivio:2017ije} were generated
at leading-order (LO) accuracy in the strong coupling constant
with \MGMCatNLO~v2.6.2  and interfaced to {\PYTHIA}~8.240 with the A14 tune for modeling of parton showering, hadronization, and the underlying
event. The PDF set used for the generation was NNPDF23LO, and the
renormalization and factorization scales were set to
half of the transverse mass, $ 0.5 \times \sqrt{\pt^2 +
m^2}$, of the ALP and the parton.  Other processes related to the coupling of the ALP to photons, vector bosons or the Higgs boson are suppressed.
Values for the ALP mass up to  $m_a = 1$~\GeV\ and values for the coupling to gluons up to $\CG=1$ were considered.
Effective scales $f_a$ in the range between 1~\TeV\ and 10~\TeV\ are explored.

Simulated samples for the production of a 125~\GeV\ Higgs boson were generated, with NLO accuracy in QCD emissions, using the \POWHEGBOXV{v2}~\cite{powheg} event generator.
The samples include gluon--gluon fusion processes ($gg \to H$ and $gg \to ZH$), vector-boson fusion (VBF) processes ($VV \to H)$, the associated production with a $W/Z$ boson in the final state ($VH$), and the associated production with a $\ttbar$ pair in the final state ($\ttbar + H$).
The simulated events were interfaced with \PYTHIA~8.212 for parton shower, hadronization and underlying-event modeling using the AZNLO tune~\cite{STDM-2012-23} with
the NNPDF30+CTEQ6L1 PDF in the case of $gg \to H$ and $gg \to ZH$,  CT10 in the case of $VH$, and  NNPDF30 PDFs in the case of $VV \to H$ and $\ttbar + H$ processes.
The $gg \to H$ sample was normalized such that it reproduces the total cross section predicted by a next-to-next-to-next-to-leading-order
(NNNLO) QCD calculation with NLO electroweak (EW) corrections applied, and $VV \to H$ and  $VH$ processes were normalized
to cross sections calculated at NNLO in QCD with NLO EW corrections. The $gg \to ZH$ sample was normalized to  cross sections calculated at NLO in QCD, and the  $\ttbar + H$
sample was normalized to  cross sections calculated at NLO in QCD with NLO EW corrections~\cite{YR4}.
In all cases, the Higgs boson invisible decay $H \to Z^* Z \to 4\,\nu$ is considered because it provides final-state topologies  consistent with those from models of new phenomena with invisibly decaying Higgs bosons.

\subsection{Background simulation}
 
After applying the final-state selection  as described in Section~\ref{sec:selection},
the primary SM background contributing to monojet event signatures is
$\Znunu$ + jets.  There are also significant contributions from $W$ + jets
events, primarily from $\Wtaunu$ + jets, with unidentified leptons in the final state.  Small contributions are expected
from $\Zll$ + jets ($\ell = e, \mu, \tau$), multijet, $t\bar{t}$,
single-top, and diboson ($WW,WZ,ZZ$) processes. Contributions from top-quark
production associated with additional vector bosons ($\ttbar + W$, $\ttbar
+ Z$, or $t + Z + q/b$ processes) are negligible and not considered in this analysis.
As discussed in detail in Section~\ref{sec:background}, the contribution from
SM background processes in the signal regions are determined using simulated samples
constrained with data in control regions.  In the following, the generation of the different simulation samples is described.

Events containing $W$ or $Z$ bosons with associated jets were simulated
using the {\SHERPA}~2.2.1~\cite{sherpa} event generator. Matrix elements (ME)
were calculated for up to two partons at NLO
and four partons at LO using
OpenLoops~\cite{Cascioli:2011va} and
Comix~\cite{Gleisberg:2008fv},
and merged with the {\SHERPA} parton shower
(PS)~\cite{Schumann:2007mg} using the ME+PS@NLO
prescription~\cite{Hoeche:2012yf}. The
NNPDF3.0NNLO~\cite{Ball:2014uwa} PDF
set was used in conjunction with a dedicated
parton-shower tuning developed by the authors of \SHERPA. The MC
predictions were initially normalized to NNLO perturbative QCD (pQCD) predictions according to
DYNNLO~\cite{Catani:2009sm,Catani:2007vq} using the  MSTW2008  90$\%$ CL
NNLO PDF set~\cite{mstw}.
 
In order to improve the description of $W$+jets and $Z$+jets processes, their MC predictions were reweighted to account for
higher-order QCD and electroweak corrections. The reweighting procedure is based on parton-level predictions for $W/Z$+jets production from Ref.~\cite{Lindert2017}, which include NNLO QCD corrections~\cite{Ridder:2015dxa,Ridder:2016nkl,Boughezal:2016isb,Boughezal:2016dtm} and NLO electroweak corrections~\cite{Denner:2011vu,Denner:2012ts,Kallweit:2015dum,Denner:2009gj} supplemented by Sudakov logarithms at two
loops~\cite{Kuhn:2004em,Kuhn:2007qc,Kuhn:2005az,Kuhn:2007cv}. These corrections are provided separately for $W$+jets, $Z\to\ell^+\ell^-$+jets and $Z\to\nu\nu$+jets processes, as a function of the vector-boson $\pt$, in order to improve the description of the measured $Z$-boson $\pt$
distribution~\cite{Aaboud:2017hbk}. The reweighting
procedure takes into account the difference between the QCD NLO
predictions as included already in {\SHERPA} and as provided by the parton-level calculations. Uncertainties in these higher-order corrections and their correlations across processes are described in Section~\ref{sec:systematics}.
 
Separate nonoverlapping samples for $W/Z$+jets production
via VBF-driven processes were generated using \HERWIGpp~(v7.1.3 for electron and $\tau$-lepton decays and v7.2 for muon decays)~\cite{Bahr:2008pv}.
The samples were produced at NLO accuracy in pQCD using VBFNLO v3.0.0~\cite{Baglio:2014uba}. The NNPDF30
PDF set was used along with the default set of tuned parameters for
parton showering, hadronization and the underlying event. The EvtGen~v1.2.0 program~\cite{EvtGen} was used to model the decays of
the bottom and charm hadrons.

For the generation of $\ttbar$ and single-top-quark events in the $Wt$-channel
and $s$-channel, the \POWHEGBOXV{v2}~\cite{powheg} event generator was used with
CT10~\cite{ct10} PDFs. Electroweak $t$-channel single-top-quark events were generated
using the  \POWHEGBOXV{v1} event generator. This event generator uses the four-flavor
scheme to calculate NLO matrix elements, with the CT10 four-flavor
PDF set.  Interference occurring beyond tree level between $Wt$ and $\ttbar$ processes was studied, considering both the diagram subtraction (DS) and diagram removal (DR) production  schemes~\cite{White:2009yt}; DR was used for the nominal background prediction, DS for the evaluation of systematic uncertainties as described in Section~\ref{sec:systematics}. The samples were normalized to NNLO pQCD predictions.
The parton shower, hadronization, and underlying event were
simulated using {\PYTHIA}~8.205 with
the A14 tune.
The top-quark mass was set to $172.5~\GeV$.
The EvtGen~v1.2.0 program was used to model the decays of
the bottom and charm hadrons.
Alternative samples were generated using \MGMCatNLO~(v2.2.1) and \POWHEGBOX
interfaced to \HERWIGpp~(v7.1.3)~\cite{Bahr:2008pv}
in order to estimate the effects of the choice of matrix-element event generator and parton-shower algorithms, and fragmentation and hadronization  effects.
The EvtGen~v1.2.0 program was used to model the decays of
the bottom and charm hadrons.
 
Diboson samples ($WW$, $WZ$,
and $ZZ$ production) were generated using {\SHERPA}~2.2.1 or {\SHERPA}~2.2.2 with NNPDF3.0NNLO,
and were normalized to NLO pQCD predictions~\cite{Campbell:2011bn}.
The EvtGen~v1.2.0 program was used to model the decays of
the bottom and charm hadrons.

 
\section{Event reconstruction}
\label{sec:reconstruction}
 
 
Jets are reconstructed from energy deposits in the calorimeters\cite{PERF-2014-07} using the \antikt
jet algorithm~\cite{Cacciari:2008gp} as provided by the \textsc{fastjet}~\cite{Fastjet} toolkit, with the radius parameter $R=0.4$. The measured jet four-momentum is calibrated using information from both simulation and
data~\cite{JETM-2018-05}.
In  addition, jets are corrected for contributions from pileup.  Jets with $\pt>\SI{20}{GeV}$ and $|\eta|<2.8$ are considered in the analysis. A combination of track-based variables developed to suppress
\pileup jets, called the
jet-vertex tagger (JVT)~\cite{ATLAS-CONF-2014-018}, is constructed. In order to remove jets originating from
pileup collisions, for central jets ($|\eta| < 2.5$) with $\pt < \SI{120}{GeV}$
a significant fraction of the tracks associated with each jet must have an origin
compatible with the primary vertex, as defined by the jet-vertex tagger.
 
Jets with $\pt>\SI{30}{GeV}$ and $|\eta|<2.5$ are identified as jets containing $b$-hadrons (\bjets) if tagged by a multivariate algorithm which uses
information about the impact parameters of inner-detector tracks matched to the
jet, the presence of displaced secondary vertices, and the reconstructed flight
paths of $b$- and $c$-hadrons inside the jet~\cite{PERF-2012-04,ATL-PHYS-PUB-2016-012}. A $60\%$ efficient
\btagging working point, as determined in a simulated sample of $\ttbar$ events,
is chosen.  This corresponds to rejection factors of approximately  $1500$, $35$
and $180$ for light-quark and gluon jets, $c$-jets, and $\tau$-leptons decaying
hadronically, respectively.

Electrons are found by combining energy deposits in the calorimeter
with tracks found in the inner detector. They are initially required
to  have $\pt > \SI{7}{GeV}$ and $|\eta|
<2.47$, and to satisfy the
`Loose' electron shower shape and track selection criteria
described in Ref.~\cite{EGAM-2018-01}, including a requirement on the match between the track and the primary vertex, which requires the  longitudinal impact parameter $|z_0| \sin \theta$ to be less than \SI{0.5}{mm}.
Overlaps between identified electrons
and jets with $\pt > \SI{30}{GeV}$ in the final state are resolved. Jets are discarded if they are not
\btagged and their separation $\Delta R = \sqrt{(\Delta \eta)^2 + (\Delta \phi)^2}$
from an identified electron is less than $0.2$. Otherwise, the electron is removed as it most likely originates from a
semileptonic $b$-hadron decay. The electrons separated by $\Delta R$ between $0.2$
and  $0.4$ from any remaining jet are removed.
 
Muon candidates are
formed by combining information from the muon spectrometer and
inner tracking detectors. They are
required to pass `Medium' identification requirements~\cite{PERF-2015-10}, and
to have  $\pt > \SI{7}{GeV}$ and  $|\eta| < 2.5$. As in the case of electrons, the muon track is required to have $|z_0| \sin \theta<\SI{0.5}{mm}$.
Jets with $\pt > \SI{30}{GeV}$ and fewer than three tracks with $\pt > \SI{500}{MeV}$ associated with them are
discarded if their separation $\Delta R$ from an identified muon is less than $0.4$.
The muon is discarded if it is matched to a jet with
$\pt>\SI{30}{GeV}$ that has at least three tracks
associated with it. If an electron and a muon share the same inner-detector track, the muon is retained and the electron is discarded in order to remove electron candidates originating from muon bremsstrahlung followed by photon conversion.
 
Hadronically decaying \taulepton candidates are formed by combining information from the calorimeters and inner tracking detectors. The \taulepton reconstruction algorithm~\cite{PERF-2013-06} is seeded by reconstructed jets with $\pt>\SI{10}{GeV}$ and $|\eta|<2.5$, and the reconstructed energies of the \taulepton candidates are corrected to the \taulepton energy scale~\cite{ATLAS-CONF-2017-029}. They are required to pass `Loose' identification requirements~\cite{ATL-PHYS-PUB-2015-045}, to have $\pt > \SI{20}{GeV}$ and $|\eta|<2.5$, excluding the transition region between the
electromagnetic barrel and endcap calorimeters ($1.37 < |\eta| < 1.52$), and to have one or three associated charged tracks. The \tauleptons close to electrons or muons ($\Delta R < 0.2$) are removed. Any jet within $\Delta R = 0.2$ of a \taulepton is removed.
 
Photons are reconstructed from clusters of energy deposited in the electromagnetic calorimeter. They are required to pass `Tight' identification requirements~\cite{EGAM-2018-01}, and to have $\pt>\SI{10}{GeV}$ and $|\eta|<2.37$. Photons are discarded if their separation $\Delta R$ from an identified muon or electron is less than $0.4$. Jets are instead discarded if their separation $\Delta R$ from an identified photon is less than $0.4$.

The vector missing transverse momentum \ptmiss is reconstructed from the negative vectorial sum of the transverse momenta of electrons, muons, \tauleptons, photons, and jets with $\pt>\SI{20}{GeV}$ and $|\eta| < 4.5$. 
Tracks compatible with the primary vertex but not associated with any of those objects are also included in the vectorial sum, as described in Section~3.4.2 of Ref.~{\cite{PERF-2016-07}}.
 
\section{Event selection}
\label{sec:selection}
 
 
This analysis is based on data
collected by ATLAS during Run~2 of the LHC,
corresponding to a total integrated luminosity of \mylumi.
The data were collected using a trigger based on a requirement on \met
as computed from calorimetry information at the final stage of the two-level trigger system~\cite{TRIG-2019-01}.
After analysis selections, the trigger was measured to be fully efficient for
events with $\met > \SI{200}{GeV}$, as determined using a data sample with
muons in the final state.
 
Events are required to have at least one reconstructed primary vertex
consistent with the beamspot envelope and containing at least two
associated tracks of $\pt>\SI{500}{MeV}$.  When more than one such vertex is
found, the vertex with the largest summed $\pt^2$ of the associated
tracks is chosen.  Events having identified muons, electrons, photons or \tauleptons 
in the final state are vetoed.
Selected events have $\met > \SI{200}{GeV}$,
a leading jet with $\pt > \SI{150}{GeV}$ and $|\eta| < 2.4$, and
up to three additional jets with $\pt > \SI{30}{GeV}$ and $|\eta| < 2.8$.
Separation in the azimuthal angle of $\Delta\phi(\text{jet},\ptmiss)> 0.4\,(0.6)$ between the missing transverse momentum direction and each
selected jet is required for events with $\met>\SI{250}{GeV}$ ($\SI{200}{GeV}<\met\leq\SI{250}{GeV}$) to reduce the multijet background
contribution, since large $\met$ can originate from jet energy
mismeasurement.
Jet quality criteria~\cite{ATLAS-CONF-2015-029} are imposed, which involve selections based
on quantities such as the pulse shape of the energy depositions measured in the
cells of the calorimeters, electromagnetic energy fraction in the calorimeter,
maximum fraction of
the jet energy collected by a single calorimeter layer, and the charged-particle fraction.\footnote{
The charged-particle fraction is defined as
$f_{{\text{ch}}}=\sum \pt^{{\text{track,jet}}}
/\pt^{{\text{jet}}}$, where $\sum \pt^{{\text{track,jet}}}$ is the scalar sum of the transverse momenta
of tracks associated with the primary vertex within a cone of size
$\Delta R=0.4$ around the jet axis, and $\pt^{{ \text{jet}}}$
is the transverse momentum of the jet as determined from calorimetric measurements.}
Loose selection criteria are applied to all jets with $\pt > \SI{30}{GeV}$
and $|\eta| < 2.8$, which remove anomalous energy depositions due to
coherent noise and electronic noise bursts in the calorimeter~\cite{DAPR-2012-01}.
Events with any jet not satisfying  the loose criteria~\cite{ATLAS-CONF-2015-029}
are discarded.

Noncollision backgrounds, for example energy depositions in the calorimeters
due to muons of beam-induced or cosmic-ray origin, are
suppressed by imposing tight selection criteria on the leading jet: the ratio
of the jet charged-particle fraction to the maximum fraction of
the jet energy collected by a single calorimeter layer,
$f_{{\text{ch}}}/f_{{\text{max}}}$, is required to be larger than $0.1$.
Jet quality requirements altogether have a negligible effect on the signal efficiency.
 
The signal region (SR) is divided into different bins of \met, which are listed in \Tab{\ref{tab:sr}}. Inclusive bins are used for a model-independent interpretation of search results, while the full set of exclusive bins are used for the interpretation within different models of new physics.
\begin{table}
\caption{Intervals and labels of the \met bins used for the signal region. Details are given in the text.}\label{tab:sr}
\begin{footnotesize}
\begin{tabular*}{\textwidth}{@{\extracolsep{\fill}}lcccccccc}\hline
Exclusive (EM)& EM0 & EM1 & EM2  & EM3      & EM4      & EM5      & EM6\\
$\met$ [\GeV] &\!200--250\!&\!250--300\!&\!300--350\!&\!350--400\!&\!400--500\!&\!500--600\!&\!600--700\\
& EM7      & EM8      & EM9      & EM10 & EM11 & EM12 &    \\
&\!700--800\!&\!800--900\!&\!900--1000\!& 1000--1100 & 1100--1200 & $>\!1200$ &\\  \hline
Inclusive (IM)& IM0 & IM1 & IM2  & IM3      & IM4      & IM5      & IM6\\
$\met$ [\GeV] & $>\!200$ & $>\!250$ & $>\!300$ & $>\!350$ & $>\!400$ & $>\!500$& $>\!600$ \\
& IM7      & IM8      & IM9      & IM10 & IM11 & IM12&   \\
& $>\!700$ & $>\!800$ & $>\!900$ & $>\!1000$ & $>\!1100$ & $>\!1200$ & \\ \hline
\end{tabular*}
\end{footnotesize}
\end{table}

 
\section{Background estimation}
\label{sec:background}
 
A semi-data-driven technique, supported by statistically independent control regions, is used to constrain the normalization of Standard Model backgrounds. The approach followed is similar to the one used in previous versions of the analysis~\cite{Aaboud:2017phn}.
 
\subsection{Control regions}\label{sec:CRs}
The estimation of the \Zjets, \Wjets, \ttbar, and \singletop backgrounds is performed using five control regions, as described below. These regions are defined in a way similar to the SR: events are selected in terms of a quantity which is -- similarly to \ptmiss in the SR -- a proxy for the transverse momentum of the system which recoils against the hadronic activity in the event. This quantity is denoted in the following by \myvecmet, and its magnitude by \mymet. The same selection criteria for jet multiplicity and leading jet \pt as in the SR are applied in the control regions, with the same requirements on the azimuthal separation of jets from \myvecmet. Control regions are binned in terms of \mymet, using the same binning as in the signal region (see \Tab{\ref{tab:sr}}). In the signal region, \mymet is equivalent to \met.
 
A control region enriched in \Wmunu events is defined by selecting events that pass the same trigger requirements as in the signal region, if they have exactly one reconstructed muon and this muon has $\pt>\SI{10}{GeV}$ and passes the requirement on the transverse impact parameter significance, $d_0/\sigma(d_0)$, to be less than $3$, and if no electrons, \tauleptons, photons or \bjets are reconstructed. In this region, $\mymet$ is defined as the magnitude of the vector sum of the missing transverse momentum and the muon transverse momentum, $|\ptmiss+\ptvec(\mu)|$, and is required to be higher than \SI{200}{GeV}. An additional requirement on the transverse mass is applied, $\SI{30}{GeV}<\mT<\SI{100}{GeV}$, where $\mT = \sqrt{2\pt(\mu)\pt(\nu) [1-\cos(\Delta\phi(\mu ,\nu))]}$ and the neutrino transverse momentum, $\ptvec(\nu)$, is taken to be the same as \ptmiss.
 
Similarly, a control region enriched in \Zmumu events is defined by selecting events that pass the same trigger requirements but have exactly two reconstructed muons, where these muons have $\pt>\SI{10}{GeV}$ and $d_0/\sigma(d_0)<3$, and the invariant mass of the dimuon system is between $66$ and \SI{116}{GeV}. In this region, $\mymet$ is defined as the magnitude of the vector sum of the missing transverse momentum and  the transverse momentum of the dimuon system, $|\ptmiss+\ptvec(\mu\mu)|$, and is required to be higher than \SI{200}{GeV}. The trigger requirements used for these two regions do not include muon information in the calculation of \met, and are fully efficient for events satisfying the selection criteria.
 
A control region enriched in \Wenu events is defined by selecting events that pass single-electron triggers, if they have exactly one reconstructed electron and this electron satisfies tight identification criteria described in Ref.~{\cite{ATL-PHYS-PUB-2015-045}}, is reconstructed outside the transition region between the electromagnetic barrel and endcap calorimeters, has $\pt>\SI{30}{GeV}$ and $d_0/\sigma(d_0)<5$, and passes the tight isolation requirements based on information from the electromagnetic calorimeter and from tracking detectors, described in Ref.~{\cite{EGAM-2018-01}}. In this region, $\mymet$ is defined as the magnitude of the vector sum of the missing transverse momentum and the electron transverse momentum, $|\ptmiss+\ptvec(e)|$, and is required to be higher than \SI{200}{GeV}. The transverse mass is required to be $\SI{30}{GeV}<\mT<\SI{100}{GeV}$. In order to further suppress backgrounds from multijet processes with jets misidentified as high-\pt electrons, the events are required to have $\met>\SI{70}{GeV}$ and $\met/\sqrt{H_\text{T}}>\SI{5}{GeV^{1/2}}$, where $H_\text{T}$ denotes the scalar sum of the \pt of the identified jets in the final state.
 
Similarly, a control region enriched in \Zee events is defined by selecting events with exactly two reconstructed electrons, where these electrons have $\pt>\SI{30}{GeV}$ and $d_0/\sigma(d_0)<5$, 
and the invariant mass of the dielectron system is between $66$ and \SI{116}{GeV}. In this region, $\mymet$ is defined as the magnitude of the vector sum of the missing transverse momentum and the transverse momentum of the dielectron system, $|\ptmiss+\ptvec(ee)|$, and is required to be higher than \SI{200}{GeV}. The single-electron trigger requirements are fully efficient for events satisfying the selection criteria for these two regions.
 
A control region enriched in \ttbar and \singletop events is defined by selecting events which pass the same cuts as for the \Wmunu and \Wenu regions, but which have at least one identified \bjet.

\Tab{\ref{tab:regions}} shows a summary of the selection criteria for all regions.
 
\begin{table}\small
\centering
\caption{Event selection criteria for the signal and control regions. Reconstructed objects are defined as explained in Section~\ref{sec:reconstruction}.}\label{tab:regions}
\begin{tabularx}{\linewidth}{l|L|L|L|L|L|L}
 
\hhline{-|-|-|-|-|-|-}
\textbf{Requirement} & \textbf{SR} & \textbf{\Wmunu} & \textbf{\Zmumu}  & \textbf{\Wenu} & \textbf{\Zee} & \textbf{Top}\\[5pt]
\hhline{=:=:=:=:=:=:=}
Primary vertex                 & \multicolumn{6}{c}{at least one with $\geq2$ associated tracks with $\pt>\SI{500}{MeV}$}                                                                                                                                                                                                                                                                                                                                                                                                                                                                                                                                                                                                                                                                                                                                                                                                                                                                                                                                                                                      \\[2pt]
\hline
Trigger                        & \multicolumn{3}{c|}{\met}                                                                                                                                                                                                                                                                                                                                       & \multicolumn{2}{c|}{single-electron}                                                                                                                                                                                                                                                                                                                                                                                                                                            & \met, single-electron                                                                                                                                                                                                                                        \\
\hline
\mymet cut & $\met > \SI{200}{GeV}$ & $|\ptmiss+\ptvec(\mu)| > \SI{200}{GeV}$ & $|\ptmiss+\ptvec(\mu\mu)|> \SI{200}{GeV}$ &
$|\ptmiss+\ptvec(e)|>\SI{200}{GeV}$ & $|\ptmiss+\ptvec(ee)|>\SI{200}{GeV}$ & $|\ptmiss+\ptvec(\mu)|>\SI{200}{GeV}$ or $|\ptmiss+\ptvec(e)|>\SI{200}{GeV}$ \\
\hline
Jets                           & \multicolumn{6}{c}{up to $4$ with $\pt>\SI{30}{GeV}, |\eta|<2.8$}\\[2pt]\hline
$|\Delta\phi(\text{jets,\myvecmet})|$ & \multicolumn{6}{c}{$>0.4$ ($>0.6$ if $\SI{200}{GeV}<\met\leq\SI{250}{GeV}$)}\\[2pt]
\hline
Leading jet                    & \multicolumn{6}{c}{$\pt>\SI{150}{GeV}, |\eta|<2.4, f_{{\text{ch}}}/f_{{\text{max}}} > 0.1$}                                                                                                                                                                                                                                                                                                                                                                                                                                                                                                                                                                                                                                                                                                                                                                                                                                                                                                                                                                        \\[2pt]
\hline
\bjets          & any             & none & any & none & any                                                                                                                                                                                                                                                                                                                                                                                                                                                                                                                                                                                                                                                                                                                                                                                        & at least one                                                                                                                                                               \\[2pt]
\hline
Electrons or muons                          & none          & exactly one muon, with $\pt>\SI{10}{GeV}$, $30<\mT<\SI{100}{GeV}$; no electron & exactly two muons, with $\pt>\SI{10}{GeV}$, $66<m_{\mu\mu}<\SI{116}{GeV}$; no electron & exactly one electron, tight, with $\pt>\SI{30}{GeV}$, $|\eta|\not\in(1.37,1.52)$, tight isolation, $30<\mT<\SI{100}{GeV}$; no muon & exactly two electrons, with $\pt>\SI{30}{GeV}$, $66<m_{ee}<\SI{116}{GeV}$; no muon & same as for \Wmunu or same as for \Wenu \\
\hline
\tauleptons     & \multicolumn{6}{c}{none}                                                                                                                                                                                                                                                                                                                                                                                                                                                                                                                                                                                                                                                                                                                                                                                                                                                                                                                                                                                                                                                                                                                       \\[2pt]
\hline
Photons                        & \multicolumn{6}{c}{none}                                                                                                                                                                                                                                                                                                                                                                                                                                                                                                                                                                                                                                                                                                                                                                                                                                                                                                                                                                                                             \\[2pt]
\hline
\end{tabularx}
\end{table}

\subsection{Multijet background}\label{sec:multijet}
The multijet background with large \met  originates mainly from the misreconstruction of the energy of a jet in the calorimeter and, to a lesser extent, is due to the presence of neutrinos in the final state from
heavy-flavor hadron decays. In this analysis, the multijet background is determined from data, using
the jet smearing method as described in Ref.~{\cite{SUSY-2011-20}}. It relies on the assumption that the \met
value of multijet events is dominated by fluctuations in the jet response in the detector, which can be measured in
the data. The method was checked using a validation region where events were selected as in the signal region,
except for a modified requirement that the minimum azimuthal distance between a jet and \ptmiss is between 0.3 and 0.4.
After event selection, the multijet background is estimated to be about $1.2\%$, $0.8\%$, $0.4\%$ and $0.3\%$ of
the total background in the exclusive signal region bins EM0, EM1, EM2 and EM3, respectively, and it is less than $0.1\%$ for the other signal region bins. A conservative 100\% uncertainty is assigned to the normalization of this background.
 
\subsection{Noncollision background}\label{sec:NCB}
After event selections are applied, the signal region may contain residual contributions from noncollision backgrounds. These backgrounds, which are not included in simulation, mainly arise when beam-halo protons intercept the LHC collimators, leading to particle cascades which produce muons. The remaining contributions are estimated following the methods set out in Ref.~\cite{DAPR-2012-01}. In particular, the jet timing, $t_j$, calculated from the
energy-weighted average of the time of the jet energy deposits, defined relative to the event time
in nominal collisions, is used. A dedicated region enhanced in beam-induced background, defined by
inverting the tight jet-quality selection imposed on the leading jet, is used to estimate the amount of non-collision background from the fraction of events with a leading-jet timing $|t_j| > \SI{5}{ns}$.
The results indicate
a contribution at the per-mille level from noncollision backgrounds in the signal region. A conservative 100\% uncertainty is assigned to the normalization of this background.
 
\subsection{Standard Model background fit}\label{sec:bkgfit}
The estimation of backgrounds in the SR is based on a simultaneous, binned likelihood fit to the \mymet distribution of the five control regions described in \Sect{\ref{sec:CRs}}. The number of events in each region and in each bin is treated as a random variable with a Poisson distribution function, with an expectation value given by the sum of the SM predictions for each background in that bin. The likelihood fit is based on the profile likelihood method~\cite{Cowan:2010js}. Systematic uncertainties are represented by Gaussian-distributed nuisance parameters, and take into account the correlation among systematic variations and across \mymet bins.
 
The normalization of all \Wjets and \Zjets processes, excluding those initiated by VBF, is multiplied by a common single floating normalization factor, which is the same across all \mymet bins. As a result, data from both $W$ and $Z$ control regions are used simultaneously to constrain the \Znunu background in the signal region. Systematic uncertainties in \Wjets and \Zjets event yields, as described in Section~\ref{sec:systematics}, cover the residual bin-by-bin differences among processes when higher-order calculations are included, taking into account the correlation of theoretical uncertainties across different processes with the calculation provided in Ref.~\cite{Lindert2017}. Similarly, one floating normalization factor is used for each of the \ttbar and \singletop backgrounds, resulting in a total of three floating background normalization factors in the fit. Compared to the previous version of the analysis, the usage of two independent normalization factors for the two main sources of top-quark backgrounds is introduced to better take into account their different expected contribution as a function of \mymet.
 
Table \ref{tab:CRfit} shows the results of the background-only fit to the control regions, when all exclusive bins are fitted simultaneously. As determined in the signal region, the normalizations of the \Wjets and \Zjets backgrounds get corrected by a multiplicative factor of \kVcronly, while the normalization of the \ttbar and \singletop backgrounds gets corrected by a multiplicative factor of \kTTcronly and \kTcronly, respectively.

\begin{table}[!tb]
\caption{Data and expected events with $\mymet>\SI{200}{GeV}$ in the five control regions (top: post-fit, bottom: pre-fit). The post-fit predictions for the SM backgrounds are obtained after the simultaneous binned likelihood fit to the five control regions, performed in the exclusive bins of \mymet (EM0--EM12).  The background predictions include both the statistical and systematic uncertainties. The individual uncertainties are correlated, and do not necessarily add in quadrature to equal the total background uncertainty. The dash ``\textendash'' denotes  contributions of less than $0.01\%$ to the total background.}
\begin{center}
\begin{footnotesize}
\begin{tabularx}{\textwidth}{@{\extracolsep{\fill}}lccccc}
\noalign{\smallskip}\hline\noalign{\smallskip}
{\normalfont \bfseries $\mathbf{\mymet>\SI{200}{GeV}}$} & \Wmunu & \Wenu & Top & \Zmumu & \Zee \\
\noalign{\smallskip}\hline\noalign{\smallskip}
Data events (\mylumi) &      1\,364\,958 &     699\,674 &     225\,606 &     196\,800 &     145\,531 \\ \hline
\noalign{\smallskip}\hline\noalign{\smallskip}
SM prediction (post-fit) &     1\,364\,800 $\pm$ 3300~~~       & 699\,700 $\pm$ 1800~~~       & 225\,600 $\pm$ 1100~~~       & 197\,000 $\pm$ 600~~       & 145\,500 $\pm$ 500~~      \\
\noalign{\smallskip}\hline\noalign{\smallskip}
Fitted \Wenu &     {\textendash}       & 578\,800 $\pm$ 3400~~~       & 16\,100 $\pm$ 900~~~       & {\textendash}       & {\textendash}      \\
Fitted \Wmunu &     1\,149\,000 $\pm$ 6000~~~       & {\textendash}       & 31\,600 $\pm$ 1700~       & {\textendash}       & {\textendash}      \\
Fitted \Wtaunu &     71\,500 $\pm$ 800       & 45\,200 $\pm$ 500~~~       & 3380 $\pm$ 180       & {\textendash}       & {\textendash}      \\
Fitted VBF \Wjets &     ~~26\,200 $\pm$ 3100       & 14\,300 $\pm$ 1700~       & 2020 $\pm$ 340       & {\textendash}       & {\textendash}      \\
Fitted \Zee &     {\textendash}       & {\textendash}       & {\textendash}       & {\textendash}       & 138\,100 $\pm$ 700~~      \\
Fitted \Zmumu &     21\,500 $\pm$ 500       & {\textendash}       & 778 $\pm$ 20       & 185\,200 $\pm$ 900~~       & {\textendash}      \\
Fitted \Ztautau &     {\textendash}       & 1900 $\pm$ 50~~       & {\textendash}       & {\textendash}       & {\textendash}      \\
Fitted \Znunu &     {\textendash}       & {\textendash}       & {\textendash}       & {\textendash}       & {\textendash}      \\
Fitted VBF \Zjets &     {\textendash}       & {\textendash}       & {\textendash}       & ~~~3300 $\pm$ 400       & ~~~2530 $\pm$ 320      \\
Fitted \singletop &    ~~22\,000 $\pm$ 6000       & 13\,200 $\pm$ 3500~       & ~~33\,000 $\pm$ 10\,000       & ~~~~~350 $\pm$ 170       & ~~~110 $\pm$ 60      \\
Fitted \ttbar &    ~~52\,000 $\pm$ 4000       & 34\,000 $\pm$ 2800~       & 137\,000 $\pm$ 9000~~~       & ~~~4000 $\pm$ 400       & ~~~1790 $\pm$ 180      \\
Expected dibosons &    ~~23\,000 $\pm$ 4000       & 12\,400 $\pm$ 1900~       & 1880 $\pm$ 340       & ~~~4000 $\pm$ 700       & ~~~2900 $\pm$ 500      \\
\noalign{\smallskip}\hline\noalign{\smallskip}
MC exp.\ SM events  &     1\,220\,000 $\pm$ 60\,000       & 623\,000 $\pm$ 32\,000       & 233\,000 $\pm$ 31\,000       & 175\,000 $\pm$ 9000       & 127\,000 $\pm$ 7000      \\
\noalign{\smallskip}\hline\noalign{\smallskip}
Fit input \Wenu &     {\textendash}       & 509\,000 $\pm$ 27\,000       & 14\,200 $\pm$ 1200~       & {\textendash}       & {\textendash}      \\
Fit input \Wmunu &     1\,010\,000 $\pm$ 50\,000       & {\textendash}       & 28\,000 $\pm$ 2300~       & {\textendash}       & {\textendash}      \\
Fit input \Wtaunu &     ~~63\,000 $\pm$ 3400       & 39\,800 $\pm$ 2100~       & 2970 $\pm$ 250       & {\textendash}       & {\textendash}      \\
Fit input VBF \Wjets &     ~~22\,000 $\pm$ 4000       & 11\,700 $\pm$ 2300~       & 1500 $\pm$ 500       & {\textendash}       & {\textendash}      \\
Fit input \Zee &     {\textendash}       & {\textendash}       & {\textendash}       & {\textendash}       & 120\,000 $\pm$ 7000      \\
Fit input \Zmumu &     ~~18\,900 $\pm$ 1100       & {\textendash}       & 689 $\pm$ 23       & 163\,000 $\pm$ 9000       & {\textendash}      \\
Fit input \Ztautau &     {\textendash}       & 1680 $\pm$ 60~~       & {\textendash}       & {\textendash}       & {\textendash}      \\
Fit input \Znunu &     {\textendash}       & {\textendash}       & {\textendash}       & {\textendash}       & {\textendash}      \\
Fit input VBF \Zjets &     {\textendash}       & {\textendash}       & {\textendash}       & ~~~2700 $\pm$ 500       & ~~~2000 $\pm$ 400      \\
Fit input \singletop &    ~~16\,000 $\pm$ 6000       & 10\,000 $\pm$ 4000~       & ~~28\,000 $\pm$ 13\,000       & ~~~~~700 $\pm$ 500       & ~~~~~280 $\pm$ 210      \\
Fit input \ttbar &    ~~60\,000 $\pm$ 7000       & 39\,000 $\pm$ 5000~       & 155\,000 $\pm$ 27\,000       & ~~~~~4600 $\pm$ 1300       & ~~~2000 $\pm$ 700      \\
Fit input dibosons &    ~~23\,000 $\pm$ 4000       & 12\,600 $\pm$ 2300~       & 1900 $\pm$ 400       & ~~~4100 $\pm$ 800       & ~~~3000 $\pm$ 600      \\
\noalign{\smallskip}\hline\noalign{\smallskip}
\end{tabularx}
\end{footnotesize}
\end{center}
\label{tab:CRfit}
\end{table}

Figures~\ref{fig:CR_1} and \ref{fig:CR_2} show the expected and observed distributions of the \mymet in the control regions. The shown expected distributions include the data-driven normalization
factors as extracted from the binned likelihood fit to the different exclusive \mymet bins in the control regions. Good agreement is observed, within statistical and systematic uncertainties, with data. As an illustration, $\chi^{2}$-statistical tests, using the binned profile likelihood fit described above, probing potential shape discrepancies between the observed and predicted $\mymet$ distributions, give  $p$-values
in a range from 0.49 (in the \Wmunu control region) to 0.96 (in the top-quark control region).
 
In order to perform model-independent tests for new physics processes, discussed in Section~\ref{sec:modelindep}, the same fit procedure is repeated in each of the inclusive bins of \mymet for signal and control regions, denoted in Table~\ref{tab:sr} by IM0--IM12.  Since in this case no shape information is available to constrain the separate contributions of \ttbar and \singletop, a single normalization factor is used for all top-quark-related processes, along with the normalization factor for $W/Z$+jets, resulting in two free background normalization factors in the fit. Additionally, the nuisance parameters related to systematic uncertainties refer to the given $\met$ inclusive region. A total of 13 separate fits are therefore performed, based on five control regions each and including two free background normalization factors. The results are expected to differ from those of the simultaneous fit to exclusive bins, due to the lack of \mymet shape information to constrain uncertainties and the normalization of backgrounds.
 
\begin{figure}
\centering
\mbox{
\begin{subfigure}{0.495\textwidth}
\centering
\includegraphics[width=\textwidth]{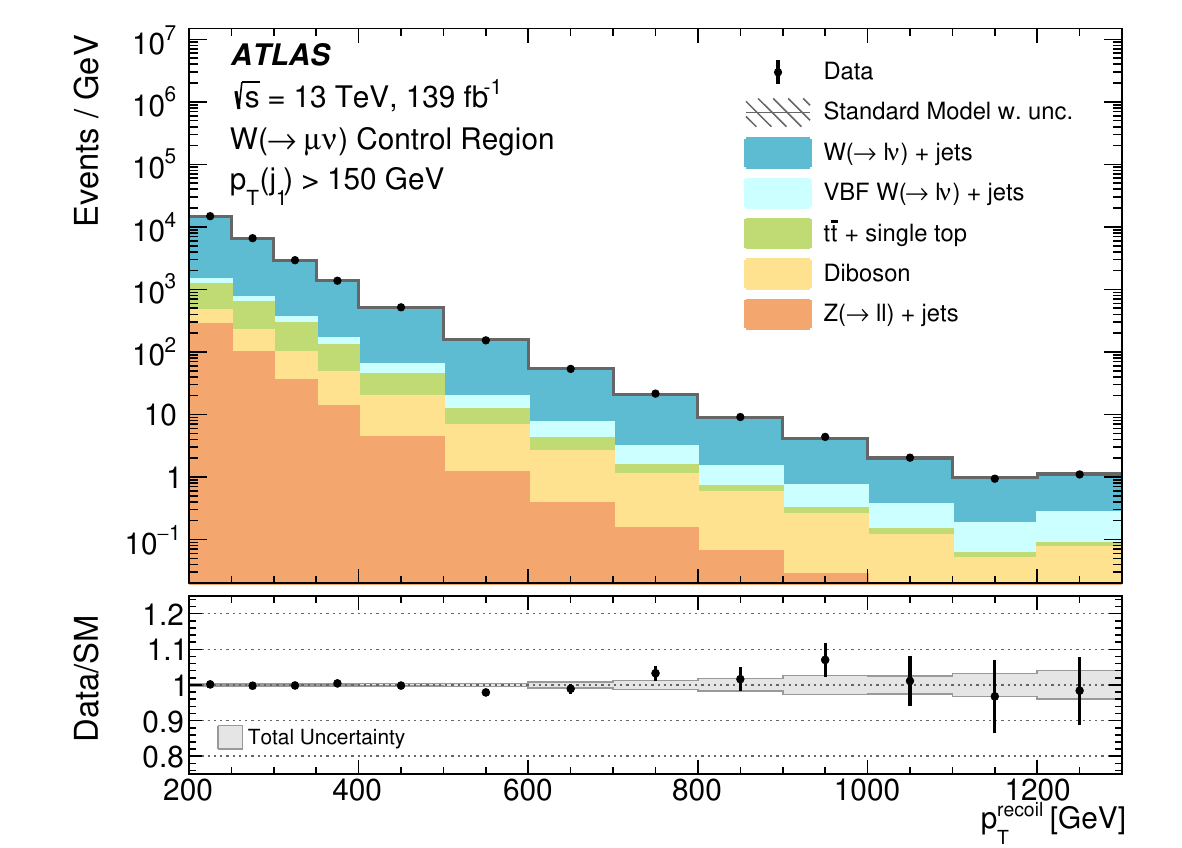}
\caption{}
\label{fig:CR_1mu0b_met}
\end{subfigure}
\begin{subfigure}{0.495\textwidth}
\centering
\includegraphics[width=\textwidth]{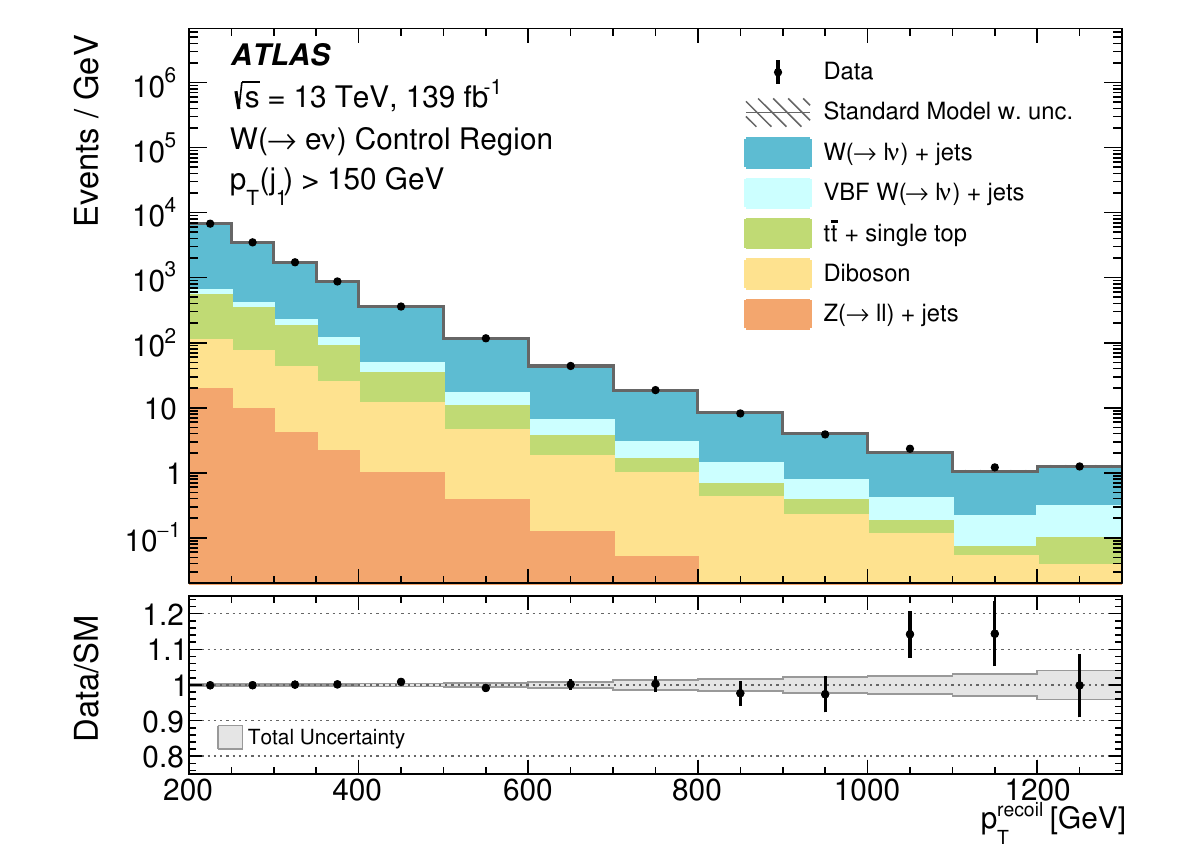}
\caption{}
\label{fig:CR_1e0b_met}
\end{subfigure}
}
\mbox{
\begin{subfigure}{0.495\textwidth}
\centering
\includegraphics[width=\textwidth]{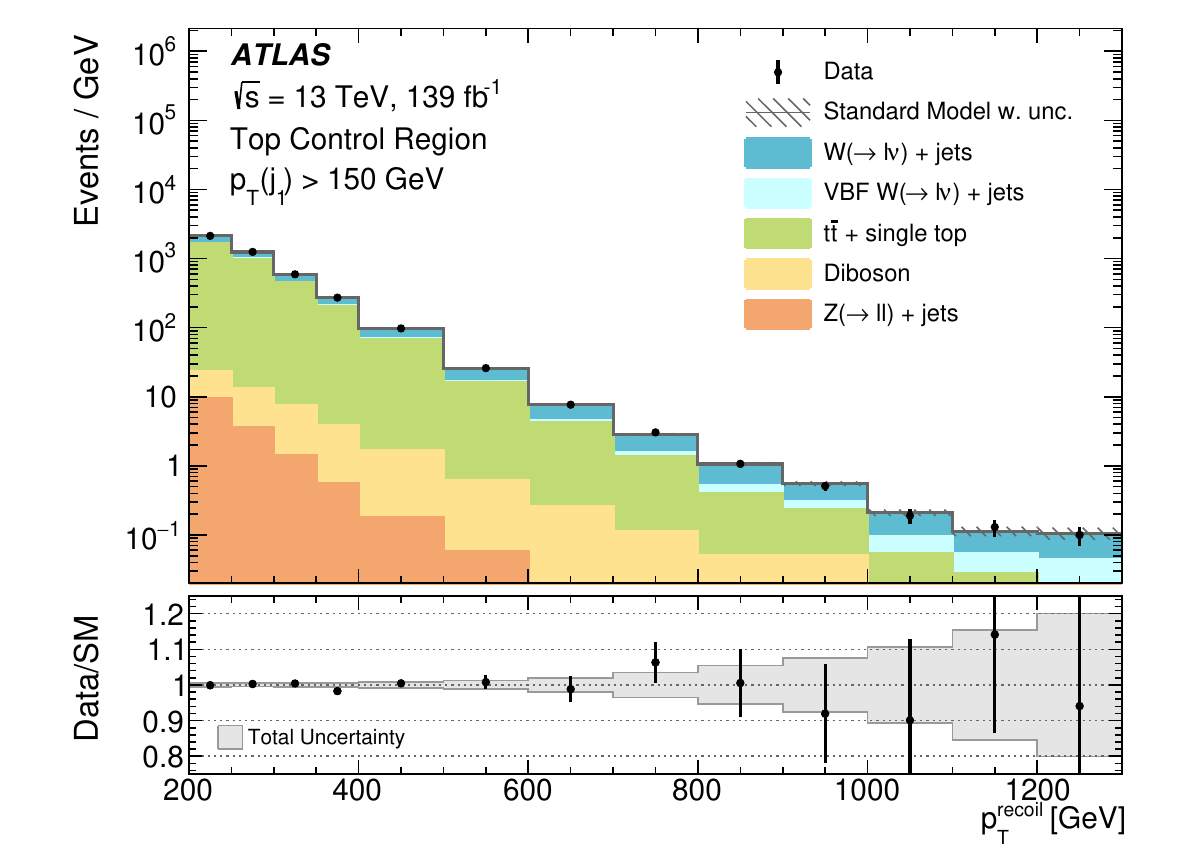}
\caption{}
\label{fig:CR_1L1b_met}
\end{subfigure}
}
\caption{The measured \mymet distributions in the \protect\subref{fig:CR_1mu0b_met} \Wmunu, \protect\subref{fig:CR_1e0b_met} \Wenu and \protect\subref{fig:CR_1L1b_met} top control regions, compared with the background predictions as estimated after the simultaneous, binned background-only fit to the data in the control regions.
The ratios of data to SM predictions after the CR fit are shown in the lower panels (black dots).
The error bands
in the ratios include the statistical and systematic uncertainties in the background predictions. Events with values beyond the range of the histogram are included in the last bin.
}
\label{fig:CR_1}
\end{figure}
 
\begin{figure}
\centering
\mbox{
\begin{subfigure}{0.495\textwidth}
\centering
\includegraphics[width=\textwidth]{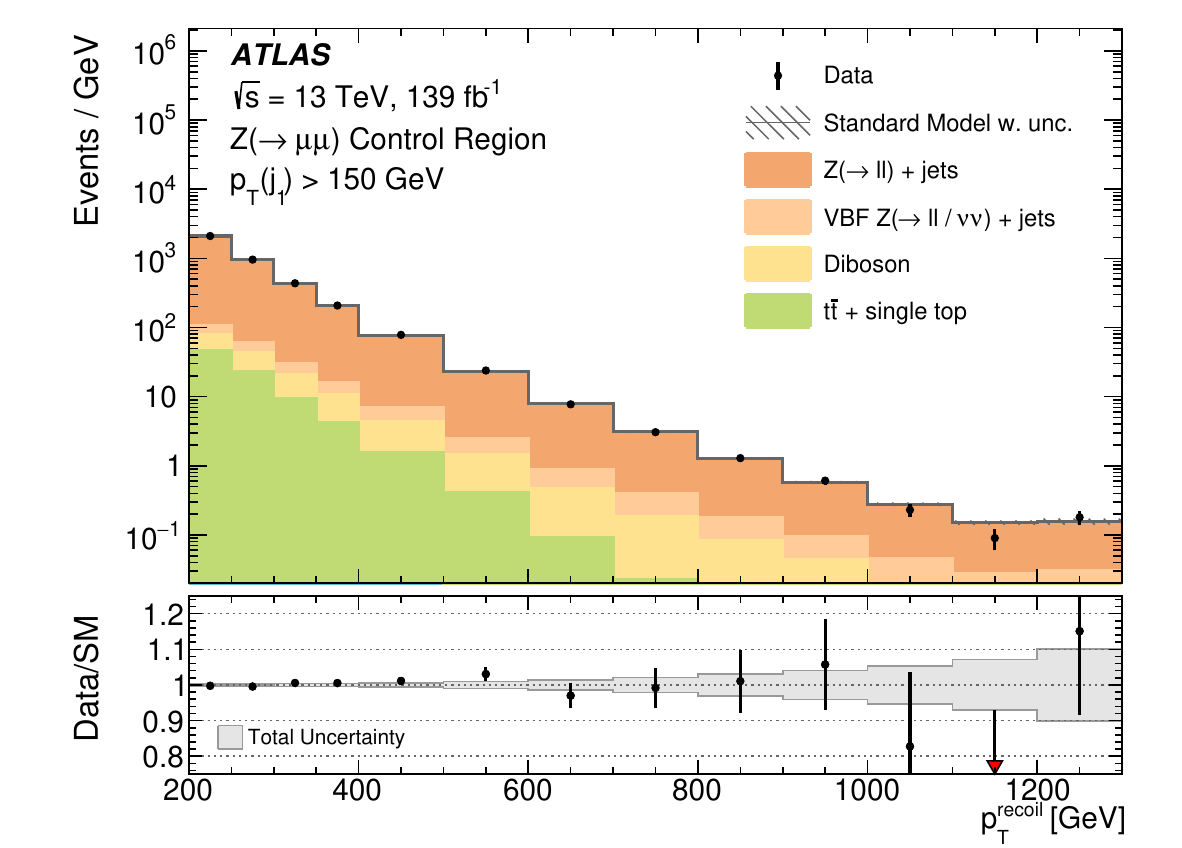}
\caption{}
\label{fig:CR_2mu_met}
\end{subfigure}
\begin{subfigure}{0.495\textwidth}
\centering
\includegraphics[width=\textwidth]{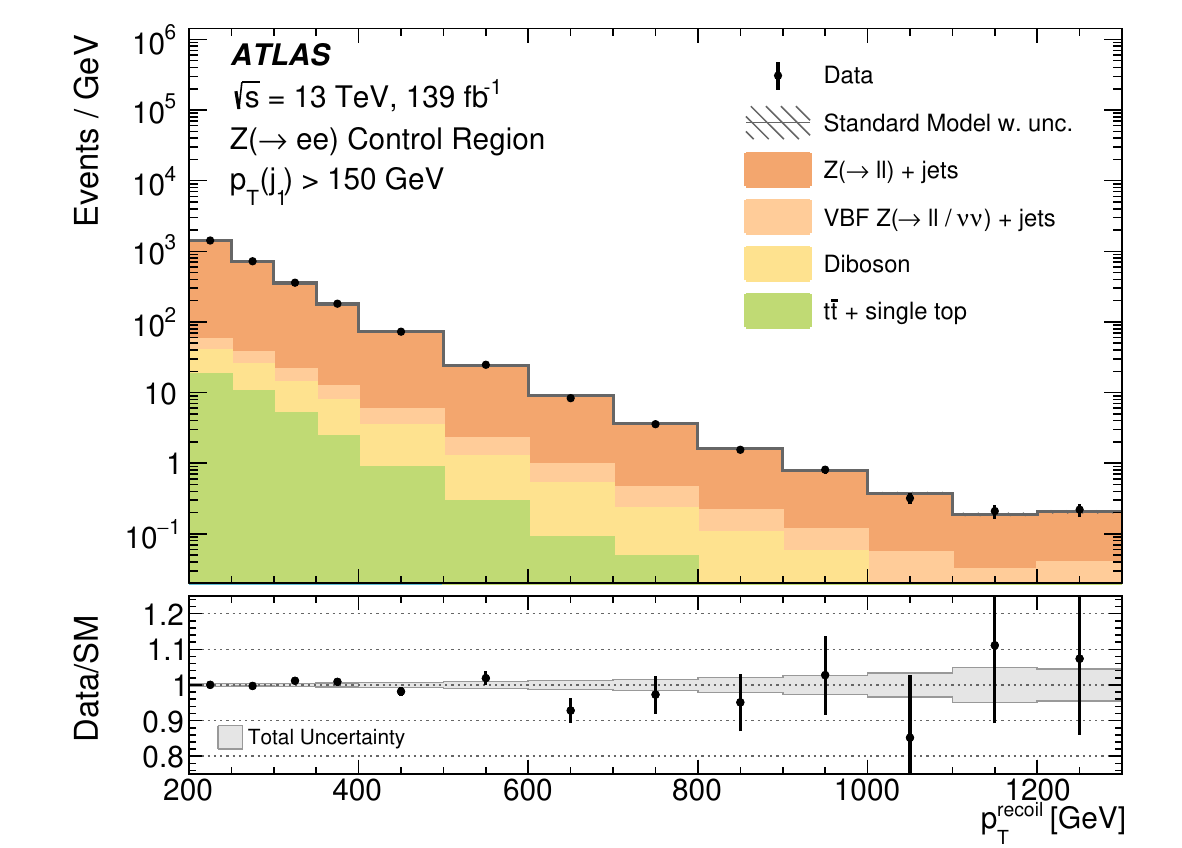}
\caption{}
\label{fig:CR_2e_met}
\end{subfigure}
}
\caption{The measured \mymet distributions  in the \protect\subref{fig:CR_2mu_met} \Zmumu and \protect\subref{fig:CR_2e_met} \Zee control regions, compared with the background predictions as estimated after the simultaneous, binned background-only fit to the data in the control regions.
The ratios of data to SM predictions after the CR fit are shown in the lower panels (black dots).
The error bands
in the ratios include the statistical and systematic uncertainties in the background predictions. Events with values beyond the range of the histogram are included in the last bin. The red arrow marker in the ratio panel indicates the point falls beyond the vertical axis range of the plot. 
}
\label{fig:CR_2}
\end{figure}

\section{Systematic uncertainties}
\label{sec:systematics}
 
 
The impact of systematic uncertainties is estimated after performing a background-only fit to data from the exclusive CRs, and evaluating the impact of the uncertainty in the total background yield in each bin of \mymet in the SR. The dominant sources of experimental uncertainty are those related to the electron, muon and jet identification and reconstruction efficiencies, while uncertainties in the \Vjets predictions give the leading contribution to theory uncertainties. More details are provided in the following sections.

\subsection{Background uncertainties}
The uncertainty in the combined 2015--2018 integrated luminosity is 1.7\%. It is derived from the calibration of the luminosity scale using $x$--$y$ beam-separation scans, following a methodology similar to that described in Ref.~\cite{DAPR-2013-01}, and using the LUCID-2 detector for the baseline luminosity measurements~\cite{LUCID2}. This uncertainty nearly cancels out in the semi-data-driven background estimation procedure, and translates into a residual uncertainty in the total background in the SR of less than $0.01\%$ ($0.05\%$) for $\mymet=\SI{200}{GeV}$ (\SI{1200}{GeV}). The uncertainty in the pileup reweighting procedure translates into a residual uncertainty in the total background in the SR of less than $0.4\%$ ($0.24\%$) for $\mymet=\SI{200}{GeV}$ (\SI{1200}{GeV}).
 
Systematic uncertainties in the jet energy scale (resolution)~\cite{JETM-2018-05} translate into uncertainties in the total background in the SR which vary between $0.17\%$ ($0.15\%)$ and $1.0\%$ ($1.3\%$) for \mymet between \SI{200}{GeV} and \SI{1200}{GeV}. The uncertainty in the modeling of the JVT requirement used to reject jets coming from pileup~\cite{PERF-2014-03} is below $0.03\%$ across the \mymet spectrum.
Uncertainties in the flavor-tagging efficiency~\cite{PERF-2016-05} translate into uncertainties in the total background in the SR between $0.1\%$ and $0.9\%$ for \mymet between \SI{200}{GeV} and \SI{1200}{GeV}. 
 
Uncertainties in the $\met$ scale (resolution) due to soft contributions to the \met calculation translate into uncertainties in the total background in the SR between $0.5\%$ ($0.34\%$) and $0.25\%$ ($0.04\%$) for \mymet between \SI{200}{GeV} and \SI{1200}{GeV}.
 
Uncertainties in the electron reconstruction and identification efficiencies are computed following the method described in Ref.~\cite{EGAM-2018-01}; the latter are treated as uncorrelated between selected and vetoed electrons, since different working points are used for their identification, as described in Section~\ref{sec:selection}. Uncertainties in the electron and photon energy scale and resolution are computed following the method described in Ref.~\cite{PERF-2017-03}. Overall, they translate into uncertainties in the total background in the SR between $0.7\%$ and $1.3\%$ for \mymet between \SI{200}{GeV} and \SI{1200}{GeV}. 
Uncertainties due to the electron isolation efficiency give a contribution of less than $0.2\%$ across the \mymet spectrum. Negligible contributions are given by the electron trigger efficiency and by the photon identification efficiency. 
Uncertainties in the muon reconstruction and identification efficiencies and in their momentum measurement are computed following the method described in Ref.~\cite{PERF-2015-10}. In order to take into account the difference between the simulated and measured identification efficiencies for muons with \pt above \SI{300}{GeV}, an additional uncertainty is included, conservatively taken as uncorrelated in muon \pt and $\eta$. Overall, these translate into uncertainties in the total background in the SR between $0.4\%$ and $1.9\%$ for \mymet between \SI{200}{GeV} and \SI{1200}{GeV}. 
Uncertainties in the \taulepton reconstruction and identification efficiencies translate into uncertainties in the total background in the SR of $0.1\%$ ($0.07\%$) for $\mymet=\SI{200}{GeV}$ ($>\SI{1200}{GeV}$).
 
Uncertainties in the higher-order QCD and electroweak parton-level calculations used to correct the MC prediction for \Vjets processes not initiated by VBF are calculated following the procedure described in Ref.~{\cite{Lindert2017}}. They are provided in the form of event weights parameterized as a function of vector-boson \pt, which are applied to derive shape and normalization uncertainties for simulated \Vjets processes in signal and control regions.
The correlations across \mymet bins are implemented as recommended in Ref.~{\cite{Lindert2017}}. The correlations across processes are implemented as follows.
 
Three sources of pure QCD uncertainties are considered. The uncertainty associated with the truncation of the perturbative expansion in the strong coupling constant is estimated by varying the QCD renormalization and factorization scales both individually and simultaneously by factors 2 and 0.5, and evaluating the change in differential cross section in bins of the vector-boson \pt for these seven combinations. The center of the resulting band is taken as the nominal estimate, and its half width is taken as its systematic uncertainty. Uncertainties in the shape of the vector-boson \pt distribution, which are relevant for the extrapolation of low-\pt measurements to high \pt, are taken into account by applying an additional uncertainty estimated from a conservative shape distortion of the aforementioned scale uncertainty.\footnote{This distortion is parameterized as a function of the vector-boson \pt between \SI{200}{GeV} and \SI{2}{TeV} and has the form $(\pt^2-p_{\text{T},0}^2)/(\pt^2+p_{\text{T},0}^2)$, where $p_{\text{T},0}=\SI{650}{GeV}$.} These two uncertainties are represented by two nuisance parameters, $\delta^{(1)}K_\mathrm{NNLO}$ and $\delta^{(2)}K_\mathrm{NNLO}$, which are taken as correlated across \Vjets processes, assuming that for $\mymet\gg m_{W,Z}$ the QCD effects on the \Wjets and \Zjets predictions are very similar. Residual differences in QCD corrections between \Wjets and \Zjets processes are estimated from the difference in QCD NNLO $k$-factors with respect to \Zjets production, and taken into account as an additional nuisance parameter, $\delta^{(3)}K_\mathrm{NNLO}$.
 
Three sources of pure EW uncertainties and one source of mixed QCD--EW uncertainties are considered. Unknown Sudakov logarithms beyond NNLO are considered as the dominant source of uncertainty at high \pt, and are treated as correlated across \Vjets processes and represented by one nuisance parameter, $\delta^{(1)}\kappa_\textrm{nNLO EW}$. An additional source of uncertainty is considered to cover possible NNLO effects not included in the calculation, conservatively defined as $5\%$ of the absolute full NLO EW correction, and is treated as uncorrelated across \Vjets processes and hence represented by the three nuisance parameters $\delta^{(2)}\kappa_\textrm{nNLO EW}^{(W)}$, $\delta^{(2)}\kappa_\textrm{nNLO EW}^{(Z\to\ell^+\ell^-)}$, and $\delta^{(2)}\kappa_\textrm{nNLO EW}^{(Z\to\nu\nu)}$. The uncertainty in the limitations of the Sudakov approximation at two loops is estimated as the difference between the NLL Sudakov approximation and the exponentiation of the full NLO EW correction, and is treated as uncorrelated between \Wjets and \Zjets processes, resulting in two nuisance parameters, $\delta^{(3)}\kappa_\textrm{nNLO EW}^{(W)}$ and $\delta^{(3)}\kappa_\textrm{nNLO EW}^{(Z)}$. Uncertainties due to the approximation of mixed QCD--EW corrections via a factorized combination of QCD and EW corrections are assumed to be proportional to the difference between the additive and multiplicative combination of QCD and EW corrections, and are treated as correlated across \Vjets processes, resulting in a single nuisance parameter, $\delta K_\textrm{NNLO mix}$.
 
Uncertainties in the parton distribution functions are treated as correlated across \Vjets processes, following the prescription of Ref.~\cite{Lindert2017}. They are estimated by the sum in quadrature over the $107$ independent Hessian PDF replicas provided by the PDF set \texttt{LUXqed\_plus\_PDF4LHC15\_nnlo}, following Eq.~(20) of Ref.~\cite{Butterworth:2015oua}. The resulting nuisance parameter is denoted by $\delta K_\mathrm{PDF}$.
 
Table~\ref{tab:VjetsNPs} summarizes the considered nuisance parameters and their assumed correlations. Before the CR-only fit, the leading impacts on the signal region background prediction for the EM0 selection come from $\delta^{(1)}K_\mathrm{NNLO}$ ($1.8\%$), $\delta^{(2)}K_\mathrm{NNLO}$ ($1.5\%$) and $\delta K_\mathrm{PDF}$ ($0.7\%$), while the largest contributions for the EM12 selection come from $\delta K_\mathrm{PDF}$ ($2.9\%$), $\delta^{(1)}K_\mathrm{NNLO}$ ($2.8\%$), $\delta^{(3)}\kappa_\textrm{nNLO EW}^{(Z)}$ ($1.6\%$) and $\delta^{(2)}K_\mathrm{NNLO}$ ($1.5\%$).   
After the CR-only fit, these uncertainties translate into uncertainties in the total background in the SR between $0.4\%$ and $2\%$ for \mymet between \SI{200}{GeV} and \SI{1200}{GeV}, dominated by QCD uncertainties at low \mymet and by EW uncertainties at high \mymet. The leading contributions for the EM0 selection are given by $\delta^{(1)}K_\mathrm{NNLO}$ ($0.20\%$), $\delta^{(3)}K_\mathrm{NNLO}$ ($0.12\%$) and $\delta^{(2)}\kappa_\textrm{nNLO EW}^{(Z\to\nu\nu)}$ ($0.11\%$), while the largest contributions for the EM12 selection come from $\delta^{(3)}\kappa_\textrm{nNLO EW}^{(Z)}$ ($1.3\%$), $\delta^{(2)}\kappa_\textrm{nNLO EW}^{(Z\to\nu\nu)}$ ($1.2\%$) and $\delta^{(2)}\kappa_\textrm{nNLO EW}^{(W)}$ ($1.0\%$). An additional systematic uncertainty is included to cover possible differences between the definition of \tauleptons at Monte Carlo generator level and the one used in the theoretical calculation from Ref.~\cite{Lindert2017}, which translates into uncertainties in the total background in the SR between $0.05\%$ and $0.1\%$ for $\mymet$ between \SI{200}{GeV} and \SI{1200}{GeV}.
 
\begin{table}\small
\centering
\caption{Uncertainties considered in the reweighting of \Vjets samples to higher-order QCD and EW parton-level calculations. For reference, the correspondence with the nuisance parameters included in Table 3 from Ref.~\cite{Lindert2017} is also indicated.}\label{tab:VjetsNPs}
\begin{tabularx}{\textwidth}{@{\extracolsep{\fill}}X|X|l}
\hhline{-|-|-}
\textbf{Source of uncertainty} & \textbf{Correlation} & \textbf{Nuisance parameter name(s) in Ref.~\cite{Lindert2017}}\\[5pt]\hhline{-|-|-}
Truncation of perturbative expansion in $\alpha_\mathrm{s}$ & Correlated across \mymet bins and \Vjets processes & $\delta^{(1)}K_\mathrm{NNLO}$\\[5pt]\hhline{-|-|-}
Shape of the vector-boson distribution and extrapolation from low \pt to high \pt & Correlated across \mymet bins and \Vjets processes & $\delta^{(2)}K_\mathrm{NNLO}$\\[5pt]\hhline{-|-|-}
Difference in QCD corrections between \Wjets and \Zjets & Correlated across \mymet bins and \Vjets processes & $\delta^{(3)}K_\mathrm{NNLO}$\\[5pt]\hhline{-|-|-}
Unknown Sudakov logarithms beyond NNLO & Correlated across \mymet bins and \Vjets processes & $\delta^{(1)}\kappa_\textrm{nNLO EW}$\\[5pt]\hhline{-|-|-}
Additional possible NNLO effects & Correlated across \mymet bins, uncorrelated between \Vjets processes & $\delta^{(2)}\kappa_\textrm{nNLO EW}^{(W)}$, $\delta^{(2)}\kappa_\textrm{nNLO EW}^{(Z\to\ell^+\ell^-)}$, $\delta^{(2)}\kappa_\textrm{nNLO EW}^{(Z\to\nu\nu)}$\\[5pt]\hhline{-|-|-}
Limitations of the Sudakov approximation at two loops & Correlated across \mymet bins, uncorrelated between \Wjets and \Zjets processes & $\delta^{(3)}\kappa_\textrm{nNLO EW}^{(W)}$, $\delta^{(3)}\kappa_\textrm{nNLO EW}^{(Z)}$\\[5pt]\hhline{-|-|-}
Interference terms between QCD and EW corrections & Correlated across \mymet bins and \Vjets processes & $\delta K_\textrm{NNLO  mix}$\\[5pt]\hhline{-|-|-}
PDF uncertainties & Correlated across \mymet bins and \Vjets processes & sum in quadrature of $\delta K_\mathrm{PDF}^{(i)}$\\[5pt]\hhline{-|-|-}
Different definition of \tauleptons between parton-level calculation and simulation & Correlated across \mymet bins and \Vjets processes & --\\\hhline{-|-|-}
\end{tabularx}
\end{table}

Uncertainties in the \Vjets processes initiated by VBF include scale and PDF uncertainties and the comparison with \SHERPA as an alternative MC generator. 
They translate into uncertainties in the total background in the SR between $0.024\%$ and $0.1\%$ for \mymet between \SI{200}{GeV} and \SI{900}{GeV}, and between $0.2\%$ and $1.1\%$ for \mymet between \SI{1000}{GeV} and \SI{1200}{GeV}.
 
Uncertainties in the theoretical predictions of the \ttbar and \singletop backgrounds are estimated for the two processes  separately by varying parton-shower parameters and the amount of initial- and final-state soft gluon radiation, by comparing predictions from different MC event generators~\cite{ATL-PHYS-PUB-2017-007} and by comparing the degree of interference between \singletop in the $Wt$-channel and \ttbar when using the DR and DS schemes described in Ref.~{\cite{ATL-PHYS-PUB-2016-020}}. In the case of \ttbar (\singletop), they translate into uncertainties in the total background in the SR between $0.06\%$ ($0.13\%$) and $0.7\%$ ($0.28\%$) for \mymet between \SI{200}{GeV} and \SI{1200}{GeV}.
 
Uncertainties in the theoretical predictions of diboson backgrounds include 
uncertainties in the QCD
renormalization, factorization and resummation scales, uncertainties due to the choice of parton distribution functions and uncertainties in the modeling of the parton showers. They translate into uncertainties in the total background in the SR between $0.01\%$ and $0.22\%$ for \mymet between \SI{200}{GeV} and \SI{1200}{GeV}.
 
Uncertainties in the multijet and noncollision backgrounds translate into uncertainties in the total background in the SR for $\mymet=\SI{200}{GeV}$ of $1\%$ and $0.2\%$, respectively, and are negligible above \SI{800}{GeV}.
 
Table~\ref{tab:systematics} summarizes the overall impact of systematic uncertainties in each bin of \mymet in the SR, as estimated from the CR-only fit. 
\begin{table}\small
\caption{Summary of the impact at low and high \mymet of systematic uncertainties on the total background in the SR, as obtained from the CR-only fit. The impact of each source of systematic uncertainty is shown as the sum in quadrature of the individual contributions represented by the corresponding nuisance parameters. The two reported values refer to the first and last bin of \mymet (EM0 and EM12). Only non-negligible contributions are shown.}
\label{tab:systematics}
\centering\begin{tabular}{ll@{$\;-\;$}l|ll@{$\;-\;$}l}
\hline
\hline
\multicolumn{6}{c}{Source of uncertainty and effect on the total SR background estimate [\%]} \\
 
\hline
Flavor tagging & 0.1 & 0.9 & $\tau$-lepton identification efficiency & 0.1 & 0.07 \\
Jet energy scale & 0.17 & 1.0 & Luminosity & 0.01 & 0.05 \\
Jet energy resolution & 0.15 & 1.3 & Noncollision background & 0.2 & 0.0 \\
Jet JVT efficiency & 0.01 & 0.03 & Multijet background & 1.0 & 0.0 \\
Pileup reweighting & 0.4 & 0.24 & Diboson theory & 0.01 & 0.22 \\
\MET resolution & 0.34 & 0.04 & Single-top theory & 0.13 & 0.28 \\
\MET scale & 0.5 & 0.25 & $t\bar{t}$ theory & 0.06 & 0.7 \\
Electron and photon energy resolution & 0.01 & 0.08 & $V$+jets $\tau$-lepton definition & 0.04 & 0.16 \\
Electron and photon energy scale & 0.3 & 0.7 & $V$+jets pure QCD corrections & 0.24 & 1.1 \\
Electron identification efficiency & 0.5 & 1.0 & $V$+jets pure EW corrections & 0.17 & 2.2 \\
Electron reconstruction efficiency & 0.15 & 0.2 & $V$+jets mixed QCD--EW corrections & 0.02 & 0.7 \\
Electron isolation efficiency & 0.04 & 0.19 & $V$+jets PDF & 0.01 & 0.7 \\
Muon identification efficiency & 0.03 & 0.9 & VBF EW $V$+jets backgrounds & 0.02 & 1.1 \\
Muon reconstruction efficiency & 0.4 & 1.5 & Limited MC statistics & 0.05 & 1.9 \\
Muon momentum scale & 0.1 & 0.7 & \multicolumn{3}{c}{}  \\
\hline
\hline
\multicolumn{6}{c}{Total background uncertainty in the Signal Region: 1.5\%--4.2\%}\\
\hline
\hline
\end{tabular}
\end{table}

\subsection{Signal uncertainties}\label{sec:systematics:signal}
 
Sources of systematic uncertainty in the predicted signal yields are considered separately for each model
of new physics, using a common set of procedures.
Experimental uncertainties include those related to the jet and $\met$ reconstruction,
energy scales and resolutions, which introduce uncertainties in the signal yields for the different models that
vary in the range between $1\%$ and $3\%$  at low $\mymet$,  and between $4\%$ and $7\%$ at large $\mymet$, depending on the parameters of the model.
The 1.7$\%$ uncertainty in the integrated luminosity is also included.
Other uncertainties related to the jet quality requirements are negligible.

Uncertainties affecting the signal acceptance in the generation of
signal samples include: uncertainties in the  modeling of the initial- and
final-state radiation and the underlying event,  determined using simulated samples with
modified parton-shower parameters (by factors of two or one half);
uncertainties due to PDFs and variations of the $\alpha_{\mathrm s}(m_Z)$
value employed, as computed from the envelope of CT10 or CT14,
MMHT2014~\cite{Harland-Lang:2014zoa} and NNPDF30 error sets; and uncertainties due to the
choice of renormalization and factorization scales, which are varied by factors of two or one half.  In addition,
theoretical uncertainties in the predicted cross sections, including PDF
and renormalization- and factorization-scale  uncertainties, are assessed and their effect is shown in terms of variations of the observed results.

In the case of WIMP production models,
the uncertainty related to the modeling of the initial- and final-state radiation translates into a $3\%$ to $6\%$ uncertainty in the signal acceptance.
The choice of different PDF sets results in up to a $10\%$ and a 20$\%$ uncertainty in the case of axial-vector and pseudoscalar models, respectively.
Varying the renormalization and factorization scales introduces $0.1\%$ to $21\%$
variations in the signal acceptance, depending on the model and the mediator and WIMP masses considered.
Renormalization and factorization  scale uncertainties introduce an uncertainty in the cross-section predictions of about  $10\%$ in the case
of the axial-vector mediator model and up to $50\%$ for the pseudoscalar mediator model.  Finally, PDF uncertainties translate into cross-section uncertainties
of about $5\%$ and $20\%$ for the axial-vector and pseudoscalar mediator models, respectively.
The larger uncertainties computed in the case of the pseudoscalar mediator model are
attributed to the presence of top-quark induced loops in the relevant  Feynman diagrams,
that are absent  in the  axial-vector mediator model.
 
Similarly, for SUSY models, the uncertainties related to the modeling
of initial- and final-state gluon radiation and the matching between matrix elements and parton showers in the simulation translate into a $7\%$ to
$8\%$ uncertainty in the signal acceptance for $\met > 200$~GeV.  Variations of the renormalization and factorization scales introduce an uncertainty of about $3\%$ in the signal acceptance.
Uncertainties in the predicted cross sections, including both renormalization/factorization scale and PDF uncertainties, increase with the squark masses and
range between $7\%$ for a mass of 100~\GeV\ and about 11$\%$ for a mass of about 1~\TeV.
 
In the case of dark-energy-inspired models,  uncertainties related to renormalization/factorization scales, PDFs, and parton-shower modeling vary the signal acceptance by  0.1\% to 3.5\%, 1\% to 16\%, and 0.1\% to 5\%, respectively, with increasing $\met$.
Renormalization/factorization scale and PDF uncertainties introduce variations in the cross-section predictions of about $30\%$ each.
 
For the ADD model,
the uncertainties related to the modeling of the initial- and final-state
gluon radiation
translate into
uncertainties in the ADD signal acceptance which
vary between $11\%$ and $13\%$ with increasing $\met$ and are approximately independent of $n$.
The  uncertainties due to the
PDFs, affecting both signal normalization and acceptance, increase from $11\%$ at $n=2$ to $43\%$ at $n=6$.
Similarly, the variations of the renormalization and factorization scales
introduce a signal yield uncertainty of $23\%$ to $36\%$, growing with increasing $n$.
 
For the ALPs production model, theoretical uncertainties related to PDFs, affecting signal normalization and acceptance, translate into uncertainties in the signal yields that vary in the range between $2\%$ and $14\%$.  Variations of the renormalization and factorization scales and matrix-elements to parton-shower matching scales, introduce uncertainties in the signal yields that vary between $5\%$ and $50 \%$. Variations in the  parton-shower modeling translate into uncertainties in the signal acceptance
in the range between $1\%$ and $20\%$, depending on the  $\met$ bin considered.

Finally, for the interpretation of an invisibly decaying Higgs boson, uncertainties related to PDFs, affecting both signal normalization and signal acceptance, translate into 0.4$\%$ to $0.8 \%$ variations in the Higgs signal yields as $\met$ increases.  Variations in the renormalization and factorization scales introduce a $10 \%$ uncertainty in the signal yields.  Uncertainties in the parton-shower modeling translate into uncertainties in the signal acceptance that vary between $3 \%$ and $9 \%$ with increasing $\met$. Uncertainties in the higher-order electroweak corrections, especially relevant for VBF and $VH$ processes, translate into uncertainties in the signal yield that vary between $1.4\%$ and $10\%$ with increasing $\met$.

\section{Results and interpretations}
\label{sec:results}
 
Figure~\ref{fig:sr1}  shows
several measured distributions in the signal region
compared with the SM predictions obtained from the fit to CRs. As discussed in Section~\ref{sec:background}, the SM predictions are normalized with normalization factors determined  from the global fit carried out in exclusive $\mymet$ bins. The fitting procedure also constrains the background uncertainties, resulting in a precise SM prediction in almost the whole $\mymet$ spectrum.
As an example, the  SM predictions are determined with a total uncertainty of $1.5\%$, $1.2\%$,  and $4.1\%$ for the EM0, EM4, and EM12 signal regions, respectively, which include correlations between uncertainties in the individual background contributions.
For illustration purposes, the ratios of data to SM predictions are shown in the lower panel, both after the CR fit and after a global background-only fit when the signal region is also included (``SR+CR fit'').
 
The number of events in the data and the individual background
predictions are presented in Tables~\ref{tab:srincl} and \ref{tab:srexcl} for inclusive and exclusive $\mymet$ bins, respectively. The results for all the signal regions are summarized in Table~\ref{tab:srsum}.
Overall, good agreement between data and SM predictions is observed. 
The compatibility of the data with a SM background hypothesis is tested using
the binned profile likelihood fit described above.
The resulting statistical tests for a background-only hypothesis, in the presence of different potential signal contributions,
give $p$-values in the range between 0.02 and 1.0, where the minimum corresponds
to a signal for stop-pair production in the $\tilde{t}_1 \to  c +  \tilde{\chi}^{0}_{1}$ decay channel with
$m_{\tilde{t}_1} = 500$~\GeV\ and $m_{\ninoone} = 420$~\GeV\ and a deviation of about $2\sigma$ from the background-only hypothesis.

\begin{figure}[!ht]
\begin{center}
\includegraphics[width=0.8\textwidth]{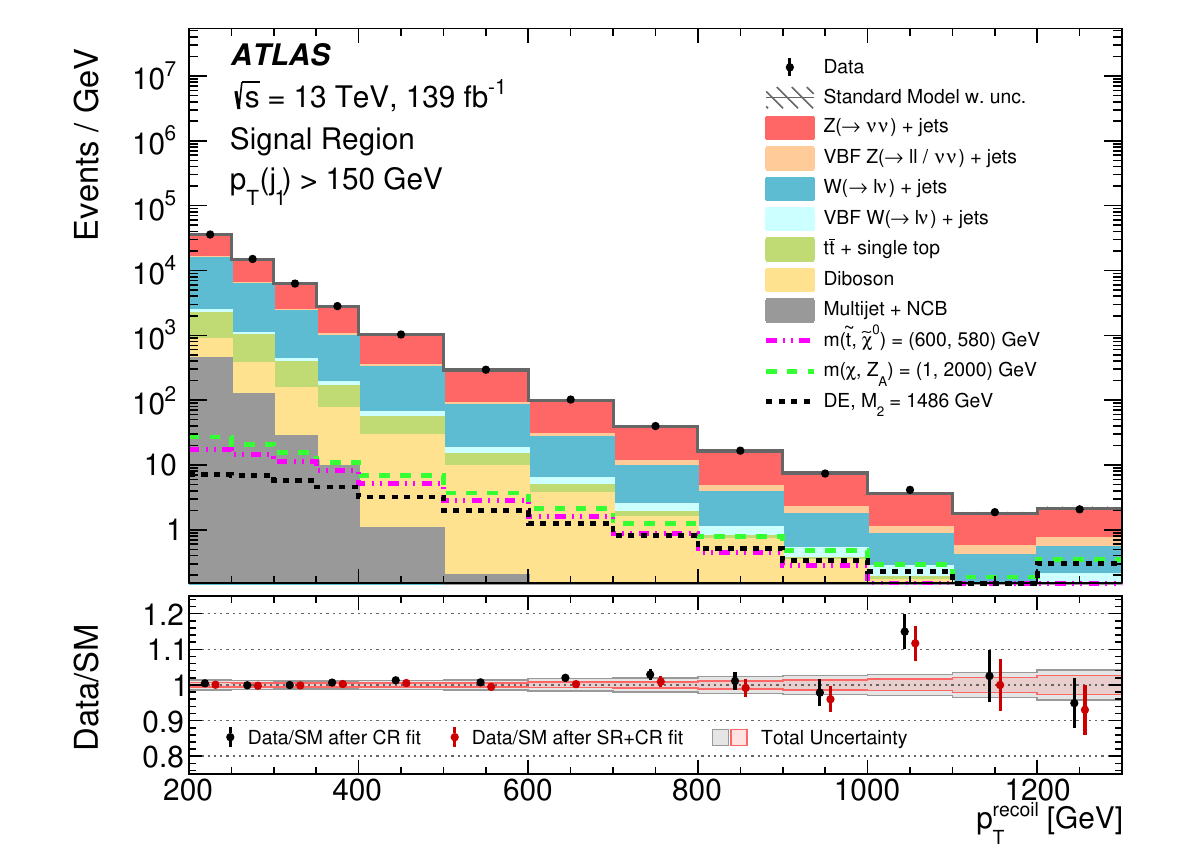}
\end{center}
\caption{
Measured distributions of $\mymet$
for the $\mymet >200~\GeV$ selection compared with the SM predictions in  the signal region.  The latter are normalized  with normalization
factors as determined by the  global fit that considers exclusive $\mymet$ control regions (``CR fit'').
For illustration purposes, the distributions of
examples of dark energy (DE), SUSY, and WIMP scenarios are included.
The ratios of data to SM predictions after the CR fit are shown in the lower panel (black dots), and compared with the same quantities when SM predictions are normalized to the results of the global background-only fit when the signal region is also included (``SR+CR fit'', red dots). The error bands in the ratio shown in the lower panel include both the statistical and systematic uncertainties in the background predictions. Events
with values beyond the range of the histogram are included in the last bin.
}
\label{fig:sr1}
\end{figure}

 
\begin{table}[!tb]
\caption{Data and SM background predictions in the signal region for several inclusive $\mymet$ selections, as determined using separate one-bin likelihood fits in the control regions.
For the SM prediction, both the statistical and systematic uncertainties are included.
In each signal region, the individual uncertainties for the different background processes can be correlated,
and do not necessarily add in quadrature to equal the total background uncertainty.
The dash ``\textendash'' denotes negligible background contributions.
For illustration, the expected event yields for particular signals for new phenomena are provided; in this case, the quoted errors include experimental uncertainties and theory uncertainties on the signal acceptance, as described in Sec.~\ref{sec:systematics:signal}.
}
\begin{center}
\begin{tiny}
\begin{tabularx}{\textwidth}{@{\extracolsep{\fill}}lcccccc}
\noalign{\smallskip}\hline\noalign{\smallskip}
{\normalfont \bfseries  Inclusive Signal Region}& {IM1} & {IM3} & {IM5} & {IM7} & {IM10} & {IM12}\\
\noalign{\smallskip}\hline\noalign{\smallskip}
Data events  (139 fb${}^{-1}$) & {1\,357\,019} & {290\,779} & {46\,855} &{7194} & {807} & {207} \\\noalign{\smallskip}\hline\noalign{\smallskip}
SM prediction            & 1\,346\,000 $\pm$ 16\,000  &  286\,000 $\pm$ 4000  &  45\,550 $\pm$ 1000  &  7070 $\pm$ 240  &  720 $\pm$ 60  &  223 $\pm$ 19 \\
\noalign{\smallskip}\hline\noalign{\smallskip}
\Wenu            & ~~72\,000 $\pm$ 5000  &  11\,900 $\pm$ 800  &  ~1400 $\pm$ 100  &  166 $\pm$ 12  &  12.4 $\pm$ 1.5 & ~~2.8 $\pm$ 0.4 \\
\Wmunu            & 105\,800 $\pm$ 2200  &  17\,600 $\pm$ 400  &  2220 $\pm$ 70~  &  305 $\pm$ 14  &  38 $\pm$ 4  &  14.6 $\pm$ 1.7 \\
\Wtaunu            & 243\,900 $\pm$ 3500  &  44\,900 $\pm$ 800  &  ~5890 $\pm$ 160  &  790 $\pm$ 40  &  66 $\pm$ 7  &  17.6 $\pm$ 2.1 \\
VBF \Wjets            & ~~~~~7900 $\pm$ 1700  &  ~~~2200 $\pm$ 600  &  ~~~450 $\pm$ 160  &  ~~80 $\pm$ 40  &  10 $\pm$ 8  &  ~~3.9 $\pm$ 3.5 \\
\Zee            & {\textendash}    &   {\textendash} &    {\textendash}    & {\textendash}    & {\textendash}   &  {\textendash}  \\
\Zmumu            & ~~~2970 $\pm$ 130      &     {\textendash}     &     {\textendash}    &   {\textendash}  &     {\textendash}    &    {\textendash}     \\
\Ztautau            & ~~~2550 $\pm$ 110     &       {\textendash}     &     {\textendash}    &    {\textendash}    &     {\textendash}    &    {\textendash}     \\
\Znunu            & ~~~816\,000 $\pm$ 10\,000  &  189\,000 $\pm$ 4000  &  32\,000 $\pm$ 1000  &  5120 $\pm$ 250  &  520 $\pm$ 50  &  157 $\pm$ 18 \\
VBF \Zjets            & ~~13\,600 $\pm$ 2800  &  ~~~~~4800 $\pm$ 1200  &  ~1210 $\pm$ 350  &  260 $\pm$ 90  &  ~~35 $\pm$ 14  &  13 $\pm$ 6 \\
$\ttbar$ and  \singletop            & ~~46\,700 $\pm$ 2100  &  ~~~6900 $\pm$ 400  &  ~610 $\pm$ 70  &  ~~45 $\pm$ 14  &      {\textendash}    &     {\textendash}    \\
Diboson            & ~~26\,000 $\pm$ 5000  &  ~~~~~7700 $\pm$ 1500  &  ~1600 $\pm$ 400  &  310 $\pm$ 80  &  ~~38 $\pm$ 12  &  13 $\pm$ 4 \\
Multijet   & ~~~~~7000 $\pm$ 7000 & ~~~~~500 $\pm$ 500 & ~~~5.3 $\pm$ 5.3 & ~~0.2 $\pm$ 0.2 &  {\textendash} &  {\textendash} \\
Noncollision background    & ~~~~~1200 $\pm$ 1200 & ~~~~~160 $\pm$ 160 & ~~~29 $\pm$ 29 & ~~6 $\pm$ 6 & {\textendash}  &  {\textendash} \\
\noalign{\smallskip}\hline\noalign{\smallskip}
SUSY, $m(\tilde{t},\,\tilde{\chi}^{0}) = (600,\,580)~\GeV$  & ~2840 $\pm$ 340  &  ~1560 $\pm$ 190  &  ~640 $\pm$ 80  &  195 $\pm$ 24~~  &  34 $\pm$ 4  &  10.8 $\pm$ 1.3 \\
DMA, $m(\chi,\,Z_A) = (1,\,2000)~\GeV$  & ~~~3960 $\pm$ 160  &  ~2150 $\pm$ 80  &  ~918 $\pm$ 34  &  330 $\pm$ 13  &  82 $\pm$ 4 & 34.5 $\pm$ 1.7\\
Dark energy, $M_{2} = 1486~\GeV$    & ~1740 $\pm$ 50  &  ~1106 $\pm$ 33  &  558 $\pm$ 27  &  235 $\pm$ 16~  &  68 $\pm$ 6  &  30.4 $\pm$ 3.3 \\
\hline
\end{tabularx}
\end{tiny}
\end{center}
\label{tab:srincl}
\end{table}

\begin{table}[!tb]
\caption{Data and SM background predictions in the signal region for several exclusive $\mymet$ selections, as determined using a binned likelihood fit in the control regions.
For the SM prediction, both the statistical and systematic uncertainties are included.
In each signal region, the individual uncertainties for the different background processes can be correlated,
and do not necessarily add in quadrature to equal the total background uncertainty.
The dash ``\textendash'' denotes negligible background contributions.
For illustration, the expected event yields for particular signals for new phenomena are provided; in this case, the quoted errors include experimental uncertainties and theory uncertainties on the signal acceptance, as described in Sec.~\ref{sec:systematics:signal}.
}
\begin{center}
\begin{tiny}
\begin{tabularx}{\textwidth}{@{\extracolsep{\fill}}lcccccc}
\noalign{\smallskip}\hline\noalign{\smallskip}
{\normalfont \bfseries  Exclusive Signal Region} & {EM2} & {EM4} & {EM6} & {EM8} & {EM9} & {EM11}\\
\noalign{\smallskip}\hline\noalign{\smallskip}
Data events  (139 fb${}^{-1}$) & {313\,912} & {102\,888} & {10\,203} & {1663} & {738} & {187} \\\noalign{\smallskip}\hline\noalign{\smallskip}
SM prediction            & 314\,000 $\pm$ 3500  &  101\,600 $\pm$ 1200  &  10\,000 $\pm$ 180  &  1640 $\pm$ 40  &  754 $\pm$ 20 &182 $\pm$ 6 \\
\noalign{\smallskip}\hline\noalign{\smallskip}
\Wenu            & ~~16\,000 $\pm$ 1000  &  ~~~3980 $\pm$ 250  &  ~~~280 $\pm$ 19  &  ~~35.8 $\pm$ 2.6  &  13.4 $\pm$ 1.0 & ~~~~3.01 $\pm$ 0.24 \\
\Wmunu            & 23\,600 $\pm$ 500  &  ~~~5940 $\pm$ 120  &  ~~~481 $\pm$ 12  &  ~~66.8 $\pm$ 2.3  &  31.2 $\pm$ 1.2 & ~~~~7.8 $\pm$ 0.4 \\
\Wtaunu            & 54\,900 $\pm$ 800  &  15\,430 $\pm$ 260  &  ~1243 $\pm$ 29  &  167 $\pm$ 6  &  77.4 $\pm$ 2.9 &  ~~15.5 $\pm$ 0.8 \\
VBF \Wjets            & ~~~2340 $\pm$ 300  &  ~~~1010 $\pm$ 150  &  ~~~140 $\pm$ 27  &  ~~29 $\pm$ 7  &  16 $\pm$ 5 &  ~~~~5.0 $\pm$ 1.9 \\
\Zee            & {\textendash}    &     {\textendash} &      {\textendash}    &     {\textendash}    &    {\textendash} &    {\textendash}     \\
\Zmumu            & ~~~597 $\pm$ 15  &  ~~~97.4 $\pm$ 2.7  &  ~~~~~4.51 $\pm$ 0.15  &  ~~~~1.49 $\pm$ 0.05  &  ~~0.60 $\pm$ 0.02 &  {\textendash}     \\
\Ztautau            & ~~~530 $\pm$ 14  &  ~115.0 $\pm$ 3.3  &  ~~~~~8.31 $\pm$ 0.28  &  ~~~~~~0.90 $\pm$ 0.04  &  ~~0.40 $\pm$ 0.02  &  ~~~~2.10 $\pm$ 0.08 \\
\Znunu            & 192\,800 $\pm$ 2100  &  ~~67\,400 $\pm$ 1000  &  ~~~7000 $\pm$ 170  &  1180 $\pm$ 40  &  534 $\pm$ 20  &  126 $\pm$ 6 \\
VBF \Zjets            & ~~~3900 $\pm$ 500  &  ~~~2170 $\pm$ 310  &  ~~~370 $\pm$ 60  &  ~~~~86 $\pm$ 17  &  ~~45 $\pm$ 10  &  ~~13.7 $\pm$ 3.3 \\
\singletop            & ~~~2800 $\pm$ 700  &  ~~~~~550 $\pm$ 180  &  ~~~15 $\pm$ 8  &      {\textendash}    &    {\textendash} &    {\textendash}     \\
$\ttbar$            & ~~~8900 $\pm$ 700  &  ~~~2000 $\pm$ 150  &  ~100 $\pm$ 8  &  ~~~~8.2 $\pm$ 1.0  &  ~~2.4 $\pm$ 0.4  &  ~~~~0.30 $\pm$ 0.05 \\
Diboson            & ~~~~~6200 $\pm$ 1000  &  ~~~2700 $\pm$ 500  &  ~~~350 $\pm$ 70  &  ~~~~71 $\pm$ 15  &  33 $\pm$ 8   &  {\textendash} \\
Multijet   & ~~~~~1100 $\pm$ 1100 & ~~~~~57 $\pm$ 57 & ~~~~~0.6 $\pm$ 0.6 & ~~~~0.1 $\pm$ 0.1 & {\textendash} &  {\textendash} \\
Noncollision background    & ~~~~~240 $\pm$ 240 & ~~~~~46 $\pm$ 46 & ~~~~~8 $\pm$ 8 & ~~~~6 $\pm$ 6 &  {\textendash} &  {\textendash}  \\
\noalign{\smallskip}\hline\noalign{\smallskip}
SUSY, $m(\tilde{t},\,\tilde{\chi}^{0}) = (600,\,580)~\GeV$ & ~~~562 $\pm$ 70 & ~516 $\pm$ 60 & ~159 $\pm$ 19 & ~~44 $\pm$ 6 & 28 $\pm$ 4 & ~~~~8.2 $\pm$ 1.1 \\
DMA, $m(\chi,\,Z_{A}) = (1,\,2000)~\GeV$ & ~~~770 $\pm$ 30 & ~~~684 $\pm$ 27 & ~212 $\pm$ 9 & ~~79 $\pm$ 4 & 47.9 $\pm$ 2.3 & ~~18.7 $\pm$ 1.1\\
Dark energy, $M_{2} = 1486~\GeV$   & ~286 $\pm$ 7 & ~320 $\pm$ 11 & ~~125 $\pm$ 7 & ~~52 $\pm$ 5 & 33.6 $\pm$ 3.2 & ~~~14.6 $\pm$ 1.8 \\
\hline
\end{tabularx}
\end{tiny}
\end{center}
\label{tab:srexcl}
\end{table}


The results
are translated into upper limits on the presence of
new phenomena, using a simultaneous likelihood fit in both the control and signal regions,
and the ${\mathrm{CL}}_{\mathrm s}$ modified frequentist approach~\cite{Read:2002hq}.   As already mentioned,
inclusive regions with minimum $\mymet$ thresholds
are used to set model-independent exclusion limits, and
the exclusive regions are used for the interpretation of the results
within different models of new physics.  For the latter, the presence of a slight excess of events at high $\mymet$ limits the reach of the observed limits, mostly for those models in which the expected signal would accumulate in the tail of the  $\mymet$ distribution.

\begin{table}[!tb]
\caption{ Data and SM background predictions in the signal region for the different selections,
as determined using a binned likelihood fit in the control regions.
For the SM predictions both the statistical and systematic uncertainties are included.
}
\begin{center}
\begin{footnotesize}
\begin{tabularx}{\textwidth}{@{\extracolsep{\fill}}lXr@{\,$\pm$\,}lXrX|lXr@{\,$\pm$\,}lXr}\hline
\multicolumn{7}{c}{{{\normalfont \bfseries Inclusive Signal Region}}} & \multicolumn{6}{c}{{{\normalfont \bfseries Exclusive Signal Region}}} \\
Region&&\multicolumn{2}{c}{Predicted}&&Observed &&Region && \multicolumn{2}{c}{Predicted}&&Observed \\ \hline
IM0 && 3\,120\,000 & 40\,000 && 3\,148\,643 && EM0 && 1\,783\,000 & 26\,000 && 1\,791\,624\\
IM1 && 1\,346\,000 & 16\,000 && 1\,357\,019 && EM1 && 753\,000 & 9000 && 752\,328\\
IM2 && 597\,000 & 8000 && 604\,691 && EM2 && 314\,000 & 3500 && 313\,912\\
IM3 && 286\,000 & 4000 && 290\,779 && EM3 && 140\,100 & 1600 && 141\,036\\
IM4 && 146\,400 & 2300 && 149\,743 && EM4 && 101\,600 & 1200 && 102\,888\\
IM5 && 45\,550 & 1000 && 46\,855 && EM5 && 29\,200 & 400 && 29\,458\\
IM6 && 16\,800 & 500 && 17\,397 && EM6 && 10\,000 & 180 && 10\,203\\
IM7 && 7070 & 240 && 7194 && EM7 && 3870 & 80 && 3986\\
IM8 && 3180 & 130 && 3208 && EM8 && 1640 & 40 && 1663\\
IM9 && 1560 & 80 && 1545 && EM9 && 754 & 20 && 738\\
IM10 && 720 & 60 && 807 && EM10 && 359 & 10 && 413\\
IM11 && 407 & 34 && 394 && EM11 && 182 & 6 && 187\\
IM12 && 223 & 19 && 207 && EM12 && 218 & 9 && 207\\
\hline
\end{tabularx}
\end{footnotesize}
\end{center}
\label{tab:srsum}
\end{table}
 

\subsection{Model-independent exclusion limits}\label{sec:modelindep}
 
Results obtained in inclusive $\mymet$ regions are translated into
model-independent observed and expected 95$\%$ CL upper limits
on the visible cross section, defined as the product of the production cross section,
acceptance and efficiency $\sigma \times A \times \epsilon$.  The limits are extracted
by dividing the 95$\%$ CL upper limit on the number of signal events by the
integrated luminosity, taking into consideration the systematic
uncertainties in the SM backgrounds  and the  uncertainty in the  integrated luminosity.
A likelihood fit is performed separately for each
of the inclusive regions IM0--IM12. The results are collected in Table~\ref{tab:indep}.
Values of $\sigma \times A \times \epsilon$
above \SI{736}{fb} (for IM0) and above \SI{0.3}{fb} (for IM12)  are excluded at 95$\%$~CL.
 
\begin{table}[!ht]
\caption{
Observed and expected 95$\%$ CL upper limits on the
number of signal events, $S_{\mathrm{obs}}^{95}$ and  $S_{\mathrm{exp}}^{95}$, and on the                                     visible cross section, defined as the product of cross section, acceptance and efficiency,
$\langle{\mathrm \sigma}\rangle_{\mathrm{obs}}^{95}$,
for the IM0--IM12 selections.}
\begin{center}
\setlength{\tabcolsep}{0.0pc}
{\footnotesize
\begin{tabularx}{\textwidth}{@{\extracolsep{\fill}}lccc}
\noalign{\smallskip}\hline\noalign{\smallskip}
{\normalfont \bfseries Selection} & $\langle{\mathrm \sigma}\rangle_{\mathrm{obs}}^{95}$~[fb]  & $S_{\mathrm{obs}}^{95}$  &
$S_{\mathrm{exp}}^{95}$ \\
\noalign{\smallskip}\hline\noalign{\smallskip}
$\mymet>\SI{200}{\GeV}$   &   $736~~~    $ &  102\,274~~~~~   & ${83\,000}^{+22\,000}_{-23\,000}$ \\\noalign{\smallskip}
$\mymet>\SI{250}{\GeV}$   &   $296~~~    $ &  41\,158~~~    & ${33\,800}^{+11\,300}_{-9400}$ \\\noalign{\smallskip}
$\mymet>\SI{300}{\GeV}$   &   $150~~~    $ &  20\,893~~~    & ${15\,400}^{+5900}_{-4300}$~~ \\\noalign{\smallskip}
$\mymet>\SI{350}{\GeV}$   &   $86~          $ &  11\,937~~~    & ~${8300}^{+3100}_{-2300}$ \\\noalign{\smallskip}
$\mymet>\SI{400}{\GeV}$   &   $52~          $ &  7214     & ~${4700}^{+1800}_{-1300}$ \\\noalign{\smallskip}
$\mymet>\SI{500}{\GeV}$   &   $21~          $ &  2918     & ${1930}^{+730}_{-540}$ \\\noalign{\smallskip}
$\mymet>\SI{600}{\GeV}$   &   $10~          $ &  1391     & ~~${940}^{+360}_{-260}$ \\\noalign{\smallskip}
$\mymet>\SI{700}{\GeV}$   &   ~~~$4.1       $ &  ~~574      & ~~${490}^{+190}_{-140}$ \\\noalign{\smallskip}
$\mymet>\SI{800}{\GeV}$   &   ~~~$2.1       $ &  ~~298      & ~~${277}^{+106}_{-77}$ \\\noalign{\smallskip}
$\mymet>\SI{900}{\GeV}$   &   ~~~$1.2       $ &  ~~164      & \,${168}^{+65}_{-47}$ \\\noalign{\smallskip}
$\mymet>\SI{1000}{\GeV}$   &   ~~~$1.3      $ &  ~~186      & \,${119}^{+45}_{-33}$ \\\noalign{\smallskip}
$\mymet>\SI{1100}{\GeV}$   &   ~~~$0.5      $ &  ~~~~73       & \,~~${75}^{+28}_{-21}$ \\\noalign{\smallskip}
$\mymet>\SI{1200}{\GeV}$   &   ~~~$0.3      $ &  ~~~~40       & \,~~${49}^{+19}_{-14}$ \\\noalign{\smallskip}
\hline
\end{tabularx}
}
\end{center}
\label{tab:indep}
\end{table}
 
 
\subsection{Model-dependent exclusion limits}
\label{sec:modeldep}
 
A simultaneous fit to the signal and control regions in the exclusive $\mymet$ bins
is performed, and used to set observed and expected 95\% CL exclusion limits on the
parameters of the different models under consideration.
Uncertainties in the signal and background predictions, and in the luminosity are considered, and
correlations between experimental systematic uncertainties in signal and background
predictions are taken into account. The contamination of
the control regions by signal events  is negligible.
 
\subsubsection{Weakly interacting massive particles}
 
As discussed in Section~\ref{sec:introduction},  simplified models are
considered  with the exchange of an
axial-vector or a pseudoscalar mediator
in the $s$-channel.
In the case of the exchange of an axial-vector mediator, and for
WIMP-pair production with $m_{Z_A}>2m_\chi$, typical
$A \times \epsilon$ values for the signal models  with a $2~\TeV$ mediator range
from $13\%$  to less than $1\%$ for the EM0 and EM12 selections, respectively, where the values refer to an initial simulated sample generated with a minimum transverse momentum of 150~GeV. Similarly, values for  $A \times \epsilon$
in the range between $13 \%$ and less than $1 \%$ are computed for the pseudoscalar mediator model
with $m_{Z_P} = 350~\GeV$ and $m_\chi = 1~\GeV$, where $1\%$ is already reached for the EM5 selection due to the softer \mymet spectrum.
 
Figure~\ref{fig:dm2}\protect\subref{fig:dm2:a}  shows the observed and expected 95\% CL exclusion
contours in the $m_{Z_A}$--$m_\chi$
parameter plane for a simplified model with an
axial-vector mediator, Dirac WIMPs, and couplings $g_q = 1/4$ and $g_\chi = 1$.
In the region $m_{Z_A} > 2 \times m_\chi$, mediator masses up to about $2.1~\TeV$ are excluded for
$m_\chi=1~\GeV$, extending significantly beyond previous  bounds.
The masses corresponding to the relic density~\cite{Boveia:2016mrp}
as determined by the Planck and WMAP
satellites~\cite{Planck2018,2013ApJS..208...19H}, within the WIMP dark-matter model and in the absence of any
interaction other than the one considered, are indicated in this figure as a line
that crosses the excluded region at $m_{Z_A} \sim 1500~\GeV$ and $m_\chi \sim 585~\GeV$.
\begin{figure}[htb]
\centering
\begin{subfigure}{0.495\textwidth}
\centering
\includegraphics[width=\textwidth]{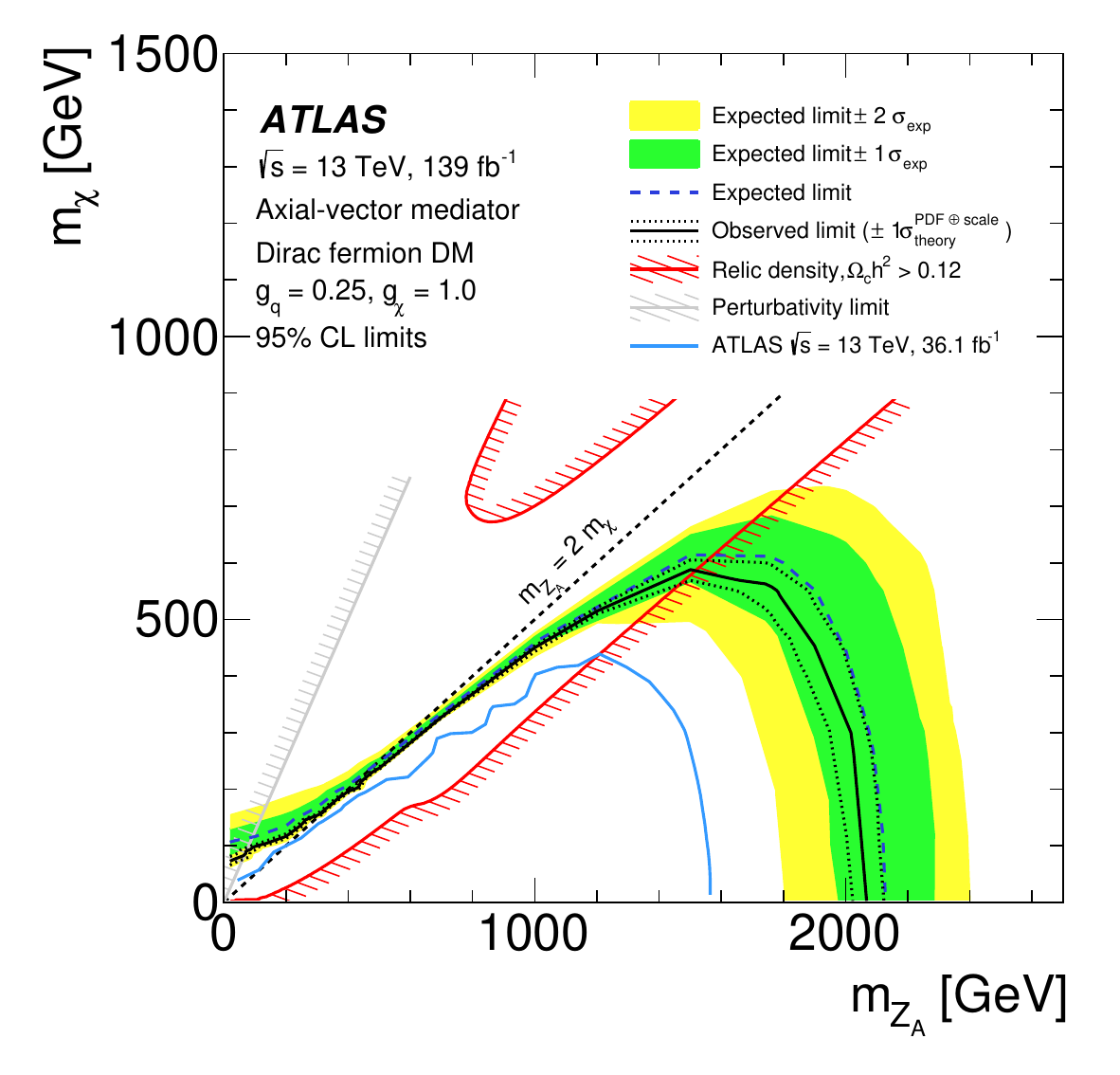}
\caption{}
\label{fig:dm2:a}
\end{subfigure}
\begin{subfigure}{0.495\textwidth}
\centering
\includegraphics[width=\textwidth]{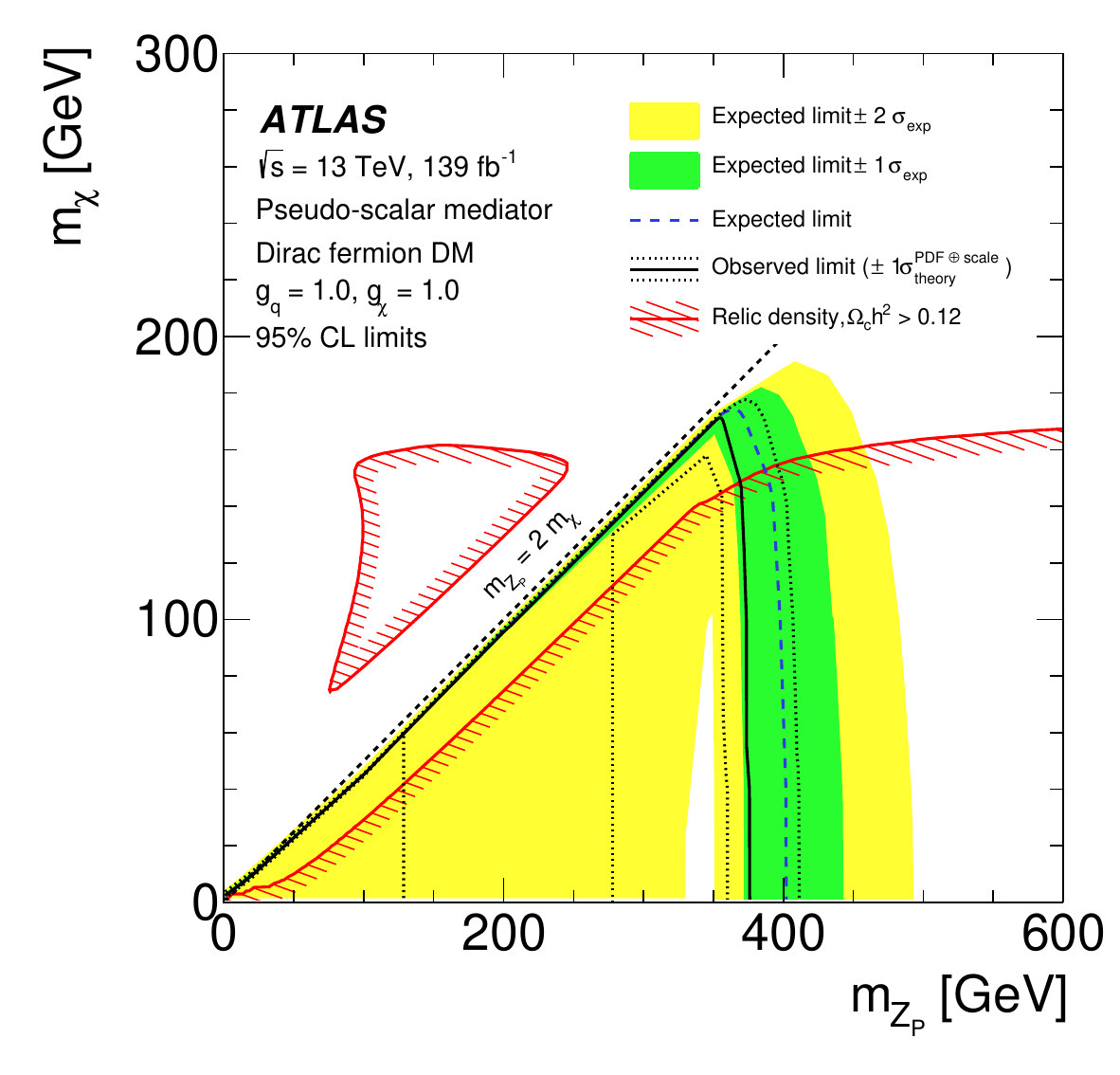}
\caption{}
\label{fig:dm2:b}
\end{subfigure}
\caption{
\protect\subref{fig:dm2:a}
95\% CL exclusion contours in the
$m_{Z_A}$--$m_\chi$
parameter plane for the axial-vector mediator model.
\protect\subref{fig:dm2:b}
95\% CL exclusion contours in the
$m_{Z_P}$--$m_\chi$
parameter plane for the pseudoscalar mediator model.
The
solid (dashed) curves show the
observed (expected) limits, while the
bands indicate the $\pm 1\sigma$  theory uncertainties in the
observed limit and the $\pm 1\sigma$  and $\pm 2\sigma$ ranges of the expected limit in the absence of a signal.
The red curves correspond to the set of points for which the expected relic density is consistent with the WMAP measurements (i.e.\ $\Omega h^2 = 0.12$), as computed with {\normalfont \scshape MadDM}~\cite{Backovic:2015cra}.
The area on the hashed side of the red contour (e.g.\ to the right of the red contour in the region $m_{Z_A} > 2 m_\chi$)
corresponds to predicted values of the relic density abundance inconsistent with the WMAP measurements.
The region excluded due to
perturbativity, defined by $m_\chi > \sqrt{\pi/2}\ m_{Z_A}$, is indicated by the gray
hatched area. The dotted lines indicate the kinematic limit for on-shell production $m_{Z_{A,P}} = 2 \times m_\chi$.
In the case of the pseudoscalar	mediator model,	the shape of the $2\sigma$ band at $m_{Z_P} \sim 350$~GeV is related to the rapid	increase of
the signal cross section at the threshold at which the mediator	 mass equals twice	the mass of the	top quark.
The same applies to the $1\sigma$ band on the observed limit.
In the case of the axial-vector mediator model,
the  results are compared with previous results from the ATLAS
Collaboration at $\sqrt{s} = 13~\TeV$ using $36.1 \ \ifb$~\cite{Aaboud:2017phn}.
}
\label{fig:dm2}
\end{figure}
 
Similarly, Figure~\ref{fig:dm2}\protect\subref{fig:dm2:b} presents observed and expected 95$\%$ CL
exclusion
contours in the $m_{Z_P}$--$m_\chi$
parameter plane for a simplified model with a
pseudoscalar mediator, Dirac WIMPs, and couplings $g_q = 1$ and $g_\chi = 1$.
For the first time,  the monojet final state in the ATLAS experiment  has enough sensitivity to
exclude a part of the parameter space in such a model.  Mediator masses below \SI{376}{GeV} are
excluded for very light WIMP candidates.

\begin{figure}[htb]
\centering
\begin{subfigure}{0.495\textwidth}
\centering
\includegraphics[width=\textwidth]{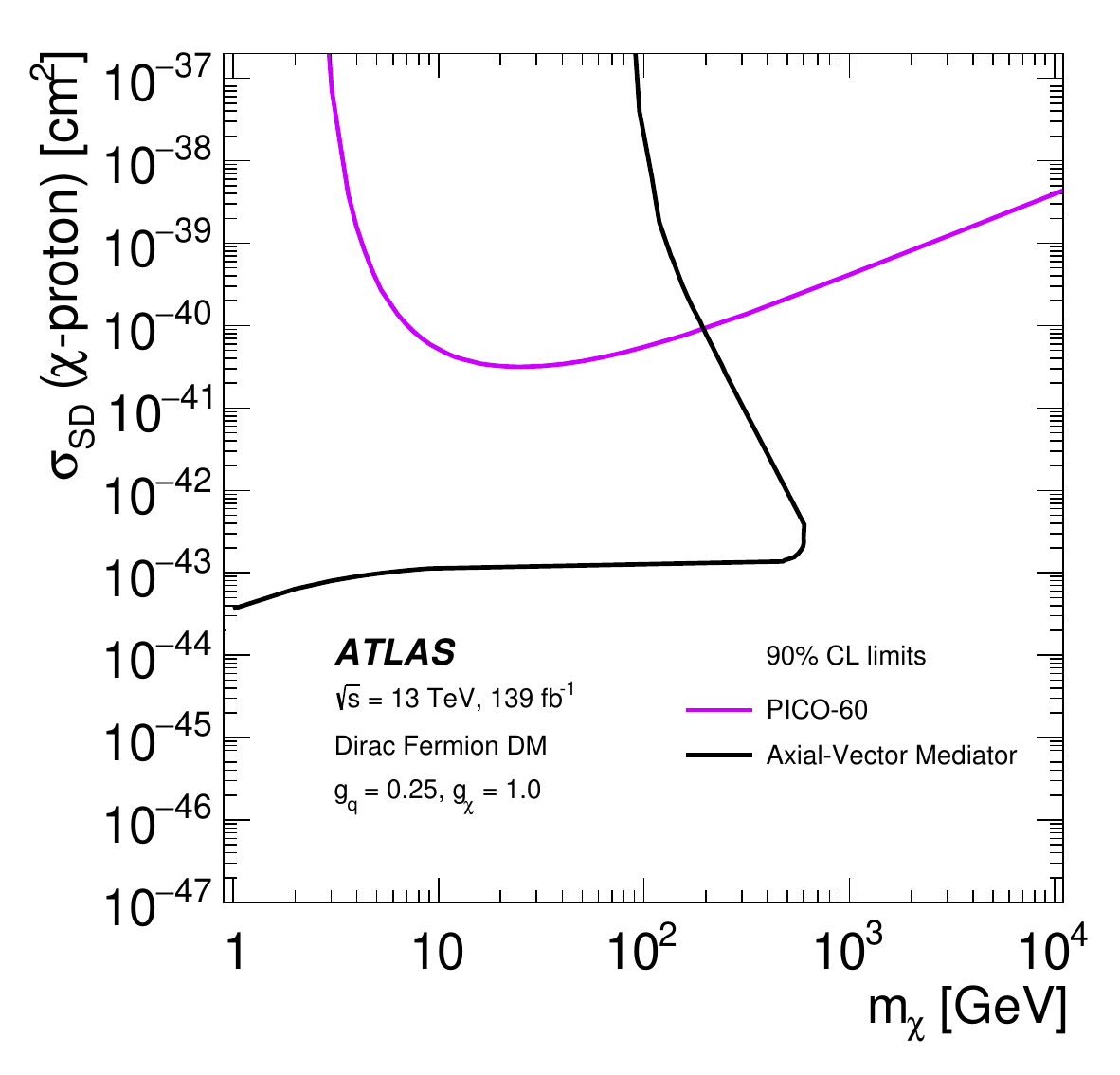}
\caption{}
\label{fig:ddirect:a}
\end{subfigure}
\begin{subfigure}{0.495\textwidth}
\centering
\includegraphics[width=\textwidth]{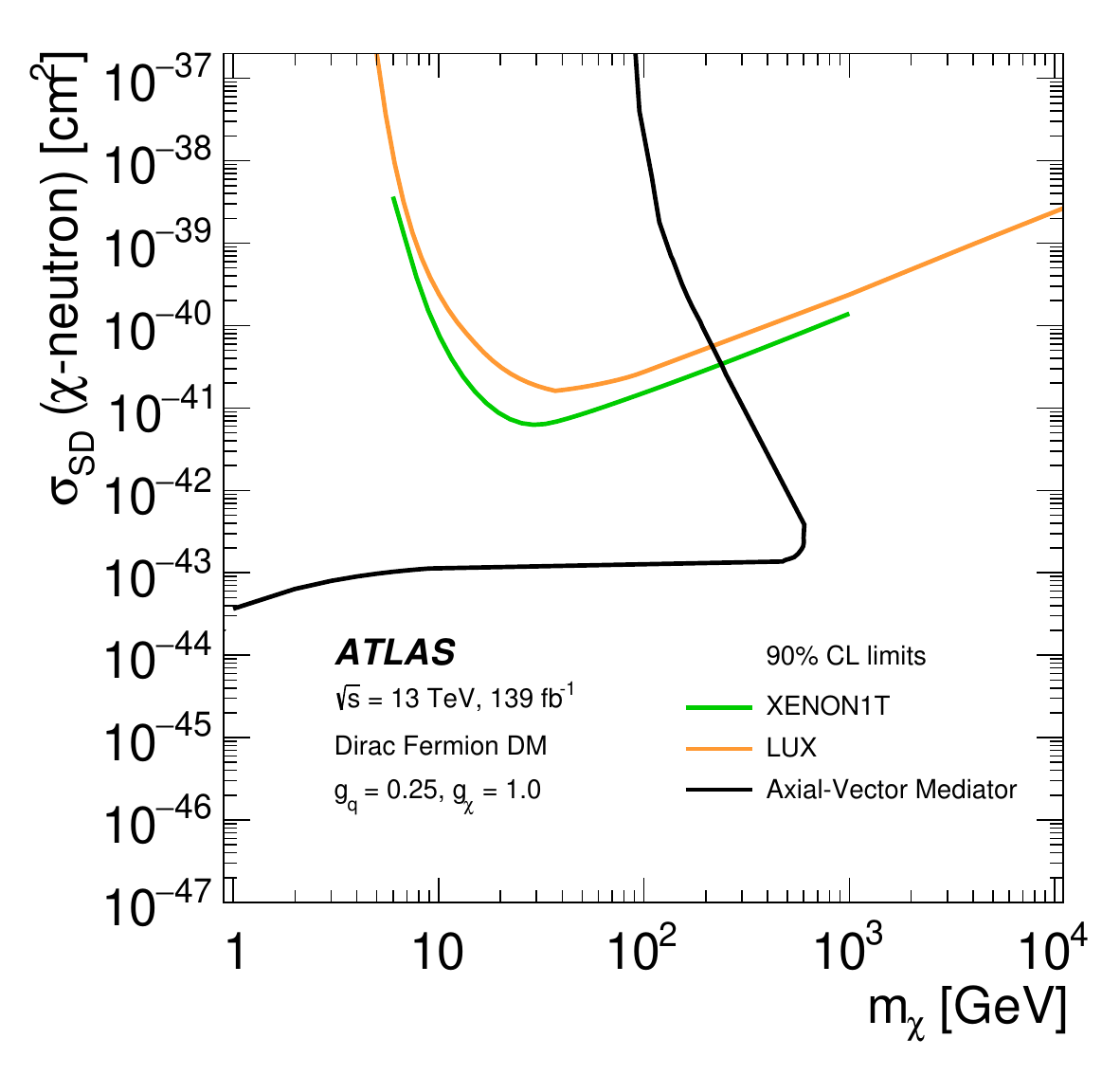}
\caption{}
\label{fig:ddirect:b}
\end{subfigure}
\caption{
A comparison of the inferred limits (black line) with the constraints from
direct-detection experiments on the spin-dependent (a)
WIMP--proton scattering cross section and (b)
WIMP--neutron scattering cross
section as a function of the WIMP mass, in the context of the simplified model with axial-vector
couplings. Unlike in the
$m_{Z_A}$--$m_\chi$
parameter plane, the limits are shown at
90\% CL. The results from this analysis, excluding the region to the left of the
contour, are compared with limits from
the
PICO~\cite{Amole:2019fdf} (purple line),  LUX~\cite{Akerib:2017kat}  (orange line),  and XENON1T~\cite{Aprile:2019dbj} (green line)
experiments.
The comparison is model-dependent and solely valid in the context of this model,
assuming minimal mediator width and the coupling values $g_q=1/4$ and $g_\chi=1$.
}
\label{fig:ddirect}
\end{figure}
 
In the case of the axial-vector mediator model,
the results are translated into 90\% CL exclusion
limits on the spin-dependent WIMP--nucleon scattering  cross section $\sigma_{\mathrm{SD}}$ as a function
of the WIMP mass, following the prescriptions from Refs.~\cite{Buchmueller:2014yoa,Boveia:2016mrp}.
Figure~\ref{fig:ddirect} shows exclusion limits for WIMP--proton and WIMP--neutron scattering cross sections as
a function of the WIMP mass, compared with the results from the PICO~\cite{Amole:2019fdf} experiment, and from the LUX~\cite{Akerib:2017kat} and
XENON1T~\cite{Aprile:2019dbj} experiments, respectively.
Stringent limits on the scattering cross section
of the order of $\SI{1.4e-43}{cm^2}$ for WIMP masses of about \SI{100}{GeV},  and $\SI{3e-44}{cm^2}$
for WIMP masses below $10~\GeV$ are
inferred from this analysis, which complement the results from direct-detection
experiments.
 
\subsubsection{Squark-pair production}
 
As in previous publications, different models of squark-pair production are considered: stop-pair production
with $\tilde{t}_1 \to  c +  \tilde{\chi}^{0}_{1}$, stop-pair production with
$\tilde{t}_1 \to  b + ff^{'} + \tilde{\chi}^{0}_{1}$, sbottom-pair production
with $\tilde{b}_1 \to b + \ninoone$, and squark-pair production with $\tilde{q}
\to q + \ninoone$ ($q =u,d,c,s$). In each case separately, the results are
translated into exclusion limits  as a function of the squark mass for different neutralino masses.
The region with stop--neutralino or sbottom-neutralino mass differences below $5~\GeV$ is not considered in the exclusion since in this regime the squarks could become long-lived. In such a compressed scenario, and for stop sbottom masses of about 600~$\GeV$, the typical value of
$A \times \epsilon$ for the selection criteria varies between 11$\%$ for EM0 and less than 1$\%$ for EM12, as computed using a sample
with a minimum missing transverse momentum of 150~\GeV. Comparable values  for $A \times \epsilon$
are obtained in the rest of the squark--neutralino mass plane.

Figure~\ref{fig:stop}\protect\subref{fig9a}  presents the results in the case of the $\tilde{t}_1 \to
c +  \tilde{\chi}^{0}_{1}$ decays.  In the compressed scenario with stop and neutralino nearly
degenerate in mass, masses up to \SI{550}{GeV} are excluded.  
Similarly, Figure~\ref{fig:stop}\protect\subref{fig9b}  shows the observed and expected 95$\%$ CL exclusion limits
as a function of the stop and neutralino masses for the  $\tilde{t}_1 \to b +
ff^{'} + \tilde{\chi}^{0}_{1}$ decay channel, assuming a branching ratio $\mathcal{B}=100\%$. For  $m_{\tilde{t}_1} - m_{\ninoone}  \sim m_b$, stop
masses up to \SI{550}{GeV} are  also excluded.
\begin{figure}[h]
\begin{center}
\begin{subfigure}{0.495\textwidth}
\centering
\includegraphics[width=\textwidth]{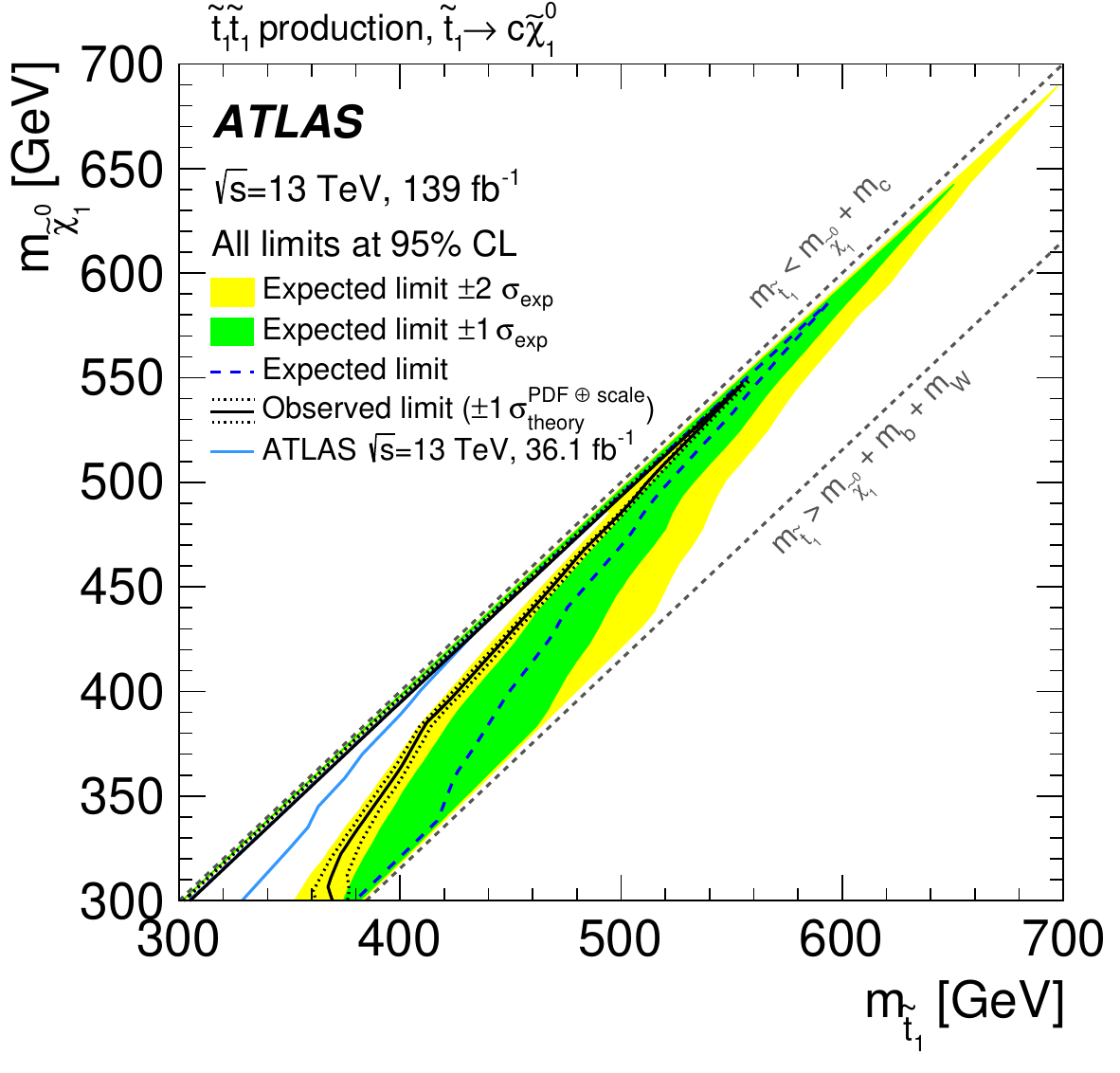}
\caption{}
\label{fig9a}
\end{subfigure}
\begin{subfigure}{0.495\textwidth}
\centering
\includegraphics[width=\textwidth]{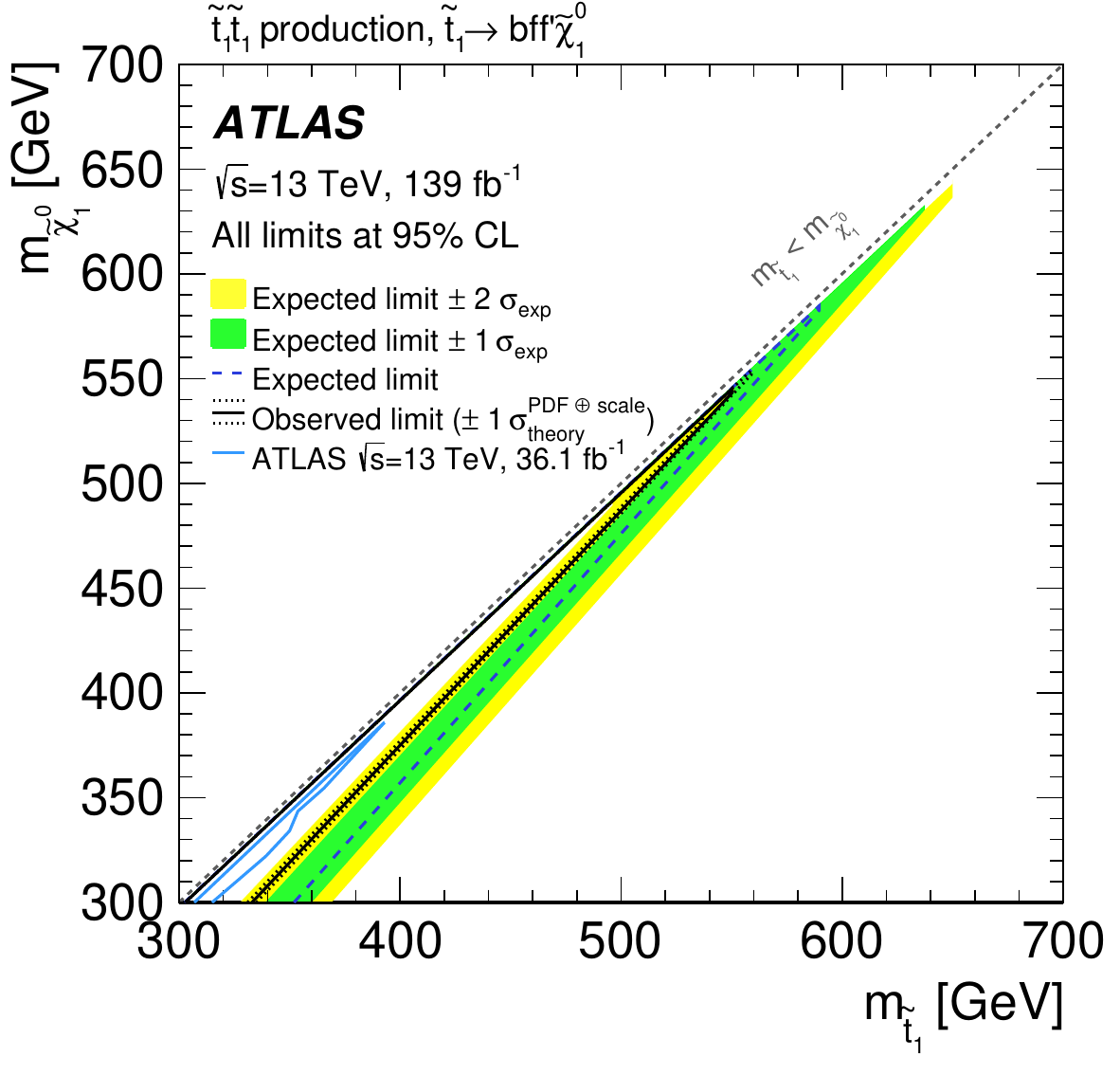}
\caption{}
\label{fig9b}
\end{subfigure}
\end{center}
\caption{
Excluded regions at the 95$\%$ CL in   the ($\tilde{t}_1,\tilde{\chi}^{0}_{1}$) mass plane
for \protect\subref{fig9a} the decay channel  $\tilde{t}_1 \to c +  \tilde{\chi}^{0}_{1}$
($\mathcal{B}=100\%$) and \protect\subref{fig9b}  the decay channel
$\tilde{t}_1 \to  b + ff^{'} + \tilde{\chi}^{0}_{1}$ ($\mathcal{B}=100\%$).
The dotted lines around the observed limits indicate the range of observed limits
corresponding to $\pm 1\sigma$ variations of the NNLO + NNLL SUSY cross-section
predictions.
The bands around the expected limits indicate
the expected $\pm1\sigma$ and $\pm 2\sigma$ ranges of limits in the absence of a signal.
The
results from this analysis are compared with previous results from the ATLAS
Collaboration at $\sqrt{s} = 13~\TeV$ using $36.1~\ifb$~\cite{Aaboud:2017phn}.
}
\label{fig:stop}
\end{figure}
Figure~\ref{fig:sbottom}\protect\subref{fig10a}  presents the  observed and expected 95$\%$ CL exclusion
limits as a function of the sbottom and neutralino masses for the  $\tilde{b}_1
\to b + \tilde{\chi}^{0}_{1}$ ($\mathcal{B}=100\%$) decay channel. In the
scenario with $m_{\tilde{b}_1} - m_{\ninoone} \sim m_b$, this analysis extends
the 95$\%$ CL exclusion limits up to a sbottom mass of \SI{545}{GeV}.
\begin{figure}[h]
\begin{center}
\begin{subfigure}{0.495\textwidth}
\centering
\includegraphics[width=\textwidth]{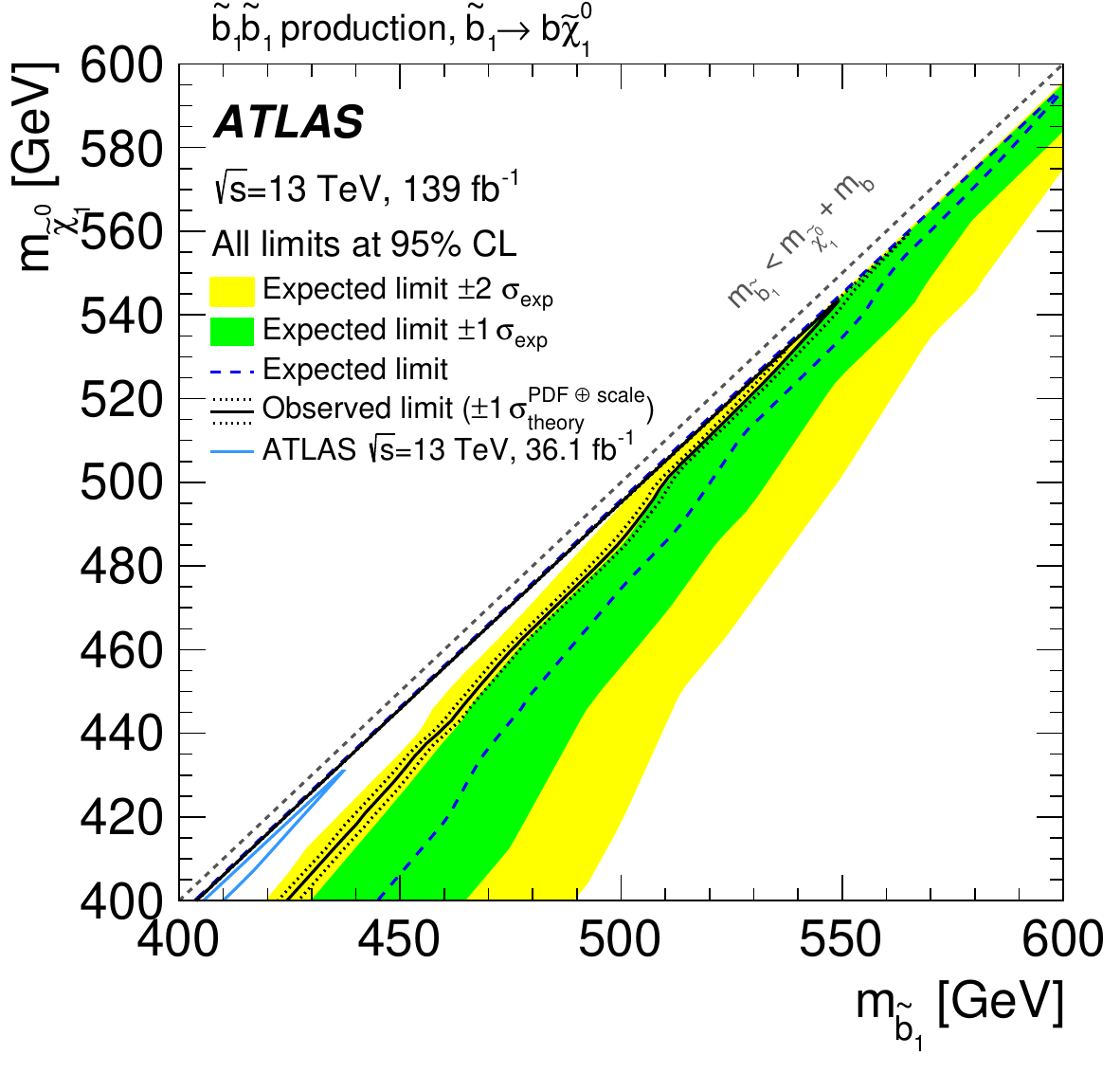}
\caption{}
\label{fig10a}
\end{subfigure}
\begin{subfigure}{0.495\textwidth}
\centering
\includegraphics[width=\textwidth]{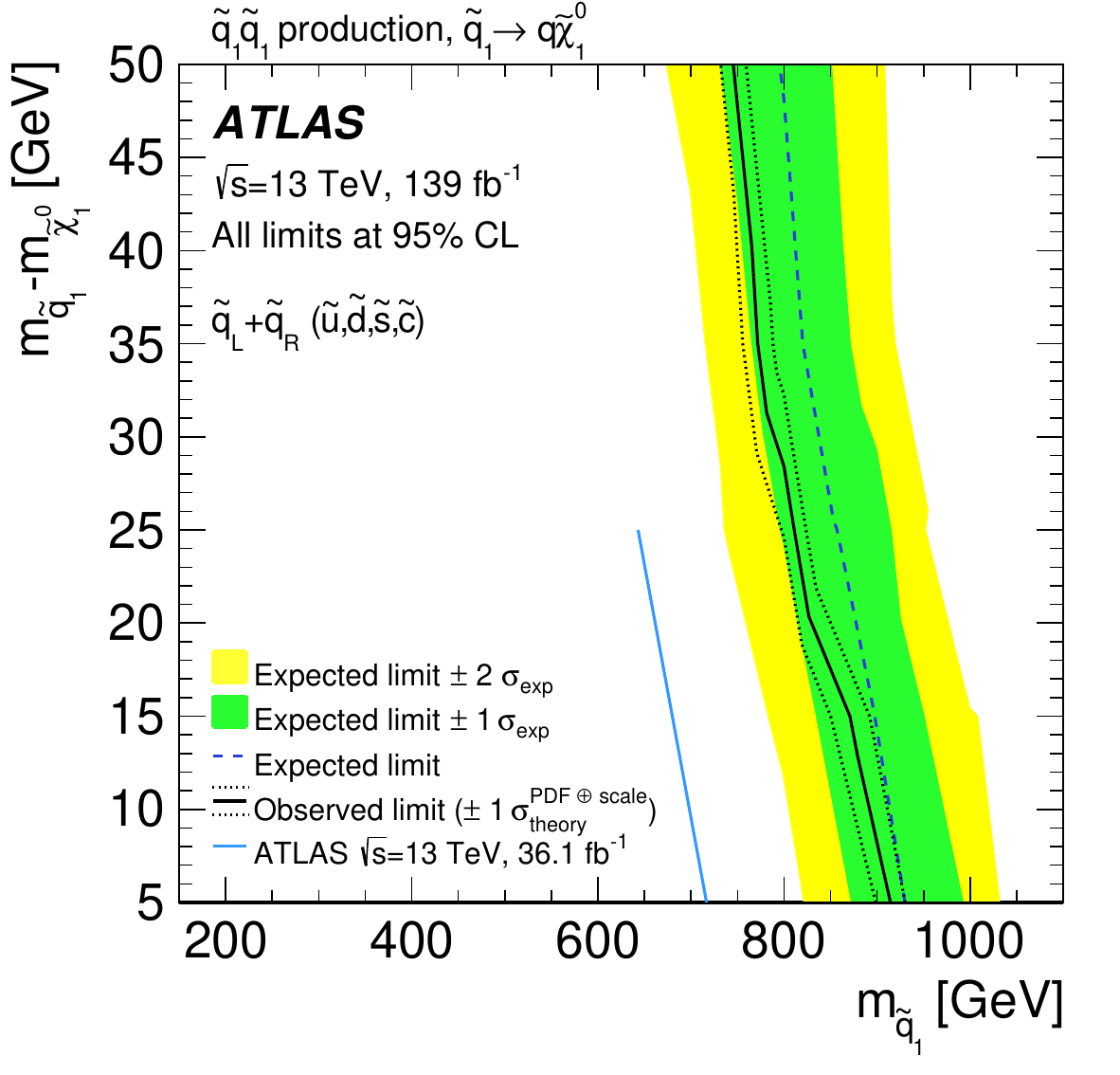}
\caption{}
\label{fig10b}
\end{subfigure}\end{center}
\caption{
\protect\subref{fig10a}  Exclusion plane at 95$\%$ CL as a function of sbottom and neutralino masses for
the decay channel  $\tilde{b}_1 \to  b +  \tilde{\chi}^{0}_{1}$ ($\mathcal{B}=100\%$).
\protect\subref{fig10b}  Exclusion region at 95$\%$ CL as a function of squark mass and the
squark--neutralino mass difference for $\tilde{q} \to q + \tilde{\chi}^{0}_{1}$ and $\tilde{q}_\mathrm{L} + \tilde{q}_\mathrm{R}$  with
($\tilde{u},\tilde{d},\tilde{c},\tilde{s}$).
The dotted lines around the observed limit indicate the range of observed limits
corresponding to $\pm 1\sigma$ variations of the NNLO + NNLL SUSY cross-section
predictions.
The bands around the expected limit indicates the expected $\pm1\sigma$ and $\pm 2\sigma$
ranges of limits in the absence of a signal.
The results from this analysis are compared with previous results from the ATLAS
Collaboration at $\sqrt{s} = 13~\TeV$ using $36.1~\ifb$~\cite{Aaboud:2017phn}.}
\label{fig:sbottom}
\end{figure}
Finally, Figure~\ref{fig:sbottom}\protect\subref{fig9b}   presents the observed and expected 95$\%$ CL
exclusion limits as a function of the squark mass and the squark--neutralino mass
difference for $\tilde{q} \to q + \ninoone$ ($q =u,d,c,s$).
In the compressed
scenario, squark masses below \SI{925}{GeV}
are excluded at 95$\%$ CL. Altogether, these results significantly improve upon the previous
exclusion limits based on $36.1~\ifb$ of total integrated
luminosity~\cite{Aaboud:2017phn}. In the very compressed scenario,  the observed
limits on the squark masses are extended by more than 100 GeV.
 
\subsubsection{Dark-energy-inspired model}
 
Exclusion limits are computed for the Horndeski dark-energy model (see Section~\ref{sec:introduction}) with $m_{\varphi}=0.1$ GeV, and
considering only the terms relevant for the monojet final state, following the work described in Ref.~\cite{Aaboud:2019yqu}.
The sensitivity remains independent of the $m_{\varphi}$ value  considered for light particles up to masses of the order of 1~\GeV.
The typical value of
$A \times \epsilon$ for the selection criteria varies between 8.2$\%$ for EM0 and less than 1$\%$ for EM12,
as computed using a sample with a minimum missing transverse momentum of 150 GeV.
Figure~\ref{fig:DE_xsec_vs_M} shows the observed and expected contours at $95\%$ CL on the $\sigma$--$M_{2}$ plane.  Values for $M_{2}$ below \SI{1486}{GeV} are excluded, which represents a significant improvement over the limits previously obtained.
The validity of the effective implementation of the model at the LHC energies was studied previously~\cite{Aaboud:2019yqu} by truncating the signal
contributions with $\sqrt{\hat{s}} < g_{*} M_2$, where $\sqrt{\hat{s}}$ is the center-of-mass energy of the hard interaction and
$g_{*}$ is a number that depends on the UV completion of the model which must satisfy $g_{*}$ $<$ 4$\pi$ in order for the couplings to be in the perturbative regime, leading to a negligible effect on the obtained exclusion limits for $g_{*} > 3.5$.

\begin{figure}[h]
\centering
\includegraphics[width=0.5\textwidth]{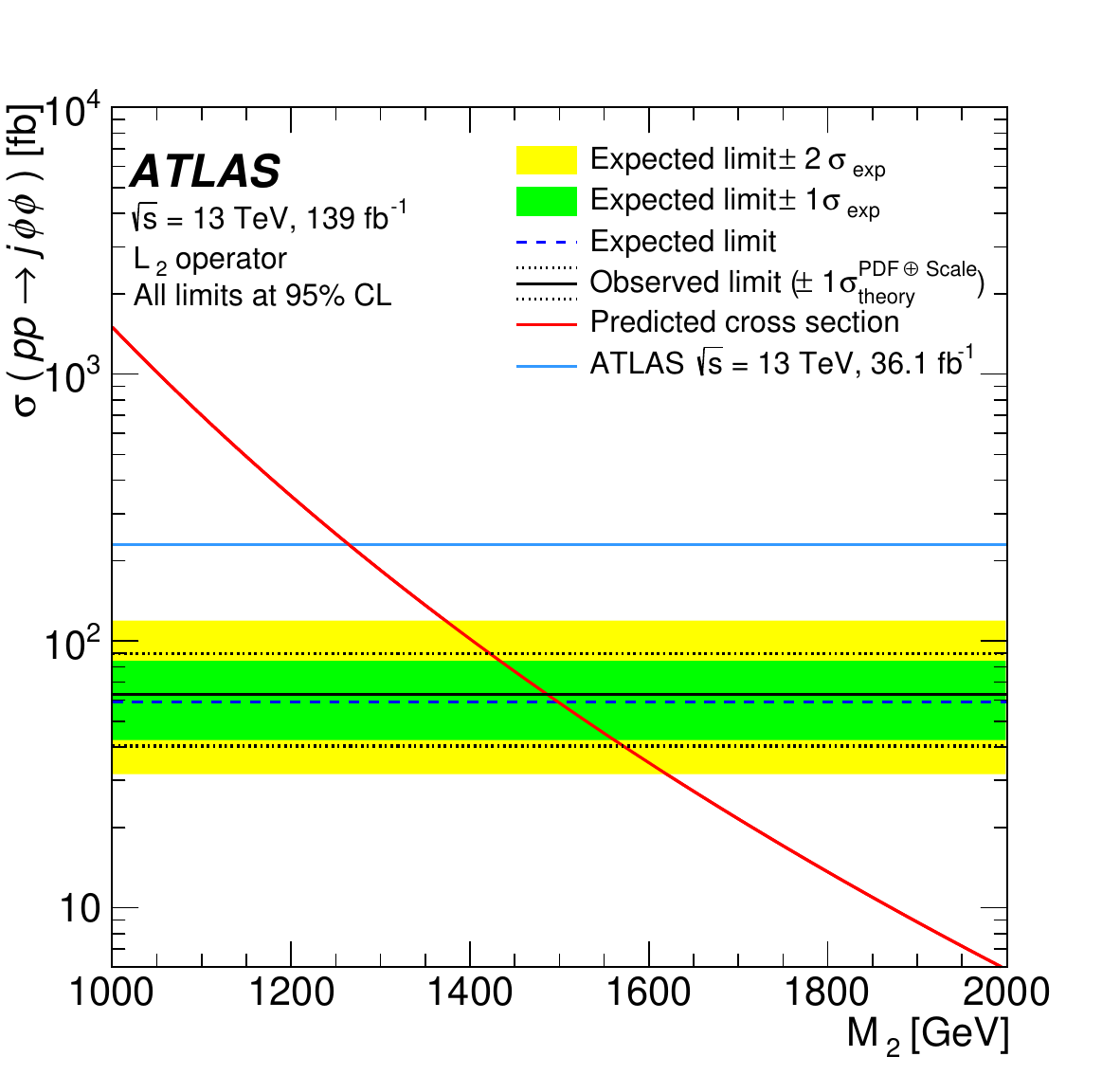}
\caption{Observed (solid line) and expected (dashed line) exclusions at $95\%$ CL on the Horndeski dark-energy model for $m_{\phi}=0.1$ GeV and $c_{i\neq 2}=0$ , $c_{2}=1$~\cite{Aaboud:2019yqu}, 
expressed in terms of the visible cross section as a function of the suppression scale $M_{2}$. The results are compared with the theoretical predictions.
The dotted lines around the observed limits indicate the range of observed limits corresponding to $\pm 1\sigma$ variations of the  cross-section predictions. The results from this analysis are compared with previous results from the ATLAS Collaboration at $\sqrt{s} = 13~\TeV$ using $36.1~\ifb$~\cite{Aaboud:2019yqu}.
}
\label{fig:DE_xsec_vs_M}
\end{figure}

\subsubsection{Large extra spatial dimensions}
 
The results are translated into limits  on the parameters of
the ADD model. As in previous analyses,
only the signal regions with $\mymet > 400~\GeV$ are employed, with
sufficient sensitivity to ADD signal.
The typical value of $A \times \epsilon$ for the selection criteria, as computed from a simulated sample with missing transverse momentum above 350~\GeV,
is of the order of 6$\%$ for EM4  and is less than $1\%$ for EM12.
Figure~\ref{fig:add} and Table~\ref{tab:add} present the results.
Values of $M_D$ below $11.2~\TeV$ at $n=2$ and below $5.9~\TeV$ at $n=6$ are excluded at
95$\%$~CL,
which improve on the exclusion limits from previous results using 36.1~$\ifb$ of $13~\TeV$ data~\cite{Aaboud:2017phn}.
As already noted in
Ref.~\cite{Aaboud:2017phn},
the  analysis partially probes the phase-space region with $\hat{s} > M_D^2$.
The  suppression of this kinematic region in computing the 95$\%$  CL lower limits on $M_D$ translates into a negligible effect on the results.

\begin{table}[!hb]
\caption{
The $95\%$ CL observed and expected lower limits on the fundamental Planck
scale in $4+n$ dimensions, $M_D$, as a function of the number of extra
dimensions $n$, considering nominal LO signal cross sections.
The impact of the $\pm 1\sigma$  theoretical uncertainty on
the observed limits and the  expected $\pm 1\sigma$ range of limits in the
absence of a signal are also given.
}
\label{tab:add_limits}
\begin{center}
\begin{footnotesize}
\begin{tabular*}{\textwidth}{@{\extracolsep{\fill}}lcc} \hline\noalign{\smallskip}
\multicolumn{3}{c}{{\normalfont \bfseries ADD Model Limits on $M_{D}$
} (95\% CL)} \\ 
& Expected [TeV]& Observed [TeV]\\ \noalign{\smallskip}\hline\noalign{\smallskip}
$n = 2$ & $11.6^{+1.0}_{-1.0}$~~ & $11.2^{+1.0}_{-1.3}$~~ \\
$n = 3$ & $8.6^{+0.6}_{-0.6}$ & $8.5^{+0.6}_{-0.8}$ \\
$n = 4$ & $7.2^{+0.4}_{-0.4}$ & $7.1^{+0.4}_{-0.6}$ \\
$n = 5$ & $6.4^{+0.3}_{-0.3}$ & $6.4^{+0.3}_{-0.5}$ \\
$n = 6$ & $5.9^{+0.2}_{-0.2}$ & $5.9^{+0.3}_{-0.4}$ \\
\hline
\end{tabular*}
\end{footnotesize}
\end{center}
\label{tab:add}
\end{table}

\begin{figure}[htb]
\begin{center}
\includegraphics[width=0.5\textwidth]{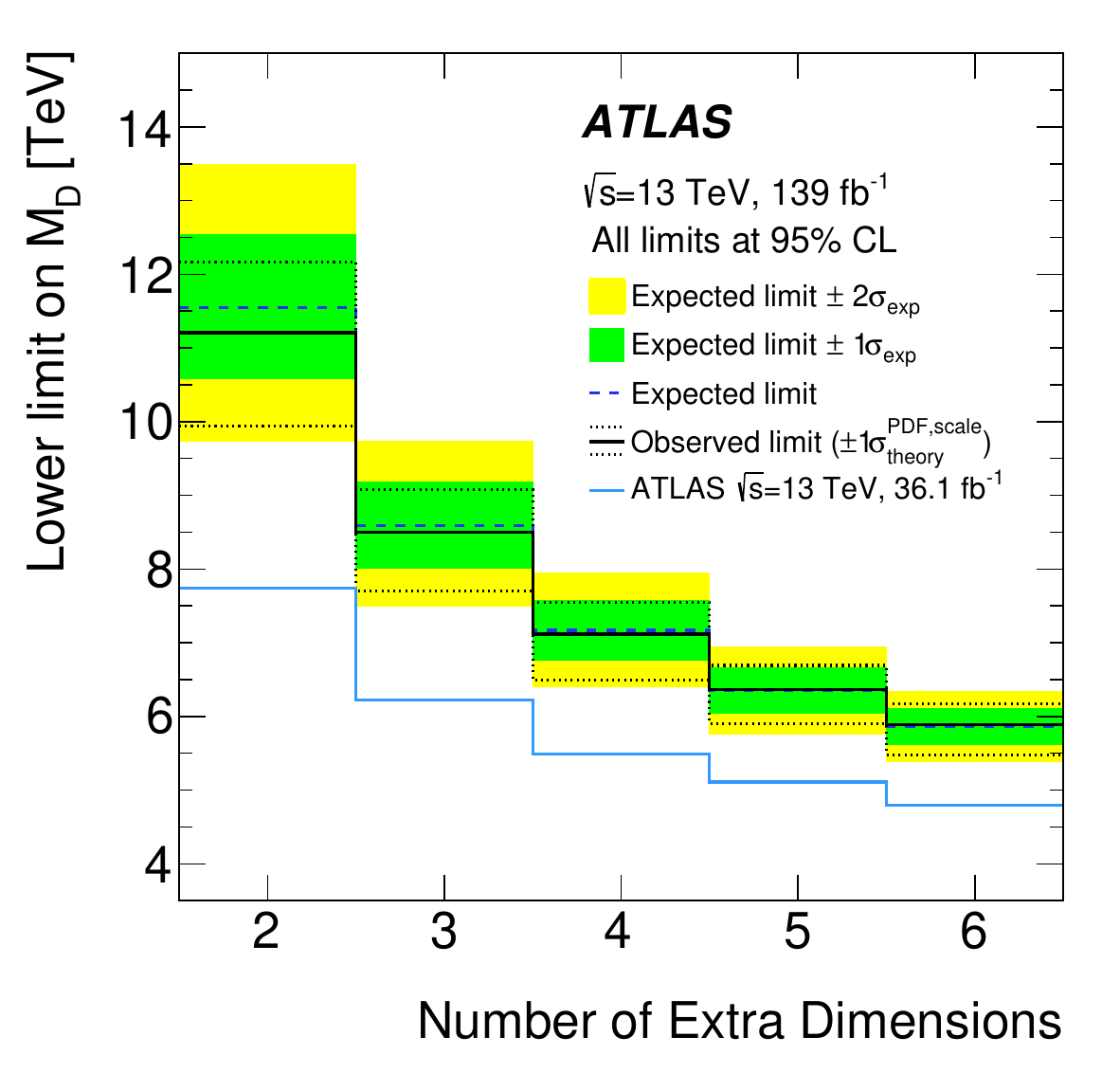}
\end{center}
\caption{
Observed and expected 95$\%$ CL lower limits on the fundamental Planck scale in $4+n$ dimensions, $M_D$, as a function of the number
of extra dimensions.
The bands indicate the $\pm 1\sigma$  theory uncertainties in the
observed limit and the $\pm 1\sigma$  and $\pm 2\sigma$ ranges of the expected limit in the absence of a signal.
The results from this analysis are compared with previous results from the ATLAS
Collaboration
using $36.1~\ifb$ of $\sqrt{s} = 13~\TeV$ data~\cite{Aaboud:2017phn}.
}
\label{fig:add}
\end{figure}
 

\subsubsection{Axion-like particles}
Results are expressed in terms of 95$\%$ CL limits on the parameters of the ALP model.  As in the case of the ADD model, the kinematic region with $\mymet > 400$~GeV provides the best sensitivity.
Figure~\ref{fig:alps} shows 95$\%$ exclusion contours in the the $\CG$--$f_a$ plane,  for an axion mass of \SI{1}{MeV}. The exclusion does not depend significantly on the axion mass for masses up to at least \SI{1}{GeV}.  The limits on $\CG$ increase linearly with $f_a$.
For  $f_a = \SI{1}{TeV}$,  couplings $\CG$ above 0.008 are excluded. Expressed in terms of the ${\CG}/{f_a}$ ratio, values above $\SI{8e-6}{GeV^{-1}}$ are
excluded at 95$\%$ CL.
As in the case of the dark energy and ADD models, the validity of the effective field implementation of the model is challenged for $\hat{s} > f_a^2$. For values of $f_a$  below \SI{2}{TeV}, the signal yields are reduced significantly when applying a suppressing weighting factor $f_a^4/\hat{s}^2$ for events with $\hat{s} > f_a^2$.  The effect is reduced to about $5\%$ for  $f_a = \SI{2}{TeV}$ and it is negligible for $f_a$ above \SI{3}{TeV}.
 
\begin{figure}[htb]
\begin{center}
\includegraphics[width=0.5\textwidth]{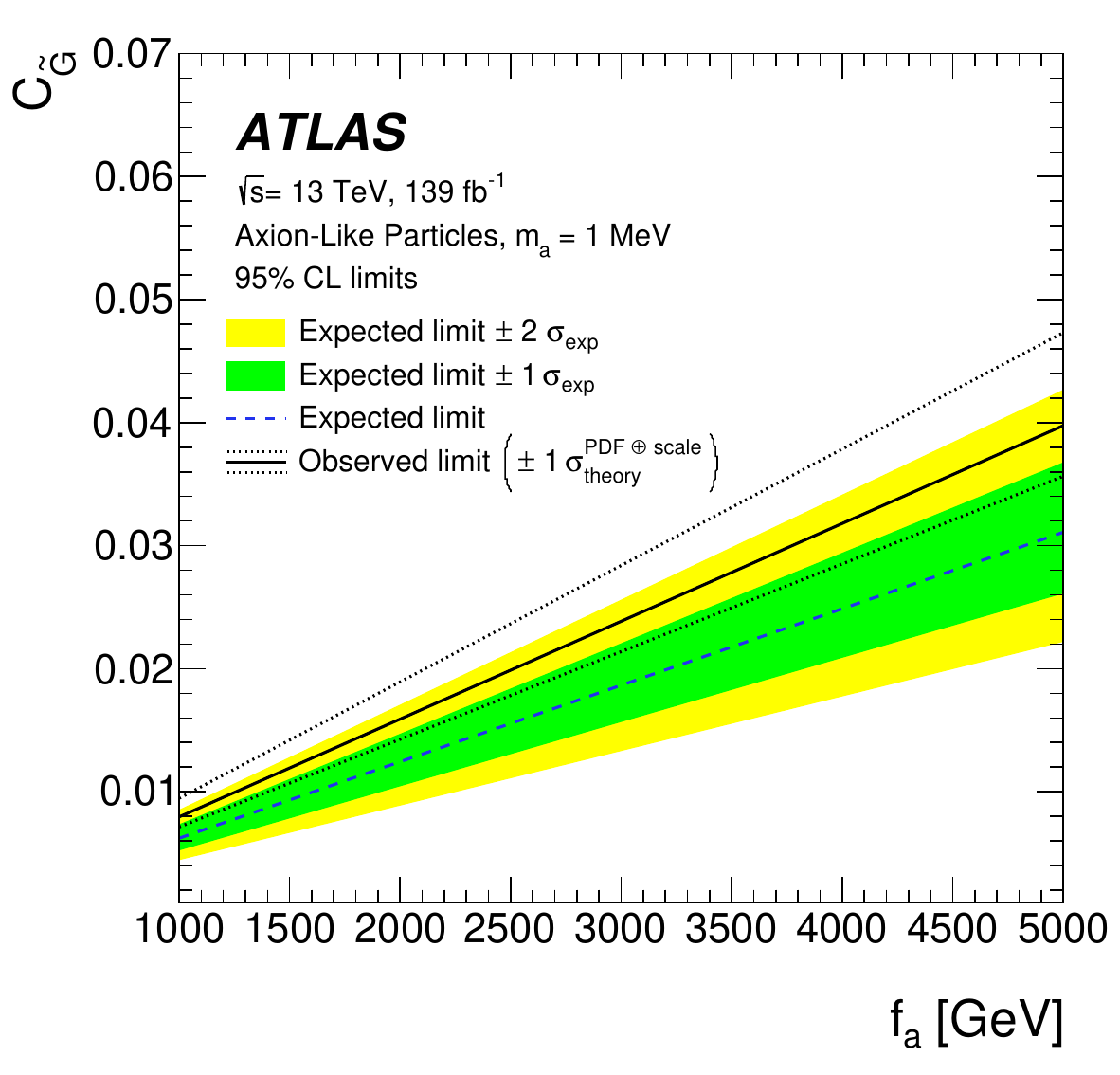}
\end{center}
\caption{
Observed and expected 95$\%$ CL upper limits on the coupling $\CG$ as a function of the effective scale $f_a$  for ALP mass of 1 MeV.
The bands indicate the $\pm 1\sigma$  theory uncertainties in the
observed limit and the $\pm 1\sigma$  and $\pm 2\sigma$ ranges of the expected limit in the absence of a signal.
The 95$\%$ CL limits are computed with no suppression  of the events with
$\hat{s} > f_a^2$.
}
\label{fig:alps}
\end{figure}
 
 
\subsubsection{Invisibly decaying Higgs boson}
The results are interpreted in terms of 95$\%$ CL upper limits on the branching ratio for an invisibly decaying Higgs boson.  The signal yields are dominated by gluon--gluon fusion production processes (about $73\%$), followed by the contributions from VBF (18$\%$),  $VH$ (8$\%$), and $\ttbar +H$ (1$\%$)  processes.  The low $\met$ region plays an important role in enhancing the sensitivity of the data to the Higgs signal and the full $\met$ spectrum is employed in computing the limits.
The observed agreement between data and the SM  background predictions in the measured $\met$ distribution leads to a 95\% CL observed (expected) exclusion limit on the invisible branching ratio of the Higgs boson of $0.34$ ($0.39_{-0.11}^{+0.16}$).
 
 
\FloatBarrier

\section{Conclusions}
\label{sec:conclusions}
 
Results are reported from a search for new phenomena in events with an
energetic jet and large missing transverse momentum in proton--proton
collisions at $\sqrt{s}=13~\TeV$ at the LHC, based on data corresponding
to an integrated luminosity of $139\,\ifb$ collected by the ATLAS
detector during 2015--2018.
Compared to previous publications, in addition to an increase of almost a factor of four in the data size,
the analysis implements a number of improvements in the signal selection and the background determination leading to enhanced sensitivity.
The measurements are in agreement with the
SM predictions.  The results are translated into model-independent
95$\%$ CL upper limits on the visible cross section for new phenomena, and these range from 736~fb to 0.3~fb with  increasing missing transverse momentum.
Improved bounds on the parameters for a variety of models for new phenomena have been derived.
In the case of simplified models for WIMP-pair
production in the $s$-channel, with Dirac fermions as dark-matter candidates,
an axial-vector mediator with masses below \SI{2.1}{TeV} is excluded at 95$\%$ CL for  very light WIMPs and coupling values $g_q = 1/4$ and $g_\chi = 1$.  For the first time, the ATLAS monojet analysis reaches sensitivity for excluding pseudoscalar mediators with masses below \SI{376}{GeV}, for very light WIMPs and coupling values $g_q = 1$ and $g_\chi = 1$.
 
Similarly, the results are interpreted in terms of a search for
squark-pair production in a compressed-mass supersymmetric scenario.
In the case of stop- and sbottom-pair production with $\tilde{t}_1 \to  c +  \tilde{\chi}^{0}_{1}$ or
$\tilde{t}_1 \to  b + ff^{'} + \tilde{\chi}^{0}_{1}$
and $\tilde{b}_1 \to  b +  \tilde{\chi}^{0}_{1}$, respectively,  squark masses
below about \SI{550}{GeV} and \SI{550}{GeV} are excluded at 95$\%$~CL, thus surpassing previous exclusions by almost \SI{100}{GeV}. In the case of squark-pair production with
$\tilde{q} \to  q +  \tilde{\chi}^{0}_{1}  \ (q=u,d,c,s)$,  squark masses below \SI{925}{GeV} are excluded.
 
The results are expressed in terms of 95$\%$ CL limits on the suppression scale $M_2$ for the Horndeski dark-energy model with $m_{\phi}=0.1$ GeV and $c_{i\neq 2}=0$ , $c_{2}=1$. Suppression scales $M_2$ below about \SI{1.5}{TeV}  are excluded.
In the case of the ADD model with large extra spatial dimensions, 95$\%$ CL lower limits on the fundamental Planck scale $M_D$ in $4+n$ dimensions vary in the range between \SI{11.2}{TeV} and \SI{5.9}{TeV} for $n=2$ and $n=6$, respectively.
In models with axion-like particles with coupling to gluons, couplings-to-effective-scale ratios $\CG/f_a$ above  $\SI{8e-6}{GeV^{-1}}$ are
excluded at 95$\%$ CL for axion masses up to \SI{1}{GeV}.
Finally,  limits are obtained for the branching ratio of an invisibly decaying Higgs boson. Branching fractions above $0.34$ are excluded at 95$\%$ CL.
 
\FloatBarrier
\section*{Acknowledgments}
 

We thank CERN for the very successful operation of the LHC, as well as the
support staff from our institutions without whom ATLAS could not be
operated efficiently.
 
We acknowledge the support of ANPCyT, Argentina; YerPhI, Armenia; ARC, Australia; BMWFW and FWF, Austria; ANAS, Azerbaijan; SSTC, Belarus; CNPq and FAPESP, Brazil; NSERC, NRC and CFI, Canada; CERN; ANID, Chile; CAS, MOST and NSFC, China; COLCIENCIAS, Colombia; MSMT CR, MPO CR and VSC CR, Czech Republic; DNRF and DNSRC, Denmark; IN2P3-CNRS and CEA-DRF/IRFU, France; SRNSFG, Georgia; BMBF, HGF and MPG, Germany; GSRT, Greece; RGC and Hong Kong SAR, China; ISF and Benoziyo Center, Israel; INFN, Italy; MEXT and JSPS, Japan; CNRST, Morocco; NWO, Netherlands; RCN, Norway; MNiSW and NCN, Poland; FCT, Portugal; MNE/IFA, Romania; JINR; MES of Russia and NRC KI, Russian Federation; MESTD, Serbia; MSSR, Slovakia; ARRS and MIZ\v{S}, Slovenia; DST/NRF, South Africa; MICINN, Spain; SRC and Wallenberg Foundation, Sweden; SERI, SNSF and Cantons of Bern and Geneva, Switzerland; MOST, Taiwan; TAEK, Turkey; STFC, United Kingdom; DOE and NSF, United States of America. In addition, individual groups and members have received support from BCKDF, CANARIE, Compute Canada, CRC and IVADO, Canada; Beijing Municipal Science \& Technology Commission, China; COST, ERC, ERDF, Horizon 2020 and Marie Sk{\l}odowska-Curie Actions, European Union; Investissements d'Avenir Labex, Investissements d'Avenir Idex and ANR, France; DFG and AvH Foundation, Germany; Herakleitos, Thales and Aristeia programmes co-financed by EU-ESF and the Greek NSRF, Greece; BSF-NSF and GIF, Israel; La Caixa Banking Foundation, CERCA Programme Generalitat de Catalunya and PROMETEO and GenT Programmes Generalitat Valenciana, Spain; G\"{o}ran Gustafssons Stiftelse, Sweden; The Royal Society and Leverhulme Trust, United Kingdom.
 
The crucial computing support from all WLCG partners is acknowledged gratefully, in particular from CERN, the ATLAS Tier-1 facilities at TRIUMF (Canada), NDGF (Denmark, Norway, Sweden), CC-IN2P3 (France), KIT/GridKA (Germany), INFN-CNAF (Italy), NL-T1 (Netherlands), PIC (Spain), ASGC (Taiwan), RAL (UK) and BNL (USA), the Tier-2 facilities worldwide and large non-WLCG resource providers. Major contributors of computing resources are listed in Ref.~\cite{ATL-SOFT-PUB-2020-001}.
 

\clearpage
 
\printbibliography

@article{Abercrombie:2015wmb,
    author = "Abercrombie, Daniel and others",
    editor = "Boveia, Antonio and Doglioni, Caterina and Lowette, Steven and Malik, Sarah and Mrenna, Stephen",
    title = "{Dark Matter Benchmark Models for Early LHC Run-2 Searches: Report of the ATLAS/CMS Dark Matter Forum}",
    eprint = "1507.00966",
    archivePrefix = "arXiv",
    primaryClass = "hep-ex",
    reportNumber = "FERMILAB-PUB-15-282-CD",
    doi = "10.1016/j.dark.2019.100371",
    journal = "Phys. Dark Univ.",
    volume = "27",
    pages = "100371",
    year = "2020"
}

@article{Harland-Lang:2014zoa,
    author = "Harland-Lang, L.A. and Martin, A.D. and Motylinski, P. and Thorne, R.S.",
    title = "{Parton distributions in the LHC era: MMHT 2014 PDFs}",
    eprint = "1412.3989",
    archivePrefix = "arXiv",
    primaryClass = "hep-ph",
    reportNumber = "LCTS-2014-47, IPPP-14-97, DCPT-14-194",
    doi = "10.1140/epjc/s10052-015-3397-6",
    journal = "Eur. Phys. J. C",
    volume = "75",
    number = "5",
    pages = "204",
    year = "2015"
}

@article{Ridder:2016nkl,
    author = "Gehrmann-De Ridder, Aude and Gehrmann, T. and Glover, E.W.N. and Huss, A. and Morgan, T.A.",
    title = "{The NNLO QCD corrections to Z boson production at large transverse momentum}",
    eprint = "1605.04295",
    archivePrefix = "arXiv",
    primaryClass = "hep-ph",
    reportNumber = "IPPP-16-39, NSF-KITP-16-067, ZU-TH-18-16",
    doi = "10.1007/JHEP07(2016)133",
    journal = "JHEP",
    volume = "07",
    pages = "133",
    year = "2016"
}

@article{Boughezal:2016isb,
    author = "Boughezal, Radja and Liu, Xiaohui and Petriello, Frank",
    title = "{Phenomenology of the Z-boson plus jet process at NNLO}",
    eprint = "1602.08140",
    archivePrefix = "arXiv",
    primaryClass = "hep-ph",
    doi = "10.1103/PhysRevD.94.074015",
    journal = "Phys. Rev. D",
    volume = "94",
    number = "7",
    pages = "074015",
    year = "2016"
}

@article{Boughezal:2016dtm,
    author = "Boughezal, Radja and Liu, Xiaohui and Petriello, Frank",
    title = "{W-boson plus jet differential distributions at NNLO in QCD}",
    eprint = "1602.06965",
    archivePrefix = "arXiv",
    primaryClass = "hep-ph",
    doi = "10.1103/PhysRevD.94.113009",
    journal = "Phys. Rev. D",
    volume = "94",
    number = "11",
    pages = "113009",
    year = "2016"
}

@article{Denner:2011vu,
    author = "Denner, Ansgar and Dittmaier, Stefan and Kasprzik, Tobias and Muck, Alexander",
    title = "{Electroweak corrections to dilepton + jet production at hadron colliders}",
    eprint = "1103.0914",
    archivePrefix = "arXiv",
    primaryClass = "hep-ph",
    doi = "10.1007/JHEP06(2011)069",
    journal = "JHEP",
    volume = "06",
    pages = "069",
    year = "2011"
}

@article{Denner:2012ts,
    author = {Denner, Ansgar and Dittmaier, Stefan and Kasprzik, Tobias and M\"uck, Alexander},
    title = "{Electroweak corrections to monojet production at the LHC}",
    eprint = "1211.5078",
    archivePrefix = "arXiv",
    primaryClass = "hep-ph",
    doi = "10.1140/epjc/s10052-013-2297-x",
    journal = "Eur. Phys. J. C",
    volume = "73",
    number = "2",
    pages = "2297",
    year = "2013"
}

@article{Kallweit:2015dum,
    author = {Kallweit, Stefan and Lindert, Jonas M. and Maierhofer, Philipp and Pozzorini, Stefano and Sch\"onherr, Marek},
    title = "{NLO QCD+EW predictions for V + jets including off-shell vector-boson decays and multijet merging}",
    eprint = "1511.08692",
    archivePrefix = "arXiv",
    primaryClass = "hep-ph",
    reportNumber = "DCPT-15-140, FR-PHENO-2015-014, IPPP-15-70, MCNET-15-23, ZU-TH-41-15, MITP-15-108",
    doi = "10.1007/JHEP04(2016)021",
    journal = "JHEP",
    volume = "04",
    pages = "021",
    year = "2016"
}

@article{Denner:2009gj,
    author = "Denner, Ansgar and Dittmaier, Stefan and Kasprzik, Tobias and Muck, Alexander",
    title = "{Electroweak corrections to W + jet hadroproduction including leptonic W-boson decays}",
    eprint = "0906.1656",
    archivePrefix = "arXiv",
    primaryClass = "hep-ph",
    reportNumber = "MPP-2009-72, PSI-PR-09-08",
    doi = "10.1088/1126-6708/2009/08/075",
    journal = "JHEP",
    volume = "08",
    pages = "075",
    year = "2009"
}

@article{Kuhn:2004em,
    author = "Kuhn, Johann H. and Kulesza, A. and Pozzorini, S. and Schulze, M.",
    title = "{Logarithmic electroweak corrections to hadronic Z+1 jet production at large transverse momentum}",
    eprint = "hep-ph/0408308",
    archivePrefix = "arXiv",
    reportNumber = "TTP04-18, SFB-CPP-04-32",
    doi = "10.1016/j.physletb.2005.01.059",
    journal = "Phys. Lett. B",
    volume = "609",
    pages = "277--285",
    year = "2005"
}

@article{Kuhn:2007qc,
    author = "Kuhn, Johann H. and Kulesza, A. and Pozzorini, S. and Schulze, M.",
    title = "{Electroweak corrections to large transverse momentum production of W bosons at the LHC}",
    eprint = "hep-ph/0703283",
    archivePrefix = "arXiv",
    reportNumber = "TTP07-07, SFB-CP-07-12, MPP-2007-35, DESY-07-041",
    doi = "10.1016/j.physletb.2007.06.028",
    journal = "Phys. Lett. B",
    volume = "651",
    pages = "160--165",
    year = "2007"
}

@article{Kuhn:2005az,
    author = "Kuhn, Johann H. and Kulesza, A. and Pozzorini, S. and Schulze, M.",
    title = "{One-loop weak corrections to hadronic production of Z bosons at large transverse momenta}",
    eprint = "hep-ph/0507178",
    archivePrefix = "arXiv",
    reportNumber = "TTP05-10, SFB-CPP-05-30",
    doi = "10.1016/j.nuclphysb.2005.08.019",
    journal = "Nucl. Phys. B",
    volume = "727",
    pages = "368--394",
    year = "2005"
}

@article{Kuhn:2007cv,
    author = "Kuhn, Johann H. and Kulesza, A. and Pozzorini, S. and Schulze, M.",
    title = "{Electroweak corrections to hadronic production of W bosons at large transverse momenta}",
    eprint = "0708.0476",
    archivePrefix = "arXiv",
    primaryClass = "hep-ph",
    reportNumber = "TTP07-19, SFB-CPP-07-43, MPP-2007-102, DESY-07-112",
    doi = "10.1016/j.nuclphysb.2007.12.029",
    journal = "Nucl. Phys. B",
    volume = "797",
    pages = "27--77",
    year = "2008"
}

@article{Ridder:2015dxa,
    author = "Gehrmann-De Ridder, A. and Gehrmann, T. and Glover, E.W.N. and Huss, A. and Morgan, T.A.",
    title = "{Precise QCD predictions for the production of a Z boson in association with a hadronic jet}",
    eprint = "1507.02850",
    archivePrefix = "arXiv",
    primaryClass = "hep-ph",
    reportNumber = "IPPP-15-44, ZU-TH-23-15",
    doi = "10.1103/PhysRevLett.117.022001",
    journal = "Phys. Rev. Lett.",
    volume = "117",
    number = "2",
    pages = "022001",
    year = "2016"
}

@article{Lindert2017,
    author = "Lindert, J.M. and others",
    title = "{Precise predictions for $V+$ jets dark matter backgrounds}",
    eprint = "1705.04664",
    archivePrefix = "arXiv",
    primaryClass = "hep-ph",
    reportNumber = "CERN-TH-2017-102, CERN-LPCC-2017-02, IPPP-17-38, FERMILAB-PUB-17-152-T, ZU--TH-12-17",
    doi = "10.1140/epjc/s10052-017-5389-1",
    journal = "Eur. Phys. J. C",
    volume = "77",
    number = "12",
    pages = "829",
    year = "2017"
}

@article{Beenakker:2016lwe,
    author = "Beenakker, Wim and Borschensky, Christoph and Krämer, Michael and Kulesza, Anna and Laenen, Eric",
    title = "{NNLL-fast: predictions for coloured supersymmetric particle production at the LHC with threshold and Coulomb resummation}",
    eprint = "1607.07741",
    archivePrefix = "arXiv",
    primaryClass = "hep-ph",
    reportNumber = "MS-TP-16-17, NIKHEF-2016-037, TTK-16-28",
    doi = "10.1007/JHEP12(2016)133",
    journal = "JHEP",
    volume = "12",
    pages = "133",
    year = "2016"
}

@article{Beenakker:2016gmf,
    author = "Beenakker, Wim and Borschensky, Christoph and Heger, Raphael and Krämer, Michael and Kulesza, Anna and Laenen, Eric",
    title = "{NNLL resummation for stop pair-production at the LHC}",
    eprint = "1601.02954",
    archivePrefix = "arXiv",
    primaryClass = "hep-ph",
    reportNumber = "MS-TP-16-01, TTK-16-01, NIKHEF-2016-002",
    doi = "10.1007/JHEP05(2016)153",
    journal = "JHEP",
    volume = "05",
    pages = "153",
    year = "2016"
}

@article{Khachatryan:2014rra,
    author = "{CMS Collaboration}",
    collaboration = "CMS",
    title = "{Search for dark matter, extra dimensions, and unparticles in monojet events in proton--proton collisions at $\sqrt{s} = 8$ TeV}",
    eprint = "1408.3583",
    archivePrefix = "arXiv",
    primaryClass = "hep-ex",
    reportNumber = "CMS-EXO-12-048, CERN-PH-EP-2014-164",
    doi = "10.1140/epjc/s10052-015-3451-4",
    journal = "Eur. Phys. J. C",
    volume = "75",
    number = "5",
    pages = "235",
    year = "2015"
}

@techreport{ATL-SOFT-PUB-2020-001,
      title         = "{ATLAS Computing Acknowledgements}",
      institution   = "CERN",
      collaboration = "ATLAS Collaboration",
      address       = "Geneva",
      number        = "ATL-SOFT-PUB-2020-001",
      month         = "May",
      year          = "2020",
      reportNumber  = "ATL-SOFT-PUB-2020-001",
      url           = "http://cds.cern.ch/record/2717821",
}

@article{YR4,
      author         = "{LHC Higgs Cross Section Working Group, D. de Florian et al.}",
      title          = "{Handbook of LHC Higgs cross sections: 4. Deciphering the
                        nature of the Higgs sector}",
      collaboration  = "LHC Higgs Cross Section Working Group",
      doi            = "10.23731/CYRM-2017-002",
      year           = "2016",
      eprint         = "1610.07922",
      archivePrefix  = "arXiv",
      primaryClass   = "hep-ph",
      reportNumber   = "FERMILAB-FN-1025-T, CERN-2017-002-M",
      SLACcitation   = "%%CITATION = ARXIV:1610.07922;%%"
}

@article{Aprile:2019dbj,
    author = "Aprile, E. and others",
    collaboration = "XENON",
    title = "{Constraining the Spin-Dependent WIMP-Nucleon Cross Sections with XENON1T}",
    eprint = "1902.03234",
    archivePrefix = "arXiv",
    primaryClass = "astro-ph.CO",
    doi = "10.1103/PhysRevLett.122.141301",
    journal = "Phys. Rev. Lett.",
    volume = "122",
    number = "14",
    pages = "141301",
    year = "2019"
}

@article{Akerib:2017kat,
    author = "Akerib, D.S. and others",
    collaboration = "LUX",
    title = "{Limits on Spin-Dependent WIMP-Nucleon Cross Section Obtained from the Complete LUX Exposure}",
    eprint = "1705.03380",
    archivePrefix = "arXiv",
    primaryClass = "astro-ph.CO",
    doi = "10.1103/PhysRevLett.118.251302",
    journal = "Phys. Rev. Lett.",
    volume = "118",
    number = "25",
    pages = "251302",
    year = "2017"
}

@article{Amole:2019fdf,
    author = "Amole, C. and others",
    collaboration = "PICO",
    title = "{Dark matter search results from the complete exposure of the PICO-60 C$_3$F$_8$ bubble chamber}",
    eprint = "1902.04031",
    archivePrefix = "arXiv",
    primaryClass = "astro-ph.CO",
    reportNumber = "FERMILAB-PUB-19-073-AE-E",
    doi = "10.1103/PhysRevD.100.022001",
    journal = "Phys. Rev. D",
    volume = "100",
    number = "2",
    pages = "022001",
    year = "2019"
}

@article{White:2009yt,
    author = "White, Chris D. and Frixione, Stefano and Laenen, Eric and Maltoni, Fabio",
    title = "{Isolating Wt production at the LHC}",
    eprint = "0908.0631",
    archivePrefix = "arXiv",
    primaryClass = "hep-ph",
    reportNumber = "NIKHEF-2009-013, CERN-TH-2009-136, ITF-UU-09-31, CP3-09-33",
    doi = "10.1088/1126-6708/2009/11/074",
    journal = "JHEP",
    volume = "11",
    pages = "074",
    year = "2009"
}

@article{Baglio:2014uba,
    author = "Baglio, J. and others",
    title = "{Release Note - VBFNLO 2.7.0}",
    eprint = "1404.3940",
    archivePrefix = "arXiv",
    primaryClass = "hep-ph",
    reportNumber = "FTUV-14-2903, IFIC-14-26, KA-TP-10-2014, LPN14-062, MAN-HEP-2014-03",
    month = "4",
    year = "2014"
}

@article{Bahr:2008pv,
    author = "Bahr, M. and others",
    title = "{Herwig++ Physics and Manual}",
    eprint = "0803.0883",
    archivePrefix = "arXiv",
    primaryClass = "hep-ph",
    reportNumber = "CERN-PH-TH-2008-038, CAVENDISH-HEP-08-03, KA-TP-05-2008, DCPT-08-22, IPPP-08-11, CP3-08-05",
    doi = "10.1140/epjc/s10052-008-0798-9",
    journal = "Eur. Phys. J. C",
    volume = "58",
    pages = "639--707",
    year = "2008"
}

@article{Khachatryan:2016whc,
      author         = "{CMS Collaboration}",
      title          = "{Searches for invisible decays of the Higgs boson in pp
                        collisions at $\sqrt{s}$ = 7, 8, and 13 TeV}",
      collaboration  = "CMS",
      journal        = "JHEP",
      volume         = "02",
      year           = "2017",
      pages          = "135",
      doi            = "10.1007/JHEP02(2017)135",
      eprint         = "1610.09218",
      archivePrefix  = "arXiv",
      primaryClass   = "hep-ex",
      reportNumber   = "CMS-HIG-16-016, CERN-EP-2016-240",
      SLACcitation   = "%%CITATION = ARXIV:1610.09218;%%"
}

@article{Backovic:2015cra,
      author         = "Backović, Mihailo and Martini, Antony and Mattelaer,
                        Olivier and Kong, Kyoungchul and Mohlabeng, Gopolang",
      title          = "{Direct detection of dark matter with MadDM v.2.0}",
      journal        = "Phys. Dark Univ.",
      volume         = "9-10",
      year           = "2015",
      pages          = "37",
      doi            = "10.1016/j.dark.2015.09.001",
      eprint         = "1505.04190",
      archivePrefix  = "arXiv",
      primaryClass   = "hep-ph",
      SLACcitation   = "%%CITATION = ARXIV:1505.04190;%%"
}

@article{Boveia:2016mrp,
    author = "Busoni, Giorgio and others",
    editor = "Boveia, Antonio and Buchmueller, Oliver and Doglioni, Caterina and Hahn, Kristian and Haisch, Ulrich and Kahlhoefer, Felix and Mangano, Michelangelo and McCabe, Christopher and Tait, Tim M.P.",
    title = "{Recommendations on presenting LHC searches for missing transverse energy signals using simplified $s$-channel models of dark matter}",
    eprint = "1603.04156",
    archivePrefix = "arXiv",
    primaryClass = "hep-ex",
    reportNumber = "CERN-LPCC-2016-001",
    doi = "10.1016/j.dark.2019.100365",
    journal = "Phys. Dark Univ.",
    volume = "27",
    pages = "100365",
    year = "2020"
}

@Article{Horndeski1974,
author="Horndeski, Gregory Walter",
title="Second-order scalar-tensor field equations in a four-dimensional space",
journal="Int. J. Theoret. Phys.",
year="1974",
month=9,
day="01",
volume="10",
number="6",
pages="363--384",
doi="10.1007/BF01807638",
url="https://doi.org/10.1007/BF01807638"
}

@article{Georgi1993, 
doi = "10.1146/annurev.ns.43.120193.001233",
author = "Georgi, H",
title = "Effective Field Theory",
journal = "Annu. Rev. Nucl. Part. Sci.",
volume = "43",
number = "1",
pages = "209--252",
year = "1993",
url = "https://doi.org/10.1146/annurev.ns.43.120193.001233",
}

@article{Riess:1998cb,
      author         = "Riess, Adam G. and others",
      title          = "{Observational Evidence from Supernovae for an
                        Accelerating Universe and a Cosmological Constant}",
      collaboration  = "Supernova Search Team",
      journal        = "Astron. J.",
      volume         = "116",
      year           = "1998",
      pages          = "1009-1038",
      doi            = "10.1086/300499",
      eprint         = "astro-ph/9805201",
      archivePrefix  = "arXiv",
      primaryClass   = "astro-ph",
      SLACcitation   = "%%CITATION = ASTRO-PH/9805201;%%"
}

@article{Perlmutter:1998np,
      author         = "Perlmutter, S. and others",
      title          = "{Measurements of $\Omega$ and $\Lambda$ from 42 High-Redshift Supernovae}",
      collaboration  = "Supernova Cosmology Project",
      journal        = "Astrophys. J.",
      volume         = "517",
      year           = "1999",
      pages          = "565-586",
      doi            = "10.1086/307221",
      eprint         = "astro-ph/9812133",
      archivePrefix  = "arXiv",
      primaryClass   = "astro-ph",
      reportNumber   = "LBNL-41801, LBL-41801",
      SLACcitation   = "%%CITATION = ASTRO-PH/9812133;%%"
}

@article{Brax:2016did,
      author         = "Brax, Philippe and Burrage, Clare and Englert, Christoph
                        and Spannowsky, Michael",
      title          = "{LHC signatures of scalar dark energy}",
      journal        = "Phys. Rev. D",
      volume         = "94",
      year           = "2016",
      number         = "8",
      pages          = "084054",
      doi            = "10.1103/PhysRevD.94.084054",
      eprint         = "1604.04299",
      archivePrefix  = "arXiv",
      primaryClass   = "hep-ph",
      reportNumber   = "IPPP-16-31, DCPT-16-62",
      SLACcitation   = "%%CITATION = ARXIV:1604.04299;%%"
}

@article{Abdallah:2015ter,
      author         = "Abdallah, Jalal and others",
      title          = "{Simplified models for dark matter searches at the LHC}",
      journal        = "Phys. Dark Univ.",
      volume         = "9-10",
      year           = "2015",
      pages          = "8",
      doi            = "10.1016/j.dark.2015.08.001",
      eprint         = "1506.03116",
      archivePrefix  = "arXiv",
      primaryClass   = "hep-ph",
      reportNumber   = "FERMILAB-PUB-15-283-CD, CERN-PH-TH-2015-139",
      SLACcitation   = "%%CITATION = ARXIV:1506.03116;%%"
}

@article{Buchmueller:2014yoa,
      author         = "Buchmueller, Oliver and Dolan, Matthew J. and Malik,
                        Sarah A. and McCabe, Christopher",
      title          = "{Characterising dark matter searches at colliders and direct detection experiments: Vector mediators}",
      journal        = "JHEP",
      volume         = "01",
      year           = "2015",
      pages          = "037",
      doi            = "10.1007/JHEP01(2015)037",
      eprint         = "1407.8257",
      archivePrefix  = "arXiv",
      primaryClass   = "hep-ph",
      reportNumber   = "IPPP-14-68, DCPT-14-136, SLAC-PUB-16033",
      SLACcitation   = "%%CITATION = ARXIV:1407.8257;%%"
}

@article{Peccei:1977hh,
      author         = "Peccei, R. D. and Quinn, Helen R.",
      title          = "{CP Conservation in the Presence of Pseudoparticles}",
      journal        = "Phys. Rev. Lett.",
      volume         = "38",
      year           = "1977",
      pages          = "1440-1443",
      doi            = "10.1103/PhysRevLett.38.1440",
      reportNumber   = "ITP-568-STANFORD",
      SLACcitation   = "%%CITATION = PRLTA,38,1440;%%"
}

@article{Brivio:2017ije,
      author         = "Brivio, I. and Gavela, M. B. and Merlo, L. and Mimasu, K.
                        and No, J. M. and del Rey, R. and Sanz, V.",
      title          = "{ALPs effective field theory and collider signatures}",
      journal        = "Eur. Phys. J. C",
      volume         = "77",
      year           = "2017",
      number         = "8",
      pages          = "572",
      doi            = "10.1140/epjc/s10052-017-5111-3",
      eprint         = "1701.05379",
      archivePrefix  = "arXiv",
      primaryClass   = "hep-ph",
      reportNumber   = "IFT-UAM-CSIC-16-141, KCL-PH-TH-2016-72, FTUAM-16-49,
                        CP3-17-04",
      SLACcitation   = "%%CITATION = ARXIV:1701.05379;%%"
}

@article{Mimasu:2014nea,
      author         = "Mimasu, Ken and Sanz, Verónica",
      title          = "{ALPs at colliders}",
      journal        = "JHEP",
      volume         = "06",
      year           = "2015",
      pages          = "173",
      doi            = "10.1007/JHEP06(2015)173",
      eprint         = "1409.4792",
      archivePrefix  = "arXiv",
      primaryClass   = "hep-ph",
      SLACcitation   = "%%CITATION = ARXIV:1409.4792;%%"
}

@article{pythia,
      author         = "Sjöstrand, Torbjorn and Mrenna, Stephen and Skands, Peter
                        Z.",
      title          = "{PYTHIA 6.4 physics and manual}",
      journal        = "JHEP",
      volume         = "05",
      pages          = "026",
      doi            = "10.1088/1126-6708/2006/05/026",
      year           = "2006",
      eprint         = "hep-ph/0603175",
      archivePrefix  = "arXiv",
      reportNumber   = "FERMILAB-PUB-06-052-CD-T, LU-TP-06-13",
      SLACcitation   = "%%CITATION = HEP-PH/0603175;%%",
}

@article{Aad:2010ah,
      author         = "{ATLAS Collaboration}",
      title          = "{The ATLAS Simulation Infrastructure}",
      collaboration  = "ATLAS",
      journal        = "Eur. Phys. J. C",
      volume         = "70",
      year           = "2010",
      pages          = "823",
      doi            = "10.1140/epjc/s10052-010-1429-9",
      eprint         = "1005.4568",
      archivePrefix  = "arXiv",
      primaryClass   = "physics.ins-det",
      SLACcitation   = "%%CITATION = ARXIV:1005.4568;%%"
}

@article{Aaboud:2019yqu,
      author         = "{ATLAS Collaboration}",
      title          = "{Constraints on mediator-based dark matter and scalar
                        dark energy models using $\sqrt s = 13$ TeV $pp$ collision
                        data collected by the ATLAS detector}",
      collaboration  = "ATLAS",
      journal        = "JHEP",
      volume         = "05",
      year           = "2019",
      pages          = "142",
      doi            = "10.1007/JHEP05(2019)142",
      eprint         = "1903.01400",
      archivePrefix  = "arXiv",
      primaryClass   = "hep-ex",
      reportNumber   = "CERN-EP-2018-334",
      SLACcitation   = "%%CITATION = ARXIV:1903.01400;%%"
}

@Article{Farrar:1978xj,
     author    = "Farrar, G. R. and Fayet, P.",
     title     = "{Phenomenology of the production, decay, and detection of
                  new hadronic states associated with supersymmetry}",
     journal   = "Phys. Lett. B",
     volume    = "76",
     year      = "1978",
     pages     = "575",
     doi       = "10.1016/0370-2693(78)90858-4",
     SLACcitation  = "%%CITATION = PHLTA,B76,575;%%"
}

@article{Trimble:1987ee,
      author         = "Trimble, Virginia",
      title          = "{Existence and Nature of Dark Matter in the Universe}",
      journal        = "Ann. Rev. Astron. Astrophys.",
      volume         = "25",
      year           = "1987",
      pages          = "425",
      doi            = "10.1146/annurev.aa.25.090187.002233",
      SLACcitation   = "%%CITATION = ARAAA,25,425;%%"
}

@article{Feng:2010gw,
      author         = "Feng, Jonathan L.",
      title          = "{Dark Matter Candidates from Particle Physics and Methods
                        of Detection}",
      journal        = "Ann. Rev. Astron. Astrophys.",
      volume         = "48",
      year           = "2010",
      pages          = "495",
      doi            = "10.1146/annurev-astro-082708-101659",
      eprint         = "1003.0904",
      archivePrefix  = "arXiv",
      primaryClass   = "astro-ph.CO",
      reportNumber   = "UCI-TR-2009-13",
      SLACcitation   = "%%CITATION = ARXIV:1003.0904;%%"
}

@article{Bertone:2004pz,
      author         = "Bertone, Gianfranco and Hooper, Dan and Silk, Joseph",
      title          = "{Particle dark matter: evidence, candidates and
                        constraints}",
      journal        = "Phys. Rept.",
      volume         = "405",
      year           = "2005",
      pages          = "279",
      doi            = "10.1016/j.physrep.2004.08.031",
      eprint         = "hep-ph/0404175",
      archivePrefix  = "arXiv",
      reportNumber   = "FERMILAB-PUB-04-047-A",
      SLACcitation   = "%%CITATION = HEP-PH/0404175;%%"
}

@article{Steigman:1984ac,
      author         = "Steigman, Gary and Turner, Michael S.",
      title          = "{Cosmological constraints on the properties of weakly
                        interacting massive particles}",
      journal        = "Nucl. Phys. B",
      volume         = "253",
      pages          = "375",
      year           = "1985",
      doi            = "10.1016/0550-3213(85)90537-1",
      reportNumber   = "FERMILAB-PUB-84-110-A, BA-84-33",
      SLACcitation   = "%%CITATION = NUPHA,B253,375;%%",
}

@article{Kolb:1990vq,
      author         = "Kolb, Edward W. and Turner, Michael S.",
      title          = "{The Early Universe}",
      journal        = "Front. Phys.",
      volume         = "69",
      pages          = "1",
      year           = "1990",
      SLACcitation   = "%%CITATION = FRPHA,69,1;%%",
}

@article{Planck2018,
      author         = "{R. Adam, et al., Planck Collaboration}",
      title          = "{Planck 2018 results. VI. Cosmological parameters}",
      collaboration  = "Planck",
%      journal        = "Astron. Astrophys.",
%      volume         = "594",
      year           = "2018",
%      pages          = "A1",
%      doi            = "10.1051/0004-6361/201527101",
      eprint         = "1807.06209",
      archivePrefix  = "arXiv",
      primaryClass   = "astro-ph.CO",
      SLACcitation   = "%%CITATION = ARXIV:1807.06209;%%"
}

@ARTICLE{2013ApJS..208...19H,
   author = {{Hinshaw}, G. and {Larson}, D. and {Komatsu}, E. and {Spergel}, D.~N. and 
        {Bennett}, C.~L. and {Dunkley}, J. and {Nolta}, M.~R. and {Halpern}, M. and 
        {Hill}, R.~S. and {Odegard}, N. and {Page}, L. and {Smith}, K.~M. and 
        {Weiland}, J.~L. and {Gold}, B. and {Jarosik}, N. and {Kogut}, A. and 
        {Limon}, M. and {Meyer}, S.~S. and {Tucker}, G.~S. and {Wollack}, E. and 
        {Wright}, E.~L.},
    title = "{Nine-year Wilkinson Microwave Anisotropy Probe (WMAP) Observations: Cosmological Parameter Results}",
  journal = {ApJS},
archivePrefix = "arXiv",
   eprint = {1212.5226},
     year = 2013,
   volume = 208,
    pages = {19},
      doi = {10.1088/0067-0049/208/2/19},
   adsurl = {http://adsabs.harvard.edu/abs/2013ApJS..208...19H},
 primaryClass = {astro-ph.CO} 
}

@article{Beenakker:1997ut,
      author         = "Beenakker, W. and Kramer, M. and Plehn, T. and Spira, M.
                        and Zerwas, P. M.",
      title          = "{Stop production at hadron colliders}",
      journal        = "Nucl. Phys. B",
      volume         = "515",
      year           = "1998",
      pages          = "3",
      doi            = "10.1016/S0550-3213(98)00014-5",
      eprint         = "hep-ph/9710451",
      archivePrefix  = "arXiv",
      reportNumber   = "DESY-97-214, CERN-TH-97-177, RAL-TR-97-056",
      SLACcitation   = "%%CITATION = HEP-PH/9710451;%%"
}

@article{Beenakker:2010nq,
      author         = "Beenakker, Wim and Brensing, Silja and Kramer, Michael
                        and Kulesza, Anna and Laenen, Eric and Niessen, Irene",
      title          = "{Supersymmetric top and bottom squark production at
                        hadron colliders}",
      journal        = "JHEP",
      volume         = "08",
      year           = "2010",
      pages          = "098",
      doi            = "10.1007/JHEP08(2010)098",
      eprint         = "1006.4771",
      archivePrefix  = "arXiv",
      primaryClass   = "hep-ph",
      reportNumber   = "CERN-PH-TH-2010-142, ITFA-2010-015, ITP-UU-10-17,
                        NIKHEF-2010-016, TTK-10-33",
      SLACcitation   = "%%CITATION = ARXIV:1006.4771;%%"
}

@article{Borschensky:2014cia,
      author         = "Borschensky, Christoph and Krämer, Michael and Kulesza,
                        Anna and Mangano, Michelangelo and Padhi, Sanjay and
                        Plehn, Tilman and Portell, Xavier",
      title          = "{Squark and gluino production cross sections in pp
                        collisions at $\sqrt{s}$ = 13, 14, 33 and 100 TeV}",
      journal        = "Eur. Phys. J. C",
      volume         = "74",
      year           = "2014",
      number         = "12",
      pages          = "3174",
      doi            = "10.1140/epjc/s10052-014-3174-y",
      eprint         = "1407.5066",
      archivePrefix  = "arXiv",
      primaryClass   = "hep-ph",
      reportNumber   = "MS-TP-14-25, CERN-PH-TH-2014-137, TTK-14-13",
      SLACcitation   = "%%CITATION = ARXIV:1407.5066;%%"
}

@article{Haisch:2013ata,
      author         = "Haisch, Ulrich and Kahlhöfer, Felix and Re, Emanuele",
      title          = "{QCD effects in mono-jet searches for dark matter}",
      journal        = "JHEP",
      volume         = "12",
      year           = "2013",
      pages          = "007",
      doi            = "10.1007/JHEP12(2013)007",
      eprint         = "1310.4491",
      archivePrefix  = "arXiv",
      primaryClass   = "hep-ph",
      reportNumber   = "OUTP-13-20P, NSF-KITP-13-241",
      SLACcitation   = "%%CITATION = ARXIV:1310.4491;%%"
}

@article{Haisch:2015ioa,
      author         = "Haisch, Ulrich and Re, Emanuele",
      title          = "{Simplified dark matter top-quark interactions at the
                        LHC}",
      journal        = "JHEP",
      volume         = "06",
      year           = "2015",
      pages          = "078",
      doi            = "10.1007/JHEP06(2015)078",
      eprint         = "1503.00691",
      archivePrefix  = "arXiv",
      primaryClass   = "hep-ph",
      reportNumber   = "OUTP-15-02P, CERN-PH-TH-2015-201",
      SLACcitation   = "%%CITATION = ARXIV:1503.00691;%%"
}

@article{powheg,
      author         = "Frixione, Stefano and Nason, Paolo and Ridolfi, Giovanni",
      title          = "{A positive-weight next-to-leading-order Monte Carlo for
                        heavy flavour hadroproduction}",
      journal        = "JHEP",
      volume         = "09",
      year           = "2007",
      pages          = "126",
      doi            = "10.1088/1126-6708/2007/09/126",
      eprint         = "0707.3088",
      archivePrefix  = "arXiv",
      primaryClass   = "hep-ph",
      reportNumber   = "BICOCCA-FT-07-12, GEF-TH-19-2007",
      SLACcitation   = "%%CITATION = ARXIV:0707.3088;%%"
}

@article{mstw,
      author         = "Martin, A. D. and Stirling, W. J. and Thorne, R. S. and
                        Watt, G.",
      title          = "{Parton distributions for the LHC}",
      journal        = "Eur. Phys. J. C",
      volume         = "63",
      year           = "2009",
      pages          = "189",
      doi            = "10.1140/epjc/s10052-009-1072-5",
      eprint         = "0901.0002",
      archivePrefix  = "arXiv",
      primaryClass   = "hep-ph",
      reportNumber   = "IPPP-08-95, DCPT-08-190, CAVENDISH-HEP-08-16",
      SLACcitation   = "%%CITATION = ARXIV:0901.0002;%%"
}

@article{Catani:2009sm,
      author         = "Catani, Stefano and Cieri, Leandro and Ferrera, Giancarlo
                        and de Florian, Daniel and Grazzini, Massimiliano",
      title          = "{Vector Boson Production at Hadron Colliders: A Fully Exclusive QCD Calculation at Next-to-Next-to-Leading Order}",
      journal        = "Phys. Rev. Lett.",
      volume         = "103",
      year           = "2009",
      pages          = "082001",
      doi            = "10.1103/PhysRevLett.103.082001",
      eprint         = "0903.2120",
      archivePrefix  = "arXiv",
      primaryClass   = "hep-ph",
      SLACcitation   = "%%CITATION = ARXIV:0903.2120;%%"
}

@article{Catani:2007vq,
      author         = "Catani, Stefano and Grazzini, Massimiliano",
      title          = "{Next-to-Next-to-Leading-Order Subtraction Formalism in Hadron Collisions and its Application to Higgs-Boson Production at the Large Hadron Collider}",
      journal        = "Phys. Rev. Lett.",
      volume         = "98",
      year           = "2007",
      pages          = "222002",
      doi            = "10.1103/PhysRevLett.98.222002",
      eprint         = "hep-ph/0703012",
      archivePrefix  = "arXiv",
      SLACcitation   = "%%CITATION = HEP-PH/0703012;%%"
}

@article{ct10,
      author         = "Lai, Hung-Liang and others",
      title          = "{New parton distributions for collider physics}",
      journal        = "Phys. Rev. D",
      volume         = "82",
      pages          = "074024",
      doi            = "10.1103/PhysRevD.82.074024",
      year           = "2010",
      eprint         = "1007.2241",
      archivePrefix  = "arXiv",
      primaryClass   = "hep-ph",
      reportNumber   = "MSUHEP-100707, SMU-HEP-10-10",
      SLACcitation   = "%%CITATION = ARXIV:1007.2241;%%"    
}

@article{sherpa,
      author         = "Gleisberg, T. and others",
      title          = "{Event generation with SHERPA 1.1}",
      journal        = "JHEP",
      volume         = "02",
      pages          = "007",
      doi            = "10.1088/1126-6708/2009/02/007",
      year           = "2009",
      eprint         = "0811.4622",
      archivePrefix  = "arXiv",
      primaryClass   = "hep-ph",
      reportNumber   = "FERMILAB-PUB-08-477-T, SLAC-PUB-13420, ZU-TH-17-08,DCPT-08-138, IPPP-08-69, EDINBURGH-2008-30, MCNET-08-14",
      SLACcitation   = "%%CITATION = ARXIV:0811.4622;%%"    
}

@article{Aaboud:2017hbk,
      author         = "{ATLAS Collaboration}",
      title          = "{Measurements of the production cross section of a $Z$
                        boson in association with jets in pp collisions at
                        $\sqrt{s} = 13$  TeV with the ATLAS detector}",
      collaboration  = "ATLAS",
      journal        = "Eur. Phys. J. C",
      volume         = "77",
      year           = "2017",
      number         = "6",
      pages          = "361",
      doi            = "10.1140/epjc/s10052-017-4900-z",
      eprint         = "1702.05725",
      archivePrefix  = "arXiv",
      primaryClass   = "hep-ex",
      reportNumber   = "CERN-EP-2016-297",
      SLACcitation   = "%%CITATION = ARXIV:1702.05725;%%"
}

@Article{EvtGen,
    author         = "{D. J. Lange}",
    title          = "{The EvtGen particle decay simulation package}",
    journal        = "Nucl. Instrum. Meth. A",
    volume         = "462",
    year           = "2001",
    pages          = "152",
    doi            = "10.1016/S0168-9002(01)00089-4",
    primaryClass   = "hep-ex",
}

@article{Campbell:2011bn,
      author         = "Campbell, John M. and Ellis, R. Keith and Williams,
                        Ciaran",
      title          = "{Vector boson pair production at the LHC}",
      journal        = "JHEP",
      volume         = "07",
      pages          = "018",
      doi            = "10.1007/JHEP07(2011)018",
      year           = "2011",
      eprint         = "1105.0020",
      archivePrefix  = "arXiv",
      primaryClass   = "hep-ph",
      reportNumber   = "FERMILAB-PUB-11-182-T",
      SLACcitation   = "%%CITATION = ARXIV:1105.0020;%%",
}

@Article{Golfand:1971iw,
     author    = "Golfand, Yu. A. and Likhtman, E. P.",
     title     = "{Extension of the Algebra of Poincare Group Generators and
                  Violation of p Invariance}",
     journal   = "JETP Lett.",
     volume    = "13",
     year      = "1971",
     pages     = "323",
     SLACcitation  = "%%CITATION = JTPLA,13,323;%%"   
}

@Article{Volkov:1973ix,
     author    = "Volkov, D. V. and Akulov, V. P.",
     title     = "{Is the neutrino a goldstone particle?}",
     journal   = "Phys. Lett. B",
     volume    = "46",
     year      = "1973",
     pages     = "109",
     doi       = "10.1016/0370-2693(73)90490-5",
     SLACcitation  = "%%CITATION = PHLTA,B46,109;%%"
}

@Article{Wess:1974tw,
     author    = "Wess, J. and Zumino, B.",
     title     = "{Supergauge transformations in four-dimensions}",
     journal   = "Nucl. Phys. B",
     volume    = "70",
     year      = "1974",
     pages     = "39",
     doi       = "10.1016/0550-3213(74)90355-1",
     SLACcitation  = "%%CITATION = NUPHA,B70,39;%%"
}

@article{ArkaniHamed:1998rs,
      author         = "Arkani-Hamed, Nima and Dimopoulos, Savas and Dvali, G.R.",
      title          = "{The hierarchy problem and new dimensions at a
                        millimeter}",
      journal        = "Phys. Lett. B",
      volume         = "429",
      pages          = "263",
      doi            = "10.1016/S0370-2693(98)00466-3",
      year           = "1998",
      eprint         = "hep-ph/9803315",
      archivePrefix  = "arXiv",
      reportNumber   = "SLAC-PUB-7769, SU-ITP-98-13",
      SLACcitation   = "%%CITATION = HEP-PH/9803315;%%",
}

@article{LUCID2,
  author="{ATLAS Collaboration}",
  title="{The new LUCID-2 detector for luminosity measurement and monitoring in ATLAS}",
  journal="{JINST}",
  volume={13},
  number={07},
  pages={P07017},
  doi="10.1088/1748-0221/13/07/P07017",
  year={2018}
}

@article{Aaboud:2017phn,
      author         = "{ATLAS Collaboration}",
      title          = "{Search for dark matter and other new phenomena in events
                        with an energetic jet and large missing transverse
                        momentum using the ATLAS detector}",
      collaboration  = "ATLAS",
      journal        = "JHEP",
      volume         = "01",
      year           = "2018",
      pages          = "126",
      doi            = "10.1007/JHEP01(2018)126",
      eprint         = "1711.03301",
      archivePrefix  = "arXiv",
      primaryClass   = "hep-ex",
      reportNumber   = "CERN-EP-2017-230",
      SLACcitation   = "%%CITATION = ARXIV:1711.03301;%%"
}

@article{Aaboud:2016tnv,
      author         = "{ATLAS Collaboration}",
      title          = "{Search for new phenomena in final states with an
                        energetic jet and large missing transverse momentum in
                        $pp$ collisions at $\sqrt{s}=13$ TeV using the ATLAS detector}",
      collaboration  = "ATLAS",
      journal        = "Phys. Rev. D",
      volume         = "94",
      year           = "2016",
      number         = "3",
      pages          = "032005",
      doi            = "10.1103/PhysRevD.94.032005",
      eprint         = "1604.07773",
      archivePrefix  = "arXiv",
      primaryClass   = "hep-ex",
      reportNumber   = "CERN-EP-2016-075",
      SLACcitation   = "%%CITATION = ARXIV:1604.07773;%%"
}

@article{Sirunyan:2017hci,
      author         = "{CMS Collaboration}",
      title          = "{Search for dark matter produced with an energetic jet or
                        a hadronically decaying W or Z boson at $ \sqrt{s}=13 $
                        TeV}",
      collaboration  = "CMS",
      journal        = "JHEP",
      volume         = "07",
      year           = "2017",
      pages          = "014",
      doi            = "10.1007/JHEP07(2017)014",
      eprint         = "1703.01651",
      archivePrefix  = "arXiv",
      primaryClass   = "hep-ex",
      reportNumber   = "CMS-EXO-16-037, CERN-EP-2017-031",
      SLACcitation   = "%%CITATION = ARXIV:1703.01651;%%"
}

@article{Aaboud:2017buf,
    author = "{ATLAS Collaboration}",
    collaboration = "ATLAS",
    title = "{Measurement of detector-corrected observables sensitive to the anomalous production of events with jets and large missing transverse momentum in $pp$ collisions at $\mathbf {\sqrt{s}=13}$  TeV using the ATLAS detector}",
    eprint = "1707.03263",
    archivePrefix = "arXiv",
    primaryClass = "hep-ex",
    reportNumber = "CERN-EP-2017-116",
    doi = "10.1140/epjc/s10052-017-5315-6",
    journal = "Eur. Phys. J. C",
    volume = "77",
    number = "11",
    pages = "765",
    year = "2017"
}

@article{ckkw-l,
      author         = "Lönnblad, Leif and Prestel, Stefan",
      title          = "{Matching tree-level matrix elements with interleaved showers}",
      journal        = "JHEP",
      volume         = "03",
      year           = "2012",
      pages          = "019",
      doi            = "10.1007/JHEP03(2012)019",
      eprint         = "1109.4829",
      archivePrefix  = "arXiv",
      primaryClass   = "hep-ph",
}

@article{Salam:1974ig,
      author         = "Salam, Abdus and Strathdee, J.",
      title          = "{Super-symmetry and non-Abelian gauges}",
      journal        = "Phys. Lett. B",
      volume         = "51",
      year           = "1974",
      pages          = "353",
      doi            = "10.1016/0370-2693(74)90226-3",
      reportNumber   = "IC/74/36",
      SLACcitation   = "%%CITATION = PHLTA,B51,353;%%"
}

@article{Wess:1974jb,
      author         = "Wess, J. and Zumino, B.",
      title          = "{Supergauge invariant extension of quantum electrodynamics}",
      journal        = "Nucl. Phys. B",
      volume         = "78",
      year           = "1974",
      pages          = "1",
      doi            = "10.1016/0550-3213(74)90112-6",
      reportNumber   = "CERN-TH-1857",
      SLACcitation   = "%%CITATION = NUPHA,B78,1;%%"
}

@article{Ferrara:1974pu,
      author         = "Ferrara, S. and Zumino, B.",
      title          = "{Supergauge invariant Yang-Mills theories}",
      journal        = "Nucl. Phys. B",
      volume         = "79",
      year           = "1974",
      pages          = "413",
      doi            = "10.1016/0550-3213(74)90559-8",
      reportNumber   = "CERN-TH-1866",
      SLACcitation   = "%%CITATION = NUPHA,B79,413;%%"
}

@article{Goldberg:1983nd,
      author         = "Goldberg, H.",
      title          = "{Constraint on the Photino Mass from Cosmology}",
      journal        = "Phys. Rev. Lett.",
      volume         = "50",
      year           = "1983",
      pages          = "1419",
      doi            = "10.1103/PhysRevLett.50.1419",
      related        = "Goldberg:1983nd-err",
      relatedstring  = "Erratum:",
      reportNumber   = "NUB-2592",
      SLACcitation   = "%%CITATION = PRLTA,50,1419;%%"
}

@article{Ellis:1983ew,
      author         = "Ellis, John and Hagelin, J.S. and Nanopoulos, Dimitri V. and Olive, Keith A. and Srednicki, M.",
      title          = "{Supersymmetric relics from the big bang}",
      booktitle      = "{IN *BATAVIA 1984, PROCEEDINGS, INNER SPACE/OUTER SPACE*,
                        458-459., In *Srednicki, M.A. (ed.): Particle physics and
                        cosmology* 223-246}",
      journal        = "Nucl. Phys. B",
      volume         = "238",
      year           = "1984",
      pages          = "453",
      doi            = "10.1016/0550-3213(84)90461-9",
      reportNumber   = "SLAC-PUB-3171",
      SLACcitation   = "%%CITATION = NUPHA,B238,453;%%"
}

@Article{Cacciari:2008gp,
     author    = "Cacciari, Matteo and Salam, Gavin P. and Soyez, Gregory",
     title     = "{The anti-\(k_{t}\) jet clustering algorithm}",
     journal   = "JHEP",
     volume    = "04",
     year      = "2008",
     pages     = "063",
     eprint    = "0802.1189",
     archivePrefix = "arXiv",
     primaryClass  =  "hep-ph",
     doi       = "10.1088/1126-6708/2008/04/063",
     SLACcitation  = "%%CITATION = 0802.1189;%%"
}

@Article{Fastjet,
      author         = "Cacciari, Matteo and Salam, Gavin P. and Soyez, Gregory",
      title          = "{FastJet user manual}",
      journal        = "Eur. Phys. J. C",
      volume         = "72",
      year           = "2012",
      pages          = "1896",
      doi            = "10.1140/epjc/s10052-012-1896-2",
      eprint         = "1111.6097",
      archivePrefix  = "arXiv",
      primaryClass   = "hep-ph",
      reportNumber   = "CERN-PH-TH-2011-297",
      SLACcitation   = "%%CITATION = ARXIV:1111.6097;%%"
}

@Article{Butterworth:2015oua,
      author         = "Butterworth, Jon and others",
      title          = "{PDF4LHC recommendations for LHC Run II}",
      journal        = "J. Phys. G",
      volume         = "43",
      year           = "2016",
      pages          = "023001",
      doi            = "10.1088/0954-3899/43/2/023001",
      eprint         = "1510.03865",
      archivePrefix  = "arXiv",
      primaryClass   = "hep-ph",
      reportNumber   = "OUTP-15-17P, SMU-HEP-15-12, TIF-UNIMI-2015-14,
                        LCTS-2015-27, CERN-PH-TH-2015-249",
      SLACcitation   = "%%CITATION = ARXIV:1510.03865;%%"
}

@article{Ball:2014uwa,
      author         = "Ball, Richard D. and others",
      title          = "{Parton distributions for the LHC run II}",
      collaboration  = "NNPDF",
      journal        = "JHEP",
      volume         = "04",
      year           = "2015",
      pages          = "040",
      doi            = "10.1007/JHEP04(2015)040",
      eprint         = "1410.8849",
      archivePrefix  = "arXiv",
      primaryClass   = "hep-ph",
      reportNumber   = "EDINBURGH-2014-15, IFUM-1034-FT, CERN-PH-TH-2013-253,
                        OUTP-14-11P, CAVENDISH-HEP-14-11",
      SLACcitation   = "%%CITATION = ARXIV:1410.8849;%%"
}

@Article{Alwall:2014hca,
      author         = "Alwall, J. and Frederix, R. and Frixione, S. and Hirschi,
                        V. and Maltoni, F. and Mattelaer, O. and Shao, H. -S. and
                        Stelzer, T. and Torrielli, P. and Zaro, M.",
      title          = "{The automated computation of tree-level and
                        next-to-leading order differential cross sections, and
                        their matching to parton shower simulations}",
      journal        = "JHEP",
      volume         = "07",
      year           = "2014",
      pages          = "079",
      doi            = "10.1007/JHEP07(2014)079",
      eprint         = "1405.0301",
      archivePrefix  = "arXiv",
      primaryClass   = "hep-ph",
      reportNumber   = "CERN-PH-TH-2014-064, CP3-14-18, LPN14-066, MCNET-14-09,
                        ZU-TH-14-14",
      SLACcitation   = "%%CITATION = ARXIV:1405.0301;%%"
}

@Article{Nason:2004rx,
      author         = "Nason, Paolo",
      title          = "{A new method for combining NLO QCD with shower Monte Carlo algorithms}",
      journal        = "JHEP",
      volume         = "11",
      pages          = "040",
      doi            = "10.1088/1126-6708/2004/11/040",
      year           = "2004",
      eprint         = "hep-ph/0409146",
      archivePrefix  = "arXiv",
}

@Article{Frixione:2007vw,
      author         = "Frixione, Stefano and Nason, Paolo and Oleari, Carlo",
      title          = "{Matching NLO QCD computations with parton shower
                        simulations: the POWHEG method}",
      journal        = "JHEP",
      volume         = "11",
      pages          = "070",
      doi            = "10.1088/1126-6708/2007/11/070",
      year           = "2007",
      eprint         = "0709.2092",
      archivePrefix  = "arXiv",
      primaryClass   = "hep-ph",
}

@Article{Alioli:2010xd,
      author         = "Alioli, Simone and Nason, Paolo and Oleari, Carlo and Re,
                        Emanuele",
      title          = "{A general framework for implementing NLO calculations in
                        shower Monte Carlo programs: the POWHEG BOX}",
      journal        = "JHEP",
      volume         = "06",
      pages          = "043",
      doi            = "10.1007/JHEP06(2010)043",
      year           = "2010",
      eprint         = "1002.2581",
      archivePrefix  = "arXiv",
      primaryClass   = "hep-ph",
}

@Article{Gleisberg:2008fv,
      author         = "Gleisberg, Tanju and H{\"o}che, Stefan",
      title          = "{Comix, a new matrix element generator}",
      journal        = "JHEP",
      volume         = "12",
      pages          = "039",
      doi            = "10.1088/1126-6708/2008/12/039",
      year           = "2008",
      eprint         = "0808.3674",
      archivePrefix  = "arXiv",
      primaryClass   = "hep-ph",
}

@Article{Schumann:2007mg,
      author         = "Schumann, Steffen and Krauss, Frank",
      title          = "{A parton shower algorithm based on Catani--Seymour dipole factorisation}",
      journal        = "JHEP",
      volume         = "03",
      pages          = "038",
      doi            = "10.1088/1126-6708/2008/03/038",
      year           = "2008",
      eprint         = "0709.1027",
      archivePrefix  = "arXiv",
      primaryClass   = "hep-ph",
}

@article{Hoeche:2012yf,
      author         = "H{\"o}che, Stefan and Krauss, Frank and Sch{\"o}nherr, Marek and
                        Siegert, Frank",
      title          = "{QCD matrix elements + parton showers. The NLO case}",
      journal        = "JHEP",
      volume         = "04",
      year           = "2013",
      pages          = "027",
      doi            = "10.1007/JHEP04(2013)027",
      eprint         = "1207.5030",
      archivePrefix  = "arXiv",
      primaryClass   = "hep-ph",
      reportNumber   = "SLAC-PUB-15191, IPPP-12-52, DCPT-12-104, LPN12-081,
                        FR-PHENO-2012-017, MCNET-12-09, --FR-PHENO-2012-017",
      SLACcitation   = "%%CITATION = ARXIV:1207.5030;%%"
}

@article{Cascioli:2011va,
      author         = "Cascioli, Fabio and Maierh{\"o}fer, Philipp and Pozzorini, Stefano",
      title          = "{Scattering Amplitudes with Open Loops}",
      journal        = "Phys. Rev. Lett.",
      volume         = "108",
      year           = "2012",
      pages          = "111601",
      doi            = "10.1103/PhysRevLett.108.111601",
      eprint         = "1111.5206",
      archivePrefix  = "arXiv",
      primaryClass   = "hep-ph",
      reportNumber   = "ZU-TH-23-11, LPN11-66",
      SLACcitation   = "%%CITATION = ARXIV:1111.5206;%%"
}

@article{Agostinelli:2002hh,
      author         = "Agostinelli, S. and others",
      title          = "{\textsc{Geant4} -- a simulation toolkit}",
      journal        = "Nucl. Instrum. Meth. A",
      volume         = "506",
      year           = "2003",
      pages          = "250",
      doi            = "10.1016/S0168-9002(03)01368-8",
      reportNumber   = "SLAC-PUB-9350, FERMILAB-PUB-03-339",
      SLACcitation   = "%%CITATION = NUIMA,A506,250;%%"
}

@Article{Read:2002hq,
    Author = {Read, Alexander L.},
    Journal = {J. Phys. G},
    Pages = {2693},
    doi = "10.1088/0954-3899/28/10/313",
    Title = "{Presentation of search results: the \(CL_S\) technique}",
    Volume = {28},
    Year = {2002}
}

@Article{Cowan:2010js,
   Author = {Cowan, Glen and Cranmer, Kyle and Gross, Eilam and Vitells, Ofer},
   Title = {{Asymptotic formulae for likelihood-based tests of new physics}},
   Journal = {Eur. Phys. J. C},
   Volume = {71},
   Year = {2011},
   Pages = {1554},
   doi = "10.1140/epjc/s10052-011-1554-0",
   Eprint = "1007.1727",
   Archiveprefix = {arXiv},
   Primaryclass = {physics.data-an},
   related = "Cowan:2010js-err",
   relatedstring  = "Erratum:",
}

@Article{PERF-2007-01,
    author         = "{ATLAS Collaboration}",
    title          = "{The ATLAS Experiment at the CERN Large Hadron Collider}",
    journal        = "JINST",
    volume         = "3",
    year           = "2008",
    pages          = "S08003",
    doi            = "10.1088/1748-0221/3/08/S08003",
    primaryClass   = "hep-ex",
}

@Article{SUSY-2011-20,
    author         = "{ATLAS Collaboration}",
    title          = "{Search for squarks and gluinos with the ATLAS detector in final states with jets and missing transverse momentum using \(4.7\,\text{fb}^{-1}\) of \(\sqrt{s} = 7\,\text{TeV}\) proton--proton collision data}",
    journal        = "Phys. Rev. D",
    volume         = "87",
    year           = "2013",
    pages          = "012008",
    doi            = "10.1103/PhysRevD.87.012008",
    reportNumber   = "CERN-PH-EP-2012-195",
    eprint         = "1208.0949",
    archivePrefix  = "arXiv",
    primaryClass   = "hep-ex",
}

@Article{DAPR-2012-01,
    author         = "{ATLAS Collaboration}",
    title          = "{Characterisation and mitigation of beam-induced backgrounds observed in the ATLAS detector during the 2011 proton--proton run}",
    journal        = "JINST",
    volume         = "8",
    year           = "2013",
    pages          = "P07004",
    doi            = "10.1088/1748-0221/8/07/P07004",
    reportNumber   = "CERN-PH-EP-2012-313",
    eprint         = "1303.0223",
    archivePrefix  = "arXiv",
    primaryClass   = "hep-ex",
}

@Article{PERF-2012-04,
    author         = "{ATLAS Collaboration}",
    title          = "{Performance of \(b\)-jet identification in the ATLAS experiment}",
    journal        = "JINST",
    volume         = "11",
    year           = "2016",
    pages          = "P04008",
    doi            = "10.1088/1748-0221/11/04/P04008",
    reportNumber   = "CERN-PH-EP-2015-216",
    eprint         = "1512.01094",
    archivePrefix  = "arXiv",
    primaryClass   = "hep-ex",
}

@Article{STDM-2012-23,
    author         = "{ATLAS Collaboration}",
    title          = "{Measurement of the \(Z/\gamma^*\) boson transverse momentum distribution in \(pp\) collisions at \(\sqrt{s} = 7\,\text{TeV}\) with the ATLAS detector}",
    journal        = "JHEP",
    volume         = "09",
    year           = "2014",
    pages          = "145",
    doi            = "10.1007/JHEP09(2014)145",
    reportNumber   = "CERN-PH-EP-2014-075",
    eprint         = "1406.3660",
    archivePrefix  = "arXiv",
    primaryClass   = "hep-ex",
}

@Article{DAPR-2013-01,
    author         = "{ATLAS Collaboration}",
    title          = "{Luminosity determination in \(pp\) collisions at \(\sqrt{s} = 8\,\text{TeV}\) using the ATLAS detector at the LHC}",
    journal        = "Eur. Phys. J. C",
    volume         = "76",
    year           = "2016",
    pages          = "653",
    doi            = "10.1140/epjc/s10052-016-4466-1",
    eprint         = "1608.03953",
    archivePrefix  = "arXiv",
    primaryClass   = "hep-ex",
}

@Article{EXOT-2013-13,
    author         = "{ATLAS Collaboration}",
    title          = "{Search for new phenomena in final states with an energetic jet and large missing transverse momentum in \(pp\) collisions at \(\sqrt{s} = 8\,\text{TeV}\) with the ATLAS detector}",
    journal        = "Eur. Phys. J. C",
    volume         = "75",
    year           = "2015",
    pages          = "299",
    doi            = "10.1140/epjc/s10052-015-3517-3",
    reportNumber   = "CERN-PH-EP-2014-299",
    eprint         = "1502.01518",
    archivePrefix  = "arXiv",
    primaryClass   = "hep-ex",
    related        = "EXOT-2013-13-err",
    relatedstring  = "Erratum:",
}

@Article{PERF-2013-06,
    author         = "{ATLAS Collaboration}",
    title          = "{Identification and energy calibration of hadronically decaying tau leptons with the ATLAS experiment in \(pp\) collisions at \(\sqrt{s} = 8\,\text{TeV}\)}",
    journal        = "Eur. Phys. J. C",
    volume         = "75",
    year           = "2015",
    pages          = "303",
    doi            = "10.1140/epjc/s10052-015-3500-z",
    reportNumber   = "CERN-PH-EP-2014-227",
    eprint         = "1412.7086",
    archivePrefix  = "arXiv",
    primaryClass   = "hep-ex",
}

@Article{PERF-2014-03,
    author         = "{ATLAS Collaboration}",
    title          = "{Performance of pile-up mitigation techniques for jets in \(pp\) collisions at \(\sqrt{s} = 8\,\text{TeV}\) using the ATLAS detector}",
    journal        = "Eur. Phys. J. C",
    volume         = "76",
    year           = "2016",
    pages          = "581",
    doi            = "10.1140/epjc/s10052-016-4395-z",
    reportNumber   = "CERN-PH-EP-2015-206",
    eprint         = "1510.03823",
    archivePrefix  = "arXiv",
    primaryClass   = "hep-ex",
}

@Article{PERF-2014-07,
    author         = "{ATLAS Collaboration}",
    title          = "{Topological cell clustering in the ATLAS calorimeters and its performance in LHC Run~1}",
    journal        = "Eur. Phys. J. C",
    volume         = "77",
    year           = "2017",
    pages          = "490",
    doi            = "10.1140/epjc/s10052-017-5004-5",
    reportNumber   = "CERN-PH-EP-2015-304",
    eprint         = "1603.02934",
    archivePrefix  = "arXiv",
    primaryClass   = "hep-ex",
}

@Article{PERF-2015-10,
    author         = "{ATLAS Collaboration}",
    title          = "{Muon reconstruction performance of the ATLAS detector in proton--proton collision data at \(\sqrt{s} = 13\,\text{TeV}\)}",
    journal        = "Eur. Phys. J. C",
    volume         = "76",
    year           = "2016",
    pages          = "292",
    doi            = "10.1140/epjc/s10052-016-4120-y",
    reportNumber   = "CERN-EP-2016-033",
    eprint         = "1603.05598",
    archivePrefix  = "arXiv",
    primaryClass   = "hep-ex",
}

@Article{PERF-2016-05,
    author         = "{ATLAS Collaboration}",
    title          = "{Measurements of \(b\)-jet tagging efficiency with the ATLAS detector using \(t\bar{t}\) events at \(\sqrt{s} = 13\,\text{TeV}\)}",
    journal        = "JHEP",
    volume         = "08",
    year           = "2018",
    pages          = "089",
    doi            = "10.1007/JHEP08(2018)089",
    reportNumber   = "CERN-EP-2018-047",
    eprint         = "1805.01845",
    archivePrefix  = "arXiv",
    primaryClass   = "hep-ex",
}

@Article{PERF-2016-07,
    author         = "{ATLAS Collaboration}",
    title          = "{Performance of missing transverse momentum reconstruction with the ATLAS detector using proton--proton collisions at \(\sqrt{s} = 13\,\text{TeV}\)}",
    journal        = "Eur. Phys. J. C",
    volume         = "78",
    year           = "2018",
    pages          = "903",
    doi            = "10.1140/epjc/s10052-018-6288-9",
    reportNumber   = "CERN-EP-2017-274",
    eprint         = "1802.08168",
    archivePrefix  = "arXiv",
    primaryClass   = "hep-ex",
}

@Article{TRIG-2016-01,
    author         = "{ATLAS Collaboration}",
    title          = "{Performance of the ATLAS trigger system in 2015}",
    journal        = "Eur. Phys. J. C",
    volume         = "77",
    year           = "2017",
    pages          = "317",
    doi            = "10.1140/epjc/s10052-017-4852-3",
    reportNumber   = "CERN-EP-2016-241",
    eprint         = "1611.09661",
    archivePrefix  = "arXiv",
    primaryClass   = "hep-ex",
}

@Article{PERF-2017-03,
    author         = "{ATLAS Collaboration}",
    title          = "{Electron and photon energy calibration with the ATLAS detector using 2015--2016 LHC proton--proton collision data}",
    journal        = "JINST",
    volume         = "14",
    year           = "2019",
    pages          = "P03017",
    doi            = "10.1088/1748-0221/14/03/P03017",
    reportNumber   = "CERN-EP-2018-296",
    eprint         = "1812.03848",
    archivePrefix  = "arXiv",
    primaryClass   = "hep-ex",
}

@Article{EGAM-2018-01,
    author         = "{ATLAS Collaboration}",
    title          = "{Electron and photon performance measurements with the ATLAS detector using the 2015--2017 LHC proton--proton collision data}",
    journal        = "JINST",
    volume         = "14",
    year           = "2019",
    pages          = "P12006",
    doi            = "10.1088/1748-0221/14/12/P12006",
    reportNumber   = "CERN-EP-2019-145",
    eprint         = "1908.00005",
    archivePrefix  = "arXiv",
    primaryClass   = "hep-ex",
}

@Article{HIGG-2018-54,
    author         = "{ATLAS Collaboration}",
    title          = "{Combination of Searches for Invisible Higgs Boson Decays with the ATLAS Experiment}",
    journal        = "Phys. Rev. Lett.",
    volume         = "122",
    year           = "2019",
    pages          = "231801",
    doi            = "10.1103/PhysRevLett.122.231801",
    reportNumber   = "CERN-EP-2019-046",
    eprint         = "1904.05105",
    archivePrefix  = "arXiv",
    primaryClass   = "hep-ex",
}

@Article{JETM-2018-05,
    author         = "{ATLAS Collaboration}",
    title          = "{Jet energy scale and resolution measured in proton--proton collisions at \(\sqrt{s} = 13\,\text{TeV}\) with the ATLAS detector}",
    year           = "2020",
    reportNumber   = "CERN-EP-2020-083",
    eprint         = "2007.02645",
    archivePrefix  = "arXiv",
    primaryClass   = "hep-ex",
}

@Article{PIX-2018-001,
    author         = "Abbott, B. and others",
    title          = "{Production and integration of the ATLAS Insertable B-Layer}",
    journal        = "JINST",
    volume         = "13",
    year           = "2018",
    pages          = "T05008",
    doi            = "10.1088/1748-0221/13/05/T05008",
    eprint         = "1803.00844",
    archivePrefix  = "arXiv",
    primaryClass   = "physics.ins-det",
}

@Article{TRIG-2019-01,
    author         = "{ATLAS Collaboration}",
    title          = "{Performance of the missing transverse momentum triggers for the ATLAS detector during Run-2 data taking}",
    journal        = "JHEP",
    volume         = "08",
    year           = "2020",
    pages          = "080",
    doi            = "10.1007/JHEP08(2020)080",
    reportNumber   = "CERN-EP-2020-050",
    eprint         = "2005.09554",
    archivePrefix  = "arXiv",
    primaryClass   = "hep-ex",
}

@Report{ATLAS-TDR-2010-19,
    author         = "{ATLAS Collaboration}",
    title          = "{ATLAS Insertable B-Layer Technical Design Report}",
    type           = "ATLAS-TDR-19; CERN-LHCC-2010-013",
    year           = "2010",
    url            = "https://cds.cern.ch/record/1291633",
}

@Article{CMS-HIG-17-023,
    author         = "{CMS Collaboration}",
    title          = "{Search for invisible decays of a Higgs boson produced through vector boson fusion in proton--proton collisions at \(\sqrt{s} = 13\,\text{TeV}\)}",
    journal        = "Phys. Lett. B",
    volume         = "793",
    year           = "2019",
    pages          = "520",
    doi            = "10.1016/j.physletb.2019.04.025",
    reportNumber   = "CERN-EP-2018-139",
    eprint         = "1809.05937",
    archivePrefix  = "arXiv",
    primaryClass   = "hep-ex",
}

@Booklet{ATLAS-CONF-2014-018,
    author         = "{ATLAS Collaboration}",
    title          = "{Tagging and suppression of pileup jets with the ATLAS detector}",
    howpublished   = "{ATLAS-CONF-2014-018}",
    url            = "https://cds.cern.ch/record/1700870",
    year           = "2014",
}

@Booklet{ATLAS-CONF-2015-029,
    author         = "{ATLAS Collaboration}",
    title          = "{Selection of jets produced in \(13~\text{TeV}\) proton--proton collisions with the ATLAS detector}",
    howpublished   = "{ATLAS-CONF-2015-029}",
    url            = "https://cds.cern.ch/record/2037702",
    year           = "2015",
}

@Booklet{ATLAS-CONF-2017-029,
    author         = "{ATLAS Collaboration}",
    title          = "{Measurement of the tau lepton reconstruction and identification performance in the ATLAS experiment using \(pp\) collisions at \(\sqrt{s} = 13~\text{TeV}\)}",
    howpublished   = "{ATLAS-CONF-2017-029}",
    url            = "https://cds.cern.ch/record/2261772",
    year           = "2017",
}

@Booklet{ATL-PHYS-PUB-2014-021,
    author         = "{ATLAS Collaboration}",
    title          = "{ATLAS Pythia~8 tunes to \(7~\text{TeV}\) data}",
    howpublished   = "{ATL-PHYS-PUB-2014-021}",
    url            = "https://cds.cern.ch/record/1966419",
    year           = "2014",
}

@Booklet{ATL-PHYS-PUB-2015-045,
    author         = "{ATLAS Collaboration}",
    title          = "{Reconstruction, Energy Calibration, and Identification of Hadronically Decaying Tau Leptons in the ATLAS Experiment for Run-2 of the LHC}",
    howpublished   = "{ATL-PHYS-PUB-2015-045}",
    url            = "https://cds.cern.ch/record/2064383",
    year           = "2015",
}

@Booklet{ATL-PHYS-PUB-2016-012,
    author         = "{ATLAS Collaboration}",
    title          = "{Optimisation of the ATLAS \(b\)-tagging performance for the 2016 LHC Run}",
    howpublished   = "{ATL-PHYS-PUB-2016-012}",
    url            = "https://cds.cern.ch/record/2160731",
    year           = "2016",
}

@Booklet{ATL-PHYS-PUB-2016-017,
    author         = "{ATLAS Collaboration}",
    title          = "{The Pythia~8 A3 tune description of ATLAS minimum bias and inelastic measurements incorporating the Donnachie--Landshoff diffractive model}",
    howpublished   = "{ATL-PHYS-PUB-2016-017}",
    url            = "https://cds.cern.ch/record/2206965",
    year           = "2016",
}

@Booklet{ATL-PHYS-PUB-2016-020,
    author         = "{ATLAS Collaboration}",
    title          = "{Studies on top-quark Monte Carlo modelling for Top2016}",
    howpublished   = "{ATL-PHYS-PUB-2016-020}",
    url            = "https://cds.cern.ch/record/2216168",
    year           = "2016",
}

@Booklet{ATL-PHYS-PUB-2017-007,
    author         = "{ATLAS Collaboration}",
    title          = "{Studies on top-quark Monte Carlo modelling with Sherpa and MG5\_aMC@NLO}",
    howpublished   = "{ATL-PHYS-PUB-2017-007}",
    url            = "https://cds.cern.ch/record/2261938",
    year           = "2017",
}
\clearpage
 
\begin{flushleft}
\hypersetup{urlcolor=black}
{\Large The ATLAS Collaboration}

\bigskip

\AtlasOrcid[0000-0002-6665-4934]{G.~Aad}$^\textrm{\scriptsize 102}$,    
\AtlasOrcid[0000-0002-5888-2734]{B.~Abbott}$^\textrm{\scriptsize 128}$,    
\AtlasOrcid[0000-0002-7248-3203]{D.C.~Abbott}$^\textrm{\scriptsize 103}$,    
\AtlasOrcid[0000-0002-2788-3822]{A.~Abed~Abud}$^\textrm{\scriptsize 36}$,    
\AtlasOrcid[0000-0002-1002-1652]{K.~Abeling}$^\textrm{\scriptsize 53}$,    
\AtlasOrcid[0000-0002-2987-4006]{D.K.~Abhayasinghe}$^\textrm{\scriptsize 94}$,    
\AtlasOrcid[0000-0002-8496-9294]{S.H.~Abidi}$^\textrm{\scriptsize 167}$,    
\AtlasOrcid[0000-0002-8279-9324]{O.S.~AbouZeid}$^\textrm{\scriptsize 40}$,    
\AtlasOrcid{N.L.~Abraham}$^\textrm{\scriptsize 156}$,    
\AtlasOrcid[0000-0001-5329-6640]{H.~Abramowicz}$^\textrm{\scriptsize 161}$
% ...
\bigskip
\\

% ...
$^{36}$CERN, Geneva; Switzerland.\\
% ...
$^{40}$Niels Bohr Institute, University of Copenhagen, Copenhagen; Denmark.\\
% ...
$^{53}$II. Physikalisches Institut, Georg-August-Universit\"{a}t G\"ottingen, G\"ottingen; Germany.\\
% ...
$^{94}$Department of Physics, Royal Holloway University of London, Egham; United Kingdom.\\
% ...
$^{102}$CPPM, Aix-Marseille Universit\'e, CNRS/IN2P3, Marseille; France.\\
$^{103}$Department of Physics, University of Massachusetts, Amherst MA; United States of America.\\
% ...
$^{128}$Homer L. Dodge Department of Physics and Astronomy, University of Oklahoma, Norman OK; United States of America.\\
% ...
$^{156}$Department of Physics and Astronomy, University of Sussex, Brighton; United Kingdom.\\
% ...
$^{161}$Raymond and Beverly Sackler School of Physics and Astronomy, Tel Aviv University, Tel Aviv; Israel.\\
% ...
$^{167}$Department of Physics, University of Toronto, Toronto ON; Canada.\\
% ...

\end{flushleft}

\FloatBarrier  
 
\end{document}